\documentclass[12pt]{article}
\usepackage{amssymb,amscd,array}
\usepackage[active]{srcltx}
\catcode `\@=11
\@addtoreset{equation}{section}

\newtheorem{thm}{Theorem}[section]

\newtheorem{lemma}[thm]{Lemma}
\newtheorem{prop}[thm]{Proposition}

\def\qed{\blacksquare}
\newcommand{\be}{\begin{equation}}
\newcommand{\ee}{\end{equation}}
\newcommand{\bea}{\begin{eqnarray}}
\newcommand{\eea}{\end{eqnarray}}
\newcommand{\R}{\mathbb{R}}

\newcommand{\C}{\mathbb{C}}

\begin{document}
\begin{titlepage}

\begin{center}
{\bf \Large{Causal Perturbative Quantum Field Theory and the Standard Model\\}}
\end{center}
\vskip 1.0truecm
\centerline{D. R. Grigore, 
\footnote{e-mail: grigore@theory.nipne.ro}}
\vskip5mm
\centerline{Department of Theoretical Physics,}
\centerline{Institute for Physics and Nuclear Engineering ``Horia Hulubei"}
\centerline{Bucharest-M\u agurele, P. O. Box MG 6, ROM\^ANIA}

\vskip 2cm
\bigskip \nopagebreak
\vskip 1cm
\begin{abstract}
\noindent
We consider the general framework of perturbative quantum field theory for the general Yang-Mills model including massless and massive vector fields and also scalar and Dirac fields. We describe the chronological products using Wick submonomials and give rigorous proofs of gauge invariance for tree and loop contributions in the second order of the perturbation theory.
\end{abstract}

\end{titlepage}

\section{Introduction}

The most natural way to arrive at the Bogoliubov axioms of perturbative quantum field theory (pQFT) is by analogy with non-relativistic 
quantum mechanics \cite{Gl}, \cite{H}. The evolution operator in non-relativistic quantum mechanics verifies
\bea
{d \over dt}U(t,s) = - i V_{\rm int}(t) U(t,s); \qquad U(s,s) = I
\eea
in terms of the interaction potential and can be expressed as follows:
\bea
U(t,s) \equiv \sum {(-i)^{n}\over n!} \int dt_{1} \cdots dt_{n}
T(t_{1},\dots,t_{n})
\label{u}
\eea
where the {\it chronological products}
$
T_{n}(t_{1},\dots,t_{n})
$
verify the following properties:
\begin{itemize}
\item
{\bf Initial condition}
\bea
T(t_{1}) = V_{\rm int}(t_{1}).
\eea
\item
{\bf Symmetry}
\be
T_{2}(t_{1},t_{2}) = ( 1 \leftrightarrow 2).
\ee
\item
{\bf Causality}:
\be
T_{2}(t_{1},t_{2}) = T_{1}(t_{1})~T_{1}(t_{2}),\qquad {\rm for} \quad t_{1} > t_{2}
\ee
and a similar formula in general.
\item
{\bf Unitary}
\bea
U(t,s)^{\dagger}~U(t,s) = I
\eea
which can be easily expressed in terms of chronological and anti-chronological products.
\item
{\bf Invariance properties}

If the interaction potential is translation invariant then we have
\be
T_{n}(t_{1} + \tau,\dots,t_{n} + \tau) = T_{n}(t_{1},\dots,t_{n}).
\ee
\end{itemize}

We can write an explicit formula
\bea
T_{2}(t_{1},t_{2}) =
\theta(t_{1} - t_{2})~V_{int}(t_{1})~V_{int}(t_{2})
+ \theta(t_{2} - t_{1})~V_{int}(t_{2})~V_{int}(t_{1}).
\eea

The purpose of perturbative quantum field theory (pQFT) is to generalize this idea in the relativistic context, especially the causality property. Basically one replaces in a consistent way
$
t_{1},\dots,t_{n}
$
by variables from the Minkowski space
$
x_{1},\dots,x_{n}
$
in a consistent way. Instead of (\ref{u}) we will have a formal series:
\bea
S \equiv \sum {i^{n}\over n!} \int dx_{1}
\cdots dx_{n}~g(x_{1} \dots g(x_{n})~
T(x_{1},\dots,x_{n})
\label{uS}
\eea
where
$
g_{1},\dots,g_{n}
$
are test functions.

In this way one arrives naturally at Bogoliubov axioms \cite{BS}, \cite{EG}, \cite{Sc1}, \cite{Sc2}. We prefer the formulation from \cite{DF} ( see also \cite{algebra} and \cite{wick+hopf}):
for every set of monomials 
$ 
A_{1}(x_{1}),\dots,A_{n}(x_{n}) 
$
in some jet variables (associated to some classical field theory) one associates the operator-valued distributions
$ 
T^{A_{1},\dots,A_{n}}(x_{1},\dots,x_{n})
$  
called chronological products; it will be convenient to use another notation: 
$ 
T(A_{1}(x_{1}),\dots,A_{n}(x_{n}))
$
and we have to generalize in the natural way the properties of
$
T(t_{1},\dots,t_{n})
$
namely:
\begin{itemize}
\item 
(skew)symmetry properties in the entries 
$ 
A_{1}(x_{1}),\dots,A_{n}(x_{n}) 
$;
\item
Poincar\'e invariance; 
\item
causality: here one has to use the natural causality from the Minkowski space, expressing the fact that a point
$x$ succeeds causally the point $y$ (the standard notations being
$
x \succeq y);
$
\item
unitarity; 
\item
the ``initial condition" which says that
$
T(A(x)) 
$
is a Wick monomial.
\end{itemize}

So we need some basic notions on free fields and Wick monomials. One can supplement these axioms by requiring 
\begin{itemize}
\item 
power counting;
\item
Wick expansion property. 
\end{itemize}
We refer to \cite{wick+hopf} for details.

It is a highly non-trivial problem to find solutions for the Bogoliubov axioms, even in the simplest case of a real scalar field. 

The procedure of Epstein and Glaser is a recursive construction for the basic objects
$ 
T(A_{1}(x_{1}),\dots,A_{n}(x_{n}))
$
and reduces the induction procedure to a distribution splitting of some distributions with causal support.  
In an equivalent way, one can reduce the induction procedure to the process of extension of distributions \cite{PS}. 

An equivalent point of view uses retarded products \cite{St1} instead of chronological products. For gauge models one has to deal with 
non-physical fields (the so-called ghost fields) and impose a supplementary axiom namely  gauge invariance, which guarantees that the 
physical states are left invariant by the chronological products.

We only remind the form of the Wick theorem which we will use here. We consider the classical field theory of a real scalar on the Minkowski space
$
{\cal M} \simeq \R^{4}
$
(with variables
$
x^{\mu}, \mu = 0,\dots,3
$
and the metric $\eta$ with
$
diag(\eta) = (1,-1,-1,-1).
$
The scalar field is described by the bundle
$
{\cal M} \times \R
$
with coordinated
$
(x^{\mu},\phi).
$
The first jet-bundle extension is
$$
J^{1}({\cal M}, \R) \simeq {\cal M} \times \R \times \R^{4}
$$
with coordinates
$
(x^{\mu}, \phi, \phi_{\mu}),~\mu = 0,\dots,3.
$

If
$
\varphi: \cal M \rightarrow \R
$
is a smooth function we can associate a new smooth function
$
j^{1}\varphi: {\cal M} \rightarrow J^{1}(\cal M, \R)
$
according to
$
j^{1}\varphi(x) = (x^{\mu}, \varphi(x), \partial_{\mu}\varphi(x)).
$

For higher order jet-bundle extensions we have to add new real variables
$
\phi_{\{\mu_{1},\dots,\mu_{r}\}}
$
considered completely symmetric in the indexes and associated to higher derivatives. We make the convention
$
\phi_{\emptyset} = \phi.
$
In classical field theory the jet-bundle extensions
$
j^{r}\varphi(x)
$
do verify Euler-Lagrange equations. To write them we need the formal derivatives defined by
\be
d_{\nu}\phi_{\{\mu_{1},\dots,\mu_{r}\}} \equiv \phi_{\{\nu,\mu_{1},\dots,\mu_{r}\}}.
\ee
One can extend in a natural way this construction to other fields: essentially the variable
$
\phi
$
will get additional index, says
$
\phi_{a}
$
and some transformation properties with respect to some symmetry group. Then the derivative variables will be denoted by
$
\phi_{a,{\{\mu_{1},\dots,\mu_{r}\}}}.
$

If $A$ is some monomial in the variables
$
\phi_{\{\mu_{1},\dots,\mu_{r}\}}
$
there is a canonical way to associate to $A$ a Wick monomial: we first define the quantum real scalar as a distribution-valued operator acting in the Fock space associated to the representation
$
[m,0]
$
of the Poincar\'e group. Using the reconstruction theorem, such a quantum field can described by the $2$-point function
\be
<\Omega, \phi^{\rm quant}(x), \phi^{\rm quant}(y) \Omega> = - i~D_{m}^{(+)}(x - y)\times {\bf 1}.
\label{2-point}
\ee
Here
\be
D_{m}(x) = D_{m}^{(+)}(x) + D_{m}^{(-)}(x)
\label{pj}
\ee
is the causal Pauli-Jordan distribution; we understand by
$
D^{(\pm)}_{m}(x)
$
the positive and negative parts of
$
D_{m}(x)
$.
The explicit formulas are:
\be
D_{m}^{(\pm)}(x) =
\pm {i \over (2\pi)^{3}}~\int dp e^{- i p\cdot x} \theta(\pm p_{0}) \delta(p^{2} - m^{2}).
\label{pjpm}
\ee
The attribute ``causal" is due to the fact that the support of
$
D_{m}
$
is inside the causal Minkowski cones:
$
V^{+} \cup V_{-}
$
where
\be
V^{+} \equiv \{ x \in {\cal M} | \quad x^{2} \geq 0,\quad x^{0} > 0\},\qquad
V^{-} \equiv \{ x \in {\cal M} | \quad x^{2} \geq 0,\quad x^{0} < 0\}.
\ee
If it is clear from the context, we can skip the superscript ``quant".
From (\ref{2-point}) we have
\be
[ \phi(x), \phi(y) ] = - i~ D_{m}(x - y) \times {\bf 1}.
\label{c-scalar}
\ee
Because we have the {\it Klein-Gordon} equations
\be
(\square + m^{2})~D_{m}^{(\pm)} = 0, \qquad (\square + m^{2})~D_{m} = 0
\label{KGpj}
\ee
we can derive also the Klein-Gordon equation for the quantum field:
\be
(\square + m^{2})~\phi(x) = 0.
\label{KG}
\ee

Let us present the more precise form of the Wick theorem that we will use in this paper \cite{wick+hopf}. For simplicity we consider the case of the real scalar field and take
$
A_{1},\cdots,A_{n}
$
to be monomials in the classical field
$
\phi
$
but not in derivatives. Then Wick property means that we can choose the chronological products such that:
\bea
[ \phi(y), T(A_{1}(x_{1}),\dots,A_{n}(x_{n}))  ]
\nonumber\\
= - i~\sum_{m=1}^{n} D_{m}(y - x_{m})~
T(A_{1}(x_{1}),\dots,{\partial\over \partial \phi}  A_{m}(x_{m}),\dots,A_{n}(x_{n}))
\label{comm-wick}
\eea
If the interaction Lagrangean is, as a classical object depending on the jet variables:
\be
A = {1\over 3!}~\phi^{3}
\ee
then it is convenient to define the Wick submonomial
\be
C \equiv {\partial\over \partial \phi} A = {1\over 2} \phi^{2}
\ee
and one can prove that the chronological products can be chosen such that
we have
\bea
T(A(x),A_{2}(x_{2}), \dots,A_{n}(x_{n})) =
:A(x)~T(A_{2}(x_{2}), \dots,A_{n}(x_{n})):
\nonumber\\
+ :\phi(x)~T(C(x),A_{2}(x_{2}), \dots,A_{n}(x_{n})):
\nonumber\\
+ :C(x)~T(\phi(x),A_{2}(x_{2}), \dots,A_{n}(x_{n})):
+ T_{0}(A(x),A_{2}(x_{2}), \dots,A_{n}(x_{n}))
\eea
where the expression
$
T_{0}
$
is of Wick type only in the entries
$
A_{2},\dots,A_{n}.
$

One can iterate this formula and obtain in the end only expressions of the type
$$
T_{0\cdots 0}(B_{1}(x_{1}),B_{2}(x_{2}), \dots,B_{n}(x_{n}))
$$
with no Wick property so they must be vacuum averages of the corresponding chronological product. For instance, in the second order of the perturbation theory:
\bea
T(A(x),A(y)) = :A(x)A(y): + T_{00}(C(x),C(y))~:\phi(x)\phi(y):
\nonumber\\
+ T_{00}(\phi(x),\phi(y))~:C(x)C(y): + T_{00}(A(x),A(y))~{\bf 1}
\label{wick}
\eea
where the expressions
$
T_{00}
$
are vacuum averages. We will also use the notations from \cite{wick+hopf}
$
T(A(x)^{(1)},A(y)^{(1)})
$
and
$
T(A(x)^{(2)},A(y)^{(2)})
$
for the second (resp. third) term in the right hand side.

In the next Section we will present the fields used for the standard model and give the general expresion of the interaction Lagrangean (in a multi-Higgs setting). In Section \ref{distributions} we describe some distributions with causal support needed for the explicit computation of the previous expression. In Section \ref{loop} we consider the second term from the previous expresion (loops contributions) for the chronological products and prove that there are no anomalies. In Section \ref{tree} we consider third term from the previous expression (tree  contributions), compute the anomalies and show what restrictions follows from their elimination.

We mainly generalize the results from \cite{standard}, \cite{fermi} and \cite{co} using the methods from \cite{wick+hopf}.
\newpage
\section{The Fields of the Standard Model\label{ym}}

The fields appearing in the standard model are: scalar fields, Dirac fields and vector field (massless and massive) of spin
$
0,1/2,1
$
respectively. We denote by
$
I_{1},\dots,I_{4}
$
some index sets. At the level of classical field theory we have the real scalar fields
$
\phi_{j}, j \in I_{3}
$
and their jet bundle extensions; when we consider their quantum counterparts we have, similarly with (\ref{2-point})
\be
<\Omega,\phi_{j}(x), \phi_{k}(y) \Omega> = - i~\delta_{jk}
D^{(+)}_{m_{j}}(x - y) \times {\bf 1}.
\label{2-scalars}
\ee
and also the Klein-Gordon equation
\be
(\square + m^{2})~\phi_{j}(x) = 0.
\label{KGs}
\ee
The Dirac field is denoted by
$
\psi_{\alpha},~\alpha = 1,\dots,4
$
and is regarded as a column matrix with four entries: it is a field living in
$
\C^{4}.
$
We will also need the Dirac matrices, which are acting in this space and verify
\be
\{ \gamma_{\mu},\gamma_{\nu}\} = 2~\eta_{\mu\nu}.
\ee
One also needs the fields
$
\bar{\psi}_{\alpha},~\alpha = 1,\dots,4
$
which can be organized as line matrices and are defined by
\be
\bar{\psi} \equiv \psi^{\dagger}\gamma_{0}.
\ee
The quantum counterpart is living in the antisymmetric Fock space associated to the representation
$
[M,1/2]
$
of the Poincar\'e group and can described by the 2-point functions
\bea
<\Omega, \psi_{\alpha}(x_{1}) \bar{\psi}_{\beta}(x_{2})\Omega> =
- i~S_{M}^{(+)}(x_{1} - x_{2})_{\alpha\beta}
\nonumber \\
<\Omega, \bar{\psi}_{\alpha}(x_{1}) \psi_{\beta}(x_{2})\Omega> =
- i~S_{M}^{(-)}(x_{2} - x_{1})_{\beta\alpha}
\label{2-dirac}
\eea
where
\be
S_{M}^{(\epsilon)} \equiv
(i~\gamma^{\mu}~\partial_{\mu} + M)~D_{M}^{(\epsilon)}.
\label{S}
\ee
We also can prove the validity of the Dirac equation
\be
i~\gamma^{\mu}~\partial_{\mu} \psi(x) = M \psi(x)~\qquad\Longleftrightarrow
\qquad
i~\partial_{\mu}\bar{\psi}(x)~\gamma^{\mu} = - M~\bar{\psi}(x)
\label{Dirac}
\ee

When there are more Dirac fields
$
\psi_{A},A = 1,\dots,N
$
we have
\bea
<\Omega, \psi_{A\alpha}(x_{1}) \bar{\psi}_{B\beta}(x_{2})\Omega> =
- i~\delta_{AB}~S_{A}^{(+)}(x_{1} - x_{2})_{\alpha\beta}
\nonumber \\
<\Omega, \bar{\psi}_{A\alpha}(x_{1}) \psi_{B\beta}(x_{2})\Omega> =
- i~\delta_{AB}~S_{A}^{(-)}(x_{2} - x_{1})_{\beta\alpha}
\label{2-diracs}
\eea
where
$
S_{A}^{(\pm)} = S_{M_{A}}^{(\pm)}, A \in I_{4}
$
and $M$ is becomes a diagonal
$
N \times N
$
matrix:
$
M_{AB} = \delta_{AB}~M_{A}.
$

To describe the vector fields one needs, according to Faddeeev and Popov, ghost fields; such fields are described at the classical level by odd Grassmann variables and we had to insert in all formulas the appropriate Fermi signs. If we consider the case of a massless vector field, then the jet variables are
$
(v^{\mu}, u, \tilde{u})
$
where
$
v^{\mu}
$
is Grassmann even and
$
u, \tilde{u}
$
are Grassmann odd variables. In this jet space we can define the {\it gauge charge} operator by
\be
d_{Q} v^{\mu} = i~d^{\mu}u,\qquad
d_{Q} u = 0,\qquad
d_{Q} \tilde{u} = - i~d_{\mu}v^{\mu}
\label{Qzero}
\ee
where
$
d^{\mu}
$
is the formal derivative. One can prove that
\be
d_{Q}^{2} \cong  0.
\label{dQ2}
\ee
where $\cong$ means ``modulo the equations of motion". The reason for this choice of
$
d_{Q}
$
comes when we consider the quantum counterparts. We define the associated Fock space by the non-zero $2$-point distributions are
\bea
<\Omega, v^{\mu}(x_{1}) v^{\nu}(x_{2})\Omega> =
i~\eta^{\mu\nu}~D_{0}^{(+)}(x_{1} - x_{2}),
\nonumber \\
<\Omega, u(x_{1}) \tilde{u}(x_{2})\Omega> = - i~D_{0}^{(+)}(x_{1} - x_{2}),
\nonumber\\
<\Omega, \tilde{u}(x_{1}) u(x_{2})\Omega> = i~D_{0}^{(+)}(x_{1} - x_{2}).
\label{2-massless-vector}
\eea

The quantum gauge charge is then defined by:
\be
~[Q, v^{\mu}] = i~\partial^{\mu}u,\qquad
\{ Q, u \} = 0,\qquad
\{Q, \tilde{u} \} = - i~\partial_{\mu}v^{\mu}, \qquad
Q~\Omega = 0.
\label{QQzero}
\ee
One can prove that
\be
Q^{2} = 0
\label{Q2}
\ee
and that the cohomology space
$
Ker(Q)/Ran(Q)
$
is naturally isomorphic to the Fock space of particles of zero mass and spin $1$ i.e. associated with the representation
$
[0,1]
$
of the Poincar\'e group.

In the case of more massless vector particles we have the variables
$
(v^{\mu}_{a}, u_{a}, \tilde{u}_{a}),~a \in I_{1}
$
where 
$
v^{\mu}_{a}
$
are Grassmann even and 
$
u_{a}, \tilde{u}_{a}
$
are Grassmann odd variables. Then the gauge charge operator is given by a generalization of (\ref{Qzero})
\be
d_{Q} v^{\mu}_{a} = i~d^{\mu}u_{a},\qquad
d_{Q} u_{a} = 0,\qquad
d_{Q} \tilde{u}_{a} = - i~d_{\mu}v^{\mu}_{a}, ~a \in I_{1}
\label{Qzeros}
\ee
so we still have (\ref{dQ2}). In the quantum case the non-zero $2$-point distributions are
\bea
<\Omega, v^{\mu}_{a} (x_{1}) v^{\nu}_{b} (x_{2})\Omega> =
i~\eta^{\mu\nu}~\delta_{ab}~D_{0}^{(+)}(x_{1} - x_{2}),
\nonumber \\
<\Omega, u_{a} (x_{1}) \tilde{u}_{b} (x_{2})\Omega> = - i~\delta_{ab}~
D_{0}^{(+)}(x_{1} - x_{2}),
\nonumber\\
<\Omega, \tilde{u_{a}}(x_{1}) u_{b} (x_{2})\Omega> = i~~\delta_{ab}~
D_{0}^{(+)}(x_{1} - x_{2})
\label{2-massless-vectors}
\eea
and
\be
[Q, v^{\mu}_{a}] = i~\partial^{\mu}u_{a},\qquad
\{ Q, u_{a} \} = 0,\qquad
\{Q, \tilde{u}_{a} \} = - i~\partial_{\mu}v^{\mu}_{a}, \qquad
Q~\Omega = 0.
\label{QQzeros}
\ee
As above, we have (\ref{Q2}).

For the case of a massive vector field we need a new ghost field which is scalar so the variables are
$
(v^{\mu}, u, \tilde{u}, \Phi)
$
where
$
v^{\mu},\Phi
$
are Grassmann even and
$
u, \tilde{u}
$
are Grassmann odd variables.

The gauge charge operator is in this case
\be
d_{Q} v^{\mu} = i~d^{\mu}u,\qquad
d_{Q} u = 0,\qquad
d_{Q} \tilde{u} = - i~(d_{\mu}v^{\mu} + m~\Phi), \qquad
d_{Q}\Phi = i~m u
\label{Qmassive}
\ee
and, as in the massless case, the gauge charge operator squares to zero - see (\ref{dQ2}).

Now let us consider the quantum counterparts. The quantum fields are
determined by the non-zero $2$-point distributions are
\bea
<\Omega, v^{\mu}(x_{1}) v^{\nu}(x_{2})\Omega> =
i~\eta^{\mu\nu}~D_{m}^{(+)}(x_{1} - x_{2}),
\nonumber \\
<\Omega, u(x_{1}) \tilde{u}(x_{2})\Omega> = - i~D_{m}^{(+)}(x_{1} - x_{2}),
\nonumber\\
<\Omega, \tilde{u}(x_{1}) u(x_{2})\Omega> = i~D_{m}^{(+)}(x_{1} - x_{2})
\nonumber\\
<\Omega, \Phi(x_{1}) \Phi(x_{2})\Omega> = - i~D_{m}^{(+)}(x_{1} - x_{2}).
\label{2-massive-vector}
\eea
The quantum gauge charge is defined by
\be
[Q, v^{\mu}] = i~\partial^{\mu}u,\qquad
\{ Q, u \} = 0,\qquad
\{Q, \tilde{u} \} = - i~(\partial_{\mu}v^{\mu} + m~\Phi)
\nonumber\\
~[Q, \Phi ] = i m u, \qquad Q\Omega = 0
\label{QQmassive}
\ee
and we have ({\ref{Q2}) so one can prove that the cohomology space
$
Ker(Q)/Ran(Q)
$
is naturally isomorphic to the Fock space of particles of mass $m$ and spin $1$ i.e. associated with the representation
$
[m,1]
$
of the Poincar\'e group.

In the case of more massive vector particles we have the variables
$
(v^{\mu}_{a}, u_{a}, \tilde{u}_{a}, \Phi_{a})~a \in I_{2}
$
where
$
v^{\mu}_{a}, \Phi_{a}
$
are Grassmann even and
$
u_{a}, \tilde{u}_{a}
$
are Grassmann odd variables. The gauge charge is
\be
d_{Q} v^{\mu}_{a} = i~d^{\mu}u_{a},\quad
d_{Q} u_{a} = 0,\quad
d_{Q} \tilde{u}_{a} = - i~(d_{\mu}v^{\mu}_{a} + m_{a} \Phi_{a}),
\quad
d_{Q}\Phi_{a} =  i~m_{a} u_{a},\qquad a \in I_{2}
\label{Qmassives}
\ee

The quantum fields are determined by the non-zero $2$-point distributions are
\bea
<\Omega, v^{\mu}_{a}(x_{1}) v^{\nu}_{b}(x_{2})\Omega> =
i~\eta^{\mu\nu}~\delta_{ab}~D_{a}^{(+)}(x_{1} - x_{2}),
\nonumber \\
<\Omega, u_{a}(x_{1}) \tilde{u}_{b}(x_{2})\Omega> = - i~\delta_{ab}~
D_{a}^{(+)}(x_{1} - x_{2}),
\nonumber\\
<\Omega, \tilde{u}_{a}(x_{1}) u_{b}(x_{2})\Omega> = i~\delta_{ab}~
D_{a}^{(+)}(x_{1} - x_{2})
\nonumber\\
<\Omega, \Phi_{a}(x_{1}) \Phi_{b}(x_{2})\Omega> = - i~\delta_{ab}~
D_{a}^{(+)}(x_{1} - x_{2}),\qquad a \in I_{2}
\label{2-massive-vectors}
\eea
where
$
D_{a} \equiv D_{m_{a}}
$
and the quantum gauge charge is
\bea
[Q, v^{\mu}_{a}] = i~\partial^{\mu}u_{a},\quad
\{ Q, u_{a} \} = 0,\quad
\{Q, \tilde{u}_{a} \} = - i~(\partial_{\mu}v^{\mu}_{a} + m~\Phi_{a}),\quad
\nonumber\\
~[Q, \Phi_{a} ] = i m u_{a}, \qquad a \in I_{2}
\nonumber\\
Q\Omega = 0
\label{QQmassives}
\eea
and we have (\ref{Q2}).

It is very convenient to group the cases
$
a \in I_{1}
$
and
$
a \in I_{2}
$
by defining the fields
$
(v^{\mu}_{a}, u_{a}, \tilde{u}_{a}, \Phi_{a})
$
for
$
a \in I_{1} \cup I_{2}
$
with the convention
$
\Phi_{a}, m_{a} = 0,~\forall a \in I_{1}.
$
Then we have in the classical framework
\be
d_{Q} v^{\mu}_{a} = i~d^{\mu}u_{a},\quad
d_{Q} u_{a} = 0,\quad
d_{Q} \tilde{u}_{a} = - i~(d_{\mu}v^{\mu}_{a} + m_{a} \Phi_{a}),
\quad
d_{Q}\Phi_{a} =  i~m_{a} u_{a},\qquad a \in I_{1} \cup I_{2}.
\label{Qmassless+massive}
\ee
so we have (\ref{dQ2}).
The quantum fields are determined by the non-zero $2$-point distributions are
\bea
<\Omega, v^{\mu}_{a}(x_{1}) v^{\nu}_{b}(x_{2})\Omega> =
i~\eta^{\mu\nu}~\delta_{ab}~D_{a}^{(+)}(x_{1} - x_{2}),
\nonumber \\
<\Omega, u_{a}(x_{1}) \tilde{u}_{b}(x_{2})\Omega> = - i~\delta_{ab}~
D_{a}^{(+)}(x_{1} - x_{2}),
\nonumber\\
<\Omega, \tilde{u}_{a}(x_{1}) u_{b}(x_{2})\Omega> = i~\delta_{ab}~
D_{a}^{(+)}(x_{1} - x_{2})
\nonumber\\
<\Omega, \Phi_{a}(x_{1}) \Phi_{b}(x_{2})\Omega> = - i~\delta_{ab}~
D_{a}^{(+)}(x_{1} - x_{2}),\qquad a \in I_{1} \cup I_{2}
\label{2-massless+massive-vectors}
\eea
and the quantum gauge charge is
\bea
[Q, v^{\mu}_{a}] = i~\partial^{\mu}u_{a},\quad
\{ Q, u_{a} \} = 0,\quad
\{Q, \tilde{u}_{a} \} = - i~(\partial_{\mu}v^{\mu}_{a} + m~\Phi_{a}),\quad
\nonumber\\
~[Q, \Phi_{a} ] = i m u_{a}, \qquad a \in I_{1} \cup I_{2},\qquad
Q\Omega = 0
\label{QQmassless+massive}
\eea
and we have (\ref{Q2}).

Now we can define the (classical) interaction Lagrangian by the relative cohomology relation:
\be
d_{Q}T \cong i~d_{\mu}T^{\mu}.
\label{G1}
\ee
We want to determine $T$, up to a relative coboundary, i.e. up to terms which are of the form
\be
T_{\rm trivial} \cong d_{Q}B + i~d_{\mu}B^{\mu}
\ee
where $\cong$ means as above ``modulo the equations of motion". If
$
A = a_{1} \cdots a_{n}
$
is a monomial in the jet bundle variables we define two additive quantities:

- the {\it canonical dimension} by postulating the
\be
\omega(b) = 1,~ \omega(f) = 3/2
\ee
for
$
b = v_{a}^{\mu}, u_{a}, \tilde{u}_{a},\Phi_{a},\phi_{j}
$
and
$
f = \psi_{A\alpha},~\bar{\psi}_{A\alpha}.
$
Also the formal derivative
$
d_{\mu}
$
increases by one unit the canonical dimension of any factor of $A$;

- the {\it ghost number} according to
\be
gh(v_{a}^{\mu}) = 0,~ gh(\Phi_{a}) = 0,~ gh(\phi_{j}) = 0,~
gh(u_{a}) = 1,~ gh(\tilde{u}_{a}) = -1.
\ee

We impose the following conditions: (a)
$T$
and
$
T^{\mu}
$
are trilinear; (b) they are Lorentz covariant; (c) they verify a restriction on the canonical dimension
$
\omega(T), \omega(T^{\mu}) \leq 4;
$
(d)
$
gh(T) = 0,~ gh(T^{\mu}) = 1;
$
(e) the gauge invariance relation (\ref{G1}) is true.
We write a generic form of $T$ as a polynomial in the scalar fields
$
\phi_{j}
$
(and formal derivatives) with the ``coefficients" depending on the gauge fields
$
(v^{\mu}_{a}, u_{a}, \tilde{u}_{a}, \Phi_{a}),~a \in I_{1} \cup I_{2}.
$

\bea
T = t + \phi_{j}~t_{j} + {1 \over 2}~\phi_{j}~\phi_{k}~t_{jk}
+ {1 \over 6}~\lambda_{jkl}\phi_{j}~\phi_{k}~\phi_{l}
+ d_{\mu}\phi_{j}~s_{j}^{\mu} + \phi_{j}~d_{\mu}\phi_{k}~s_{jk}^{\mu}
\nonumber\\
T^{\mu} = t^{\mu} + \phi_{j}~t_{j}^{\mu}
+ {1 \over 2}~\phi_{j}~\phi_{k}~t_{jk}^{\mu}
+ d_{\nu}\phi_{j}~s_{j}^{\mu\nu} + \phi_{j}~d_{\nu}\phi_{k}~s_{jk}^{\mu\nu}
\nonumber\\
T^{\mu\nu} = t^{\mu\nu}
\label{TI}
\eea
with the expressions
$
t^{I}, t^{I}_{j},\dots
$
independent of the scalar fields
$
\phi_{k}
$
and
$
\lambda_{jkl}
$
are constants completely symmetric in all indices. Moreover we assume that
$
t_{jk}, t_{jk}^{\mu}
$
are symmetric in
$
j \leftrightarrow k.
$
Indeed, one starts from the generic form
\bea
T^{I} = t^{I} + \phi_{j}~t^{I}_{j}
+ {1 \over 2}~\phi_{j}~\phi_{k}~t^{I}_{jk}
+ {1 \over 6}~\lambda_{jkl}\phi_{j}~\phi_{k}~\phi_{l}
+ d_{\mu}\phi_{j}~s_{j}^{I,\mu} + \phi_{j}~d_{\mu}\phi_{k}~s_{jk}^{I,\mu}
\nonumber\\
+ {1 \over 2}~\phi_{j}~\phi_{k}~\phi_{k,\mu}~\lambda^{I,\mu}_{jk}
+ d_{\mu}d_{\nu}\phi_{j}~s^{I,\mu\nu}_{j}
\eea
and use the constraints (a) - (d) imposed above to prove that we have in fact the preceding generic form. Now one simplifies the expressions
$
s_{j}^{\mu,\nu},~s_{jk}^{\mu,\nu}.
$
First, one notices that the anti-symmetric part
$
s_{j}^{[\mu\nu]}
$
produces a total divergence
\bea
\phi_{j,\nu}~s_{j}^{[\mu\nu]} = d_{\nu} (\phi_{j}~s_{j}^{[\mu\nu]}) -
\phi_{j}~d_{\nu}s_{j}^{[\mu\nu]}
\nonumber
\eea
so by eliminating the total derivative and redefining
$
t_{j}
$
we can make
$
s_{j}^{\mu,\nu}
$
symmetric in
$
\mu \leftrightarrow \nu.
$
The analysis of
$
s_{jk}^{\mu,\nu}
$
is more subtle. We consider the coefficient of
$
\phi_{j,\mu}~\phi_{k,\nu}
$
from the relation (\ref{G1}) and obtain
\bea
s_{jk}^{[\mu\nu]} = j \leftrightarrow k, \qquad
s_{jk}^{\{\mu\nu\}} = - (j \leftrightarrow k).
\nonumber
\eea
So we have
\bea
\phi_{j}~\phi_{k,\nu}~s_{jk}^{[\mu\nu]} =
{1\over 2}~d_{\nu} (\phi_{j}~\phi_{k}s_{jk}^{[\mu\nu]})
- {1\over 2}~\phi_{j}~\phi_{k}~d_{\nu}s_{jk}^{[\mu\nu]}.
\nonumber
\eea
By eliminating the total derivative and redefining
$
t^{\mu}_{jk}
$
we can make
$
s_{jk}^{\mu,\nu}
$
symmetric in
$
\mu \leftrightarrow \nu.
$
If we consider the restrictions on the canonical dimension and on ghost number one has in fact
$
s_{j}^{\mu\,\nu},~s_{jk}^{\mu,\nu} \sim \eta^{\mu\nu}
$
so (\ref{TI}) becomes
\bea
T = t + \phi_{j}~t_{j} + {1 \over 2}~\phi_{j}~\phi_{k}~t_{jk}
+ {1 \over 6}~\lambda_{jkl}\phi_{j}~\phi_{k}~\phi_{l}
+ d_{\mu}\phi_{j}~s_{j}^{\mu} + \phi_{j}~d_{\mu}\phi_{k}~s_{jk}^{\mu}
\nonumber\\
T^{\mu} = t^{\mu} + \phi_{j}~t_{j}^{\mu}
+ {1 \over 2}~\phi_{j}~\phi_{k}~t_{jk}^{\mu}
+ d^{\mu}\phi_{j}~s_{j} + \phi_{j}~d^{\mu}\phi_{k}~s_{jk}
\nonumber\\
T^{\mu\nu} = t^{\mu\nu}
\label{TI1}
\eea
Then the gauge invariance condition (\ref{G1})
becomes equivalent to:
\be
d_{Q}t^{I} \cong i~d_{\mu}t^{I\mu}
\label{G21}
\ee
\bea
d_{Q}t_{j} \cong i~( d_{\mu}t_{j}^{\mu} - m_{j}^{2}~s_{j})
\nonumber\\
d_{Q}t_{j}^{\mu} \cong 0
\label{G22}
\eea
\bea
d_{Q}t_{jk} \cong i~( d_{\mu}t_{jk}^{\mu}
- m_{k}^{2}~s_{jk} - m_{j}^{2}~s_{kj})
\nonumber\\
d_{Q}t_{jk}^{\mu} \cong 0
\label{G23}
\eea
\bea
d_{Q}s_{j}^{\mu} \cong i~( t_{j}^{\mu} + d^{\mu}s_{j})
\nonumber\\
d_{Q}s_{j} \cong 0
\label{G24}
\eea
\bea
d_{Q}s_{jk}^{\mu} \cong i~( t_{jk}^{\mu} + d^{\mu}s_{jk})
\nonumber\\
d_{Q}s_{jk} \cong 0.
\label{G25}
\eea
Next we list the generic forms for the expressions independent on the scalar fields
$
\phi_{j}.
$
We have
\be
t = \sum t^{\alpha}
\label{t}
\ee
where
\bea
t^{1} = f_{abc}^{1}~v_{a}^{\mu}~v_{b}^{\nu}~d_{\nu}v_{c\mu}
\nonumber\\
t^{2} = f_{abc}^{2}~v_{a}^{\mu}~u_{b}~d_{\mu}\tilde{u}_{c}
\nonumber\\
t^{3} = f_{abc}^{3}~d_{\mu} v_{a}^{\mu}~u_{b}~\tilde{u}_{c}
\nonumber\\
t^{4} = f_{abc}^{4}~\Phi_{a}~d_{\mu}\Phi_{b}~v_{c}^{\mu}
\nonumber\\
t^{5} = f_{abc}^{5}~\Phi_{a}~v_{b\mu}~v_{c}^{\mu}
\nonumber\\
t^{6} = f_{abc}^{6}~\Phi_{a}~\tilde{u}_{b}~u_{c}
\nonumber\\
t^{7} = {1\over 3}~f_{abc}^{7}~\Phi_{a}~\Phi_{b}~\Phi_{c}
\nonumber\\
t^{8} = f_{abc}^{8}~v_{a}^{\mu}~d_{\mu}u_{b}~\tilde{u}_{c}
\nonumber\\
t^{9} = f_{abc}^{9}~v_{a}^{\mu}~v_{b\mu}~d_{\nu}v_{c}^{\nu}
\nonumber\\
t^{10} = {1\over 2}~f_{abc}^{10}~\Phi_{a}~\Phi_{b}~d_{\nu}v_{c}^{\nu}
\nonumber\\
t^{11} = j^{\mu}_{a}~v_{a}^{\mu}, \qquad
j^{\mu}_{a} =
\bar{\psi} t^{\epsilon}_{a} \otimes \gamma^{\mu}\gamma_{\epsilon} \psi
\nonumber\\
t^{12} = j_{a}~\Phi_{a}, \qquad
j_{a} =
\bar{\psi} s^{\epsilon}_{a} \otimes \gamma_{\epsilon} \psi
\label{t1}
\eea
where we suppose that
\be
f_{abc}^{5} = (b \leftrightarrow c),\qquad
f_{abc}^{7} = f_{\{abc\}}^{7}
\label{s0}
\ee
and we have used the following notations in the Dirac sector:
\be
\gamma_{5} = i~\gamma_{0}~\gamma_{1}\gamma_{2}\gamma_{3}, \qquad
\gamma_{\epsilon} = {1\over 2} (I + \epsilon~\gamma_{5})
\label{gama5}
\ee
The expressions
$
t^{\epsilon}_{a}, s^{\epsilon}_{a}
$
are matrices in
$
\C^{4}
$
of the type:
\be
t^{\epsilon}_{a} = t_{a} + \epsilon~t^{\prime}_{a}, \qquad
s^{\epsilon}_{a} = s_{a} + \epsilon~s^{\prime}_{a}, \qquad
\forall a \in I_{1} \cup I_{2}.
\label{ttt}
\ee
One can eliminate some terms
\bea
t^{8} = f_{abc}^{8}~[ d_{\mu} (v_{a}^{\mu}~u_{b}~\tilde{u}_{c})
- d_{\mu}v_{a}^{\mu}~u_{b}~\tilde{u}_{c}
- v_{a}^{\mu}~u_{b}~d_{\mu}\tilde{u}_{c}]
\nonumber\\
t^{9} = f_{abc}^{9}~[ d_{\nu} (v_{a}^{\mu}~v_{b\mu}~v_{c}^{\nu})
- 2 d_{\nu}v_{a\mu}~v_{b}^{\mu}~v_{c}^{\nu} ]
\nonumber
\eea
so if we add a total divergence and redefine
$
t^{j}, j = 1,2,3
$
we can eliminate
$
t^{8}
$
and
$
t^{9}.
$

Also if we consider
\be
B_{1} = b^{1}_{abc}~\tilde{u}_{a}~u_{b}~\tilde{u}_{c},
\qquad b^{1}_{abc} = - (a \leftrightarrow c)
\label{b1}
\ee
we have the coboundary term
\bea
d_{Q}B_{1} = - 2 i~b^{1}_{abc}~
(d_{\mu} v_{a}^{\mu} + m_{a} \Phi_{a})~u_{b}~\tilde{u}_{c}
\nonumber
\eea
and we redefine
$
t^{6}
$
we can make
\be
f_{abc}^{3} = (a \leftrightarrow c).
\label{s1}
\ee
Next we have
\bea
f_{\{ab\}c}^{4}~\Phi_{a}~\Phi_{b,\mu}~v_{c}^{\mu}
= {1\over 2}~[ d_{\mu} (\Phi_{a}~\Phi_{b}~v_{c}^{\mu})
- \Phi_{a}~\Phi_{b}~d_{\mu}v_{c}^{\mu}]
\nonumber
\eea
so if we redefine
$
t^{10}
$
we can make
\be
f_{abc}^{4} = - (a \leftrightarrow b).
\label{s2}
\ee
Finally we consider
\be
B_{2} = b^{2}_{abc}~\Phi_{a}~\Phi_{b}~\tilde{u}_{c},
\qquad b^{1}_{abc} = (a \leftrightarrow b)
\label{b2}
\ee
and the coboundary
\bea
d_{Q}B_{2} = i~b^{2}_{abc}~
[ 2 m_{a}~u_{a} \Phi_{b}~\tilde{u}_{c}
- \Phi_{a}~\Phi_{b}~(d_{\mu}v_{c}^{\mu} + m_{c} \Phi_{c}) ]
\nonumber
\eea
so if we redefine
$
t^{6}
$
and
$
t^{7}
$
we can eliminate
$
t^{10}.
$

We proceed in the same way with the expression
\be
t^{\mu} = \sum t^{\alpha\mu}
\label{t3}
\ee
where
\bea
t^{1,\mu} = g_{abc}^{1}~u_{a}~v_{b\nu}~d^{\nu}v_{c}^{\mu}
\nonumber\\
t^{2,\mu} = g_{abc}^{2}~u_{a}~v_{b\nu}~d^{\mu}v_{c}^{\nu}
\nonumber\\
t^{3,\mu} = {1\over 2}~g_{abc}^{3}~u_{a}~u_{b}~d^{\mu}\tilde{u}_{c}
\nonumber\\
t^{4,\mu} = g_{abc}^{4}~u_{a}~d^{\mu}u_{b}~\tilde{u}_{c}
\nonumber\\
t^{5,\mu} = g_{abc}^{5}~\Phi_{a}~d^{\mu}\Phi_{b}~u_{c}
\nonumber\\
t^{6,\mu} = {1\over 2}~g_{abc}^{6}~\Phi_{a}~\Phi_{b}~d^{\mu}u_{c}
\nonumber\\
t^{7,\mu} = g_{abc}^{7}~\Phi_{a}~v_{b}^{\mu}~u_{c}
\nonumber\\
t^{7,\mu} = g_{abc}^{8}~d^{\mu}u_{a} v_{b\nu}~v_{c}^{\nu}
\nonumber\\
t^{9,\mu} = k^{\mu}_{a}~u_{a}, \qquad
k^{\mu}_{a} =
\bar{\psi} k^{\epsilon}_{a} \otimes \gamma^{\mu}\gamma_{\epsilon} \psi
\label{t4}
\eea
where we take
\be
g_{abc}^{3} = - (a \leftrightarrow b), \qquad
g_{abc}^{6} = (a \leftrightarrow b), \qquad
g_{abc}^{8} = (b \leftrightarrow c).
\label{s0mu}
\ee
We eliminated the term
\bea
g_{abc}^{8}~d^{\mu}u_{a} v_{b\nu}~v_{c}^{\nu} =
g_{abc}^{8}~[ d^{\mu} ( u_{a} v_{b\nu}~v_{c}^{\nu} )
- 2 u_{a}~v_{b\nu}~d^{\mu}v_{c}^{\nu} ]
\nonumber
\eea
by a redefinition of
$
t^{2,\mu}.
$
We also have
\bea
g_{[ab]c}^{4}~u_{a}~d^{\mu}u_{b}~\tilde{u}_{c} =
{1\over 2}~g_{[ab]c}^{4}~[ d^{\mu} ( u_{a}~u_{b}~\tilde{u}_{c})
- u_{a}~u_{b}~d^{\mu}\tilde{u}_{c} ]
\nonumber
\eea
so if we redefine
$
t^{3,\mu}
$
we can fix
\be
g_{abc}^{4} = (a \leftrightarrow b).
\label{s3}
\ee
Similarly we have
\bea
g_{\{ab\}c}^{5}~\Phi_{a}~d^{\mu}\Phi_{b}~u_{c} =
{1\over 2}~g_{\{ab\}c}^{5}~[ d^{\mu} ( \Phi_{a}~\Phi_{b}~u_{c})
- \Phi_{a}~\Phi_{b}~d^{\mu}u_{c} ]
\nonumber
\eea
so if we redefine
$
t^{6,\mu}
$
we can fix
\be
g_{abc}^{5} = - (a \leftrightarrow b).
\label{s4}
\ee
Now we can write explicitly
\be
d_{Q}t - i~d_{\mu}t^{\mu} \cong 0
\ee
Various Wick polynomials give various equations:
\be
f_{abc}^{1} - g_{abc}^{1} = 0 \qquad
(d_{\mu}u_{a}~v_{b\nu}~d^{\nu}v_{c}^{\mu})
\label{G31}
\ee
\be
f_{bac}^{1} - g_{abc}^{2} = 0 \qquad
(d_{\mu}u_{a}~v_{b\nu}~d^{\mu}v_{c}^{\nu})
\label{G32}
\ee
\be
f_{abc}^{1} + ( a \leftrightarrow b) = 0 \qquad
(v_{a}^{\mu}~v_{b}^{\nu}~d_{\mu}d_{\nu}u_{c})
\label{G33}
\ee
\be
f_{bac}^{2} + g_{abc}^{3} + g_{abc}^{4} = 0 \qquad
(u_{a}~d_{\mu}u_{b}~d^{\mu}\tilde{u}_{c})
\label{G34}
\ee
\be
f_{bac}^{2} - g_{abc}^{1} = 0 \qquad
(u_{a}~v_{b}^{\mu}~d_{\mu}d_{\nu}v_{c}^{\nu})
\label{G35}
\ee
\be
f_{bac}^{3} + ( b \leftrightarrow c) = 0 \qquad
(u_{a}~d_{\mu}v_{b}^{\mu}~d_{\nu}v_{c}^{\nu})
\label{G36}
\ee
\be
g_{abc}^{1} + ( b \leftrightarrow c) = 0 \qquad
(u_{a}~d_{\mu}v_{b\nu}~d^{\nu}v_{c}^{\mu})
\label{G37}
\ee
\be
g_{abc}^{2} + ( b \leftrightarrow c) = 0 \qquad
(u_{a}~d_{\mu}v_{b\nu}~d^{\mu}v_{c}^{\nu})
\label{G38}
\ee
\be
f_{abc}^{4} - g_{abc}^{5} - g_{abc}^{6} = 0 \qquad
(\Phi_{a}~d_{\mu}\Phi_{b}~d^{\mu}u_{c})
\label{G39}
\ee
\be
k_{a}^{\mu} - j_{a}^{\mu} = 0\qquad
(d_{\mu}u_{a});
\qquad
d_{\mu}k_{a}^{\mu} - m_{a}~j_{a} = 0\qquad
(u_{a})
\label{G310}
\ee
\be
m_{c}~f_{bac}^{2} + m_{a}~f_{acb}^{4} - g_{cba}^{7} = 0 \qquad
(u_{a}~v_{b}^{\mu}~d_{\mu}\Phi_{c}), \quad c \in I_{2}
\label{G311}
\ee
\be
m_{c}~f_{bac}^{3} - f_{cba}^{6} - g_{cba}^{7} = 0 \qquad
(u_{a}~d_{\mu}v_{b}^{\mu}~\Phi_{c}), \quad c \in I_{2}
\label{G312}
\ee
\be
m_{b}~f_{abc}^{4} + 2 f_{abc}^{5} - g_{acb}^{7} = 0 \qquad
(\Phi_{a}~d_{\mu}u_{b}~v_{c}^{\mu}), \quad a \in I_{2}
\label{G313}
\ee
\be
- m_{a}^{2}~f_{abc}^{3} - m_{a}~f_{acb}^{6}
+ {1\over 2}~m_{c}^{2}~g_{abc}^{3} + m_{b}^{2}~g_{abc}^{4} =
(a \leftrightarrow b) \qquad
(u_{a}~u_{b}~\tilde{u}_{c})
\label{G314}
\ee
\be
m_{a}^{2}~f_{abc}^{5} + m_{c}^{2}~g_{abc}^{2} + ( b \leftrightarrow c)  = 0, \qquad
(u_{a}~v_{b\mu}~v_{c}^{\mu}),\quad a \in I_{2}
\label{G315}
\ee
\be
- m_{b}~f_{abc}^{6} + {1\over 2}~m_{a}~f_{abc}^{7}
+ m_{b}^{2}~g_{abc}^{5} + {1\over 2}~m_{a}^{2}~g_{bca}^{6}
+ ( b \leftrightarrow c)  = 0 \qquad
(u_{a}~\Phi_{b}~\Phi_{c}), \quad b,c \in I_{2}
\label{G316}
\ee

From (\ref{G33}) it follows that
$
f_{abc}^{1}
$
is antisymmetric in
$
a \leftrightarrow b
$
and from (\ref{G31}) + (\ref{G37}) that it is antisymmetric in
$
b \leftrightarrow c
$
so the expression
\be
f_{abc} \equiv f_{abc}^{1}
\label{f}
\ee
is completely antisymmetric. Now (\ref{G31}) and (\ref{G32}) lead to
\be
g_{abc}^{1} = f_{abc}, \qquad g_{abc}^{2} = - f_{abc}
\label{g12}
\ee
and (\ref{G38}) becomes an identity. From (\ref{G35}) we get
\be
f_{abc}^{2} = - f_{abc}.
\label{f2}
\ee
The equation (\ref{G34}) becomes
\be
g_{abc}^{3} + g_{abc}^{4} =  - f_{abc}
\label{G34a}
\ee
so the antisymmetric part in
$
a \leftrightarrow b
$
gives
\be
g_{abc}^{3} = - f_{abc}
\label{g3}
\ee
and it remains
\be
g_{abc}^{4} =  0.
\label{g4}
\ee
From (\ref{G36}) and (\ref{s1}) it follows
\be
f_{abc}^{3} = 0.
\label{f3}
\ee
Let us define
\be
f_{abc}^{\prime} = f_{abc}^{4}
\label{fprime}
\ee
which is, according to (\ref{s2}) antisymmetric in
$
a \leftrightarrow b.
$
If we take the antisymmetric (resp. symmetric) part in
$
a \leftrightarrow b
$
of the relation (\ref{G39}) one gets
\be
g_{abc}^{5} = f_{abc}^{\prime}
\label{g5}
\ee
\be
g_{abc}^{6} = 0.
\label{g6}
\ee
Now (\ref{G311}) leads to
\be
g_{abc}^{7} = - m_{a}~f_{abc} - m_{c}~f_{acb}^{\prime}
\label{g7}
\ee
so (\ref{G312}) gives
\be
f_{abc}^{6} = - g_{abc}^{7}  = m_{a}~f_{abc} + m_{c}~f_{acb}^{\prime}.
\label{f6}
\ee
We go now to  (\ref{G313}) and get
\be
m_{b}~f_{abc}^{\prime} + 2 f_{abc}^{5} =
- m_{a}~f_{acb} - m_{b}~f_{abc}^{\prime}.
\ee
We consider the antisymmetric (resp. symmetric) part in
$
b \leftrightarrow c
$
and obtain
\be
m_{b}~f_{abc}^{\prime} - m_{c}~f_{acb}^{\prime} =
m_{a}~f_{abc}
\label{ff}
\ee
and
\be
f_{abc}^{5} =
- {1\over 2}~( m_{b}~f_{abc}^{\prime} + m_{c}~f_{acb}^{\prime} ).
\label{f5}
\ee
This leads to simplified formulas for
$
g_{abc}^{7}
$
and
$
f_{abc}^{6}
$
namely
\be
g_{abc}^{7} = - m_{b}~f_{abc}^{\prime}
\label{g7a}
\ee
\be
f_{abc}^{6} = m_{b}~f_{abc}^{\prime}.
\label{f6a}
\ee
The relations (\ref{G314}) and (\ref{G315}) are now identities and   (\ref{G316}) gives
\be
f_{abc}^{7} = 0.
\label{f7}
\ee
Summing up, we obtain
\bea
t = f_{abc} \left( {1\over 2}~v_{a\mu}~v_{b\nu}~F_{c}^{\nu\mu}
- v_{a}^{\mu}~u_{b}~d_{\mu}\tilde{u}_{c}\right)
\nonumber\\
+ f_{abc}^{\prime}~(\Phi_{a}~d_{\mu}\Phi_{b}~v_{c}^{\mu}
- m_{b} \Phi_{a}~v_{b\mu}~v_{c}^{\mu}
+ m_{b} \Phi_{a}~\tilde{u}_{b}~u_{c} )
\nonumber\\
+ j_{a}^{\mu}~v_{a\mu} + j_{a}~\Phi_{a}
\label{t11}
\eea
and
\bea
t^{\mu} = f_{abc} \left( u_{a\mu}~v_{b\nu}~F_{c}^{\nu\mu}
- {1\over 2}~u_{a}~u_{b}~d^{\mu}\tilde{u}_{c}\right)
\nonumber\\
+ f_{abc}^{\prime}~(\Phi_{a}~d^{\mu}\Phi_{b}~u_{c}
- m_{b} \Phi_{a}~v_{b}^{\mu}~u_{c} )
\nonumber\\
+ j_{a}^{\mu}~u_{a}
\label{tmu1}
\eea
where
\be
F^{\mu\nu}_{a} \equiv d^{\mu}v^{\nu}_{a} - d^{\nu}v^{\mu}_{a}, 
\quad \forall a \in I_{1} \cup I_{2}.
\ee 
We also have the symmetry properties:
$
f_{abc} = f_{[abc]},~f_{abc}^{\prime} = f_{[ab]c}^{\prime}
$
and the relation (\ref{ff}); from (\ref{G310}) follows
\be
d_{\mu}j_{a}^{\mu} - m_{a}~j_{a} = 0\
\label{G310a}
\ee
If we apply the operator
$
d_{Q}
$
to
$
t^{\mu}
$
we easily obtain
\be
t^{\mu\nu} = {1\over 2}~u_{a}~u_{b}~F_{c}^{\mu\nu}.
\label{tmunu}
\ee

We go now in the scalar sector and determine
\be
t_{j} = \sum t_{j}^{\alpha}
\ee
where the generic terms are:
\bea
t_{j}^{1} = h_{jbc}^{1}~d_{\mu}\Phi_{b}~v_{c}^{\mu}
\nonumber\\
t_{j}^{2} = h_{jbc}^{2}~\Phi_{b}~d_{\mu}v_{c}^{\mu}
\nonumber\\
t_{j}^{3} = h_{jbc}^{3}~v_{b\mu}~v_{c}^{\mu}
\nonumber\\
t_{j}^{4} = h_{jbc}^{4}~\tilde{u}_{b}~u_{c}
\nonumber\\
t_{j}^{5} = {1\over 2}~h_{jbc}^{5}~\Phi_{b}~\Phi_{c}
\nonumber\\
t_{j}^{6} = \bar{\Psi} s_{j}^{\epsilon} \otimes \gamma_{\epsilon} \Psi
\label{tj}
\eea
and we can take
\be
h_{jbc}^{3} = ( b \leftrightarrow c), \quad
h_{jbc}^{5} = ( b \leftrightarrow c).
\ee

We also have the generic forms
\be
t^{\mu}_{j} = \sum t_{j}^{\mu,\alpha}
\ee
where
\bea
t_{j}^{\mu,1} = k_{jbc}^{1}~d^{\mu}\Phi_{b}~u_{c}
\nonumber\\
t_{j}^{\mu,2} = k_{jbc}^{2}~\Phi_{b}~d^{\mu}u_{c}
\nonumber\\
t_{j}^{\mu,3} = h_{jbc}^{3}~v_{b}^{\mu}~u_{c}
\eea
and
\be
s_{j} = l_{jbc}~\Phi_{b}~u_{c}.
\ee

We notice that
\bea
\phi_{j}~t_{j}^{2} =
i d_{Q}(h_{jbc}^{2}~\phi_{j}~\Phi_{b}~\tilde{u}_{c})
+ h_{jbc}^{2}~\phi_{j}~m_{b}~u_{b}~\tilde{u}_{c}
+ h_{jbc}^{2}~m_{c}~\phi_{j}~m_{b}~\Phi_{b}~\Phi_{c}
\nonumber
\eea
so we can make
\be
h_{jbc}^{2} = 0
\label{h2}
\ee
by redefining
$
h_{jbc}^{4}
$
and
$
h_{jbc}^{5}.
$

Also
\bea
\phi_{j}~t^{\mu,2}_{j} =
- i d_{Q}(k_{jbc}^{2}~\phi_{j}~\Phi_{b}~v_{c}^{\mu})
- k_{jbc}^{2}~\phi_{j}~m_{b}~u_{b}~v_{c}^{\mu}
\nonumber
\eea
so we can make
\be
k_{jbc}^{2} = 0
\label{k2}
\ee
by redefining
$
k_{jbc}^{3}.
$

The first relation (\ref{G22}) becomes equivalent to
\be
h_{jbc}^{1} - k_{jbc}^{1} = 0\qquad
(d_{\mu}\Phi_{b}~d^{\mu}u_{c}), \quad b \in I_{2}
\label{G221}
\ee
\be
m_{b}~h_{jbc}^{1} + 2 h_{jbc}^{3} - k_{jbc}^{3} = 0\qquad
(d_{\mu}u_{b}~v_{c}^{\mu})
\label{G222}
\ee
\be
h_{jcb}^{4} + k_{jcb}^{3} = 0\qquad
(u_{b}~d_{\mu}v_{c}^{\mu})
\label{G223}
\ee
\be
- m_{b}~h_{jbc}^{4} + m_{c}~h_{jcb}^{5}
+ m_{b}^{2}~k_{jbc}^{1} +  m_{j}^{2}~l_{jbc} = 0
\qquad
(\Phi_{b}~u_{c}), \quad b \in I_{2}
\label{G224}
\ee
and the second relation (\ref{G22}) gives
\be
m_{b}~k_{jbc}^{1} + k_{jbc}^{3} = 0\qquad
(d^{\mu}u_{b}~u_{c}).
\label{G225}
\ee
Also the second relation (\ref{G24}) leads to
\be
m_{b}~l_{jbc} - m_{c}~l_{jcb} = 0\qquad  \forall b,c \in I_{2}
\label{G241}
\ee
\be
l_{jbc} = 0 \qquad \forall b \in I_{2}, c \in I_{1}.
\label{G242}
\ee
We make the notation
\be
f_{jbc}^{\prime} \equiv h_{jbc}^{1}
\label{fprimej}
\ee
and we have from (\ref{G221}) and (\ref{G225})
\be
k_{jbc}^{1} = f_{jbc}^{\prime}
\label{k1}
\ee
\be
k_{jbc}^{3} = - m_{b}~f_{jbc}^{\prime}.
\label{k3}
\ee
Then (\ref{G223}) and leads to
\be
h_{jbc}^{4} = m_{b}~f_{jbc}^{\prime}
\label{h4}
\ee
and (\ref{G222}) to
\be
h_{jbc}^{3} = - {1\over 2}~( m_{b}~f_{jbc}^{\prime} + m_{c}~f_{jcb}^{\prime}).
\label{h3}
\ee
Also, from (\ref{G224}) to
\be
h_{jbc}^{5} = - {m_{j}^{2} \over m_{c}}~l_{jbc}\qquad \forall b,c \in I_{2}.
\label{h5}
\ee
One can use (\ref{G241}) to prove that the expression
$
h_{jbc}^{5}
$
is symmetric in
$
b \leftrightarrow c
$
as it should be.

We still need the generic form of
\be
s_{j}^{\mu} = p_{jbc}~\Phi_{b}~v_{c}^{\mu}
\label{sj}
\ee
and the first relation (\ref{G24}) becomes equivalent to
\be
f_{jbc}^{\prime} + l_{jbc} = 0\qquad (d^{\mu}\Phi_{b}~u_{c})
\label{G2411}
\ee
\be
l_{jbc} - p_{jbc} = 0\qquad (\Phi_{b}~d^{\mu}u_{c})
\label{G2422}
\ee
\be
m_{b}~f_{jbc}^{\prime} + m_{c}~p_{jbc} = 0\qquad (v_{b}^{\mu}~u_{c})
\label{G243}
\ee
for all
$
b \in I_{2}.
$
If we use (\ref{G242}) it follows that we have
\be
l_{jbc} = - f_{jbc}^{\prime}, \qquad \forall b, c \in I_{2},
\label{l}
\ee
and
\be
f_{jbc}^{\prime} = 0, \qquad \forall b \in I_{2}, c \in I_{1}.
\label{fprime2}
\ee
Then from (\ref{G2411}) we have
\bea
p_{jbc} = - f_{jbc}^{\prime}, \qquad \forall b, c \in I_{2},
\nonumber\\
p_{jbc} = 0, \qquad \forall b \in I_{2}, c \in I_{1}.
\label{p}
\eea
From (\ref{G241}) or (\ref{G243}) we get
\be
m_{b}~f_{jbc}^{\prime} - m_{c}~f_{jcb}^{\prime} = 0,
\qquad \forall b, c \in I_{2}.
\label{G241a}
\ee
We use (\ref{l}) in (\ref{h5}) and we obtain
\be
h_{jbc}^{5} = {m_{j}^{2} \over m_{c}}~f_{jbc}^{\prime}
\qquad \forall b,c \in I_{2}.
\label{h5a}
\ee
We introduce the notation
\be
f_{jbc}^{\prime\prime} \equiv h_{jbc}^{5}
\ee
and summing up, we have
\bea
t_{j} = f_{jbc}^{\prime}~( d_{\mu}\Phi_{b}~v_{c}^{\mu}
- m_{b}~v_{b}^{\mu}~v_{c\mu} + m_{b}~\tilde{u}_{b}~u_{c})
+ {1\over 2}~f_{jbc}^{\prime\prime}~\Phi_{b}~\Phi_{c} + d_{j}
\nonumber\\
d_{j} \equiv \bar{\psi} s_{j}^{\epsilon} \otimes \gamma_{\epsilon} \psi
\label{tj1}
\eea
\be
t_{j}^{\mu} =
f_{jbc}^{\prime}~( d^{\mu}\Phi_{b}~u_{c} - m_{b}~v_{b}^{\mu}~u_{c}).
\label{tjmu1}
\ee
\be
s_{j}^{\mu} = - f_{jbc}^{\prime}~\Phi_{b}~v_{c}^{\mu}
\label{sj1}
\ee
Finally we have the generic forms
\be
t_{jk} = q_{jkc}^{1}~\Phi_{c}
\label{tjk}
\ee
\be
t_{jk}^{\mu} = q_{jkc}^{2}~d^{\mu}u_{c}
\label{tmujk}
\ee
\be
s_{jk} = q_{jkc}^{3}~u_{c}
\label{sjk}
\ee
\be
s_{jk}^{\mu} = q_{jkc}^{4}~v_{c}^{\mu}
\label{smujk}
\ee
where we impose
\be
q_{jkc}^{1} = j \leftrightarrow k, \qquad
q_{jkc}^{2} = j \leftrightarrow k, \qquad
q_{jkc}^{3} = - (j \leftrightarrow k).
\ee
Let us note that
\bea
\phi_{j}~\phi_{k}~t_{jk}^{\mu} =
- i d_{Q}(q_{jkc}^{2}~\phi_{j}~\phi_{k}~v_{c}^{\mu})
\nonumber
\eea
so we can take
\be
q_{jkc}^{2} = 0.
\label{q2}
\ee

Then the relation (\ref{G23}) becomes equivalent to
\be
m_{c}~q_{jkc}^{1} + m_{k}^{2}~q_{jkc}^{3} + m_{j}^{2}~q_{kjc}^{3} = 0\qquad
(u_{c}).
\label{G231}
\ee
From the first relation (\ref{G25}) we have
\be
q_{jkc}^{4} - q_{jkc}^{3} = 0 \qquad (d^{\mu}u_{c}).
\label{G251}
\ee
We make the notations
\be
f_{jkc}^{\prime\prime} \equiv q_{jkc}^{1}
\label{fpp}
\ee
and
\be
f_{jkc}^{\prime} \equiv q_{jkc}^{3}.
\label{fp2}
\ee
and we finally have from (\ref{G251})
\be
q_{jkc}^{4} = f_{jkc}^{\prime}
\ee
and from (\ref{G231})
\be
f_{jkc}^{\prime\prime} =
- {1 \over m_{c}}~( m_{k}^{2}~q_{jkc}^{3} + m_{j}^{2}~q_{kjc}^{3}), \qquad \forall c \in I_{2}
\label{fjkc}
\ee
which is symmetric in
$
j \leftrightarrow k
$
as it should be.

We can now summarize the result in the following
\begin{thm}
The generic form of the interaction Lagrangean of the standard model is:
\bea
T = f_{abc} \left( {1\over 2}~v_{a\mu}~v_{b\nu}~F_{c}^{\nu\mu}
+ u_{a}~v_{b}^{\mu}~d_{\mu}\tilde{u}_{c}\right)
\nonumber\\
+ f_{abc}^{\prime}~(\Phi_{a}~d_{\mu}\Phi_{b}~v_{c}^{\mu}
- m_{b}~\Phi_{a}~v_{b\mu}~v_{c}^{\mu}
+ m_{b}~\Phi_{a}~\tilde{u}_{b}~u_{c} )
+ j_{a}^{\mu}~v_{a\mu} + j_{a}~\Phi_{a}
\nonumber\\
+ \phi_{j}~f_{jbc}^{\prime}~( d_{\mu}\Phi_{b}~v_{c}^{\mu}
- m_{b}~v_{b}^{\mu}~v_{c\mu} + m_{b}~\tilde{u}_{b}~u_{c})
+ {1\over 2}~f_{jbc}^{\prime\prime}~\phi_{j}~\Phi_{b}~\Phi_{c}
+ \phi_{j}~d_{j}
\nonumber\\
- f_{jbc}^{\prime}~d_{\mu}\phi_{j}~\Phi_{b}~v_{c}^{\mu}
+ f_{jkc}^{\prime}~\phi_{j}~d_{\mu}\phi_{k}~v_{c}^{\mu}
+ {1\over 2}~f_{jkc}^{\prime\prime}~\phi_{j}~\phi_{k}~\Phi_{c}
+ {1\over 3!}~\lambda_{jkl}~\phi_{j}~\phi_{k}~\phi_{l}.
\label{Tint}
\eea
In this case we can take
\bea
T^{\mu} = f_{abc} \left( u_{a\mu}~v_{b\nu}~F_{c}^{\nu\mu}
- {1\over 2}~u_{a}~u_{b}~d^{\mu}\tilde{u}_{c}\right) + j_{a}^{\mu}~u_{a}
\nonumber\\
+ f_{abc}^{\prime}~(\Phi_{a}~d^{\mu}\Phi_{b}~u_{c}
- m_{b} \Phi_{a}~v_{b}^{\mu}~u_{c} )
\nonumber\\
+ \phi_{j}~f_{jbc}^{\prime}~( d^{\mu}\Phi_{b}~u_{c} - m_{b}~v_{b}^{\mu}~u_{c})
- f_{jbc}^{\prime}~d^{\mu}\phi_{j}\Phi_{b}~u_{c}
+ f_{jkc}^{\prime}~\phi_{j}~d^{\mu}\phi_{k}~u_{c}
\label{Tmu}
\eea
and
\be
T^{\mu\nu} = {1\over 2}~u_{a}~u_{b}~F_{c}^{\mu\nu}.
\label{Tmunu}
\ee
Here the sums over
$
a, b, \dots
$
are running in
$
I_{1} \cup I_{2}
$
and the sums over
$
j, k \dots
$
over
$
I_{3}.
$
We have also made the conventions
\be
f_{abc}^{\prime} = 0, \quad s_{a}^{\epsilon} = 0 \qquad \forall a \in I_{1}.
\label{fs}
\ee
The constants appearing in these expressions are subject to the following relations:
\be
f_{abc} = f_{[abc]},
\label{sf1}
\ee
\be
f_{abc}^{\prime} = f_{[ab]c}^{\prime}
\label{sf2}
\ee
\be
m_{b}~f_{abc}^{\prime} - m_{c}~f_{acb}^{\prime} =
m_{a}~f_{abc}\qquad \forall a \in I_{2}
\label{ff1}
\ee
\be
m_{b}~f_{jbc}^{\prime} - m_{c}~f_{jcb}^{\prime} = 0,
\qquad \forall b\in I_{2}, \quad c \in I_{1} \cup I_{2}
\label{ff2}
\ee
\be
d_{\mu}j_{a}^{\mu} - m_{a}~j_{a} = 0 \quad \Longleftrightarrow \quad
i~(M~t_{a}^{\epsilon} - t_{a}^{- \epsilon}~M) = m_{a}~s_{a}^{\epsilon}, \quad
\forall a \in I_{1} \cup I_{2}
\label{j}
\ee
\be
f_{jbc}^{\prime\prime} = {m_{j}^{2} \over m_{c}}~f_{jbc}^{\prime}
\qquad \forall b,c \in I_{2}.
\label{ff3}
\ee
\be
f_{jkc}^{\prime\prime} = - {1 \over m_{c}}~
( m_{k}^{2}~f^{\prime}_{jkc} + m_{j}^{2}~f^{\prime}_{kjc})
= {1 \over m_{c}} (m_{j}^{2} - m_{k}^{2})~f^{\prime}_{jkc},
\qquad \forall c \in I_{2}.
\label{ff4}
\ee
\end{thm}

One can write in a compact way the previous expressions.
If we define
\be
f^{\prime}_{cja} \equiv - f^{\prime}_{jca}
\ee
then we have
\bea
T = f_{abc} \left( {1\over 2}~v_{a\mu}~v_{b\nu}~F_{c}^{\nu\mu}
+ u_{a}~v_{b}^{\mu}~d_{\mu}\tilde{u}_{c}\right)
\nonumber\\
+ f_{\alpha\beta c}^{\prime}~\Phi_{\alpha}~d_{\mu}\Phi_{\beta}~v_{c}^{\mu}
- f_{\alpha bc}^{\prime}~m_{b}~\Phi_{\alpha}~(v_{b\mu}~v_{c}^{\mu}
- \tilde{u}_{b}~u_{c} )
+ j_{a}^{\mu}~v_{a\mu} + j_{\alpha}~\Phi_{\alpha}
\nonumber\\
+ {1\over 3!}~f^{\prime\prime}_{\alpha\beta\gamma}~
\Phi_{\alpha}~\Phi_{\beta}~\Phi_{\gamma}
\label{Tint1}
\eea
\bea
T^{\mu} = f_{abc} \left( u_{a\mu}~v_{b\nu}~F_{c}^{\nu\mu}
- {1\over 2}~u_{a}~u_{b}~d^{\mu}\tilde{u}_{c}\right) + j_{a}^{\mu}~u_{a}
\nonumber\\
+ f_{\alpha\beta c}^{\prime}~\Phi_{\alpha}~d^{\mu}\Phi_{\beta}~u_{c}
- f_{\alpha bc}^{\prime}~m_{b}~\Phi_{a}~v_{b}^{\mu}~u_{c}
\label{Tmu1}
\eea
where now the sums over
$
\alpha, \beta, \dots
$
are running in
$
I_{1} \cup I_{2} \cup I_{3}
$
and we have to rewrite conveniently the various restrictions on the constants from the previous theorem. In the following we will prefer to work with the expressions from the previous theorem.

The expression (\ref{Tint}) gives a Wick polynomial
$
T^{\rm quant}
$
formally the same, but: 
(a) the jet variables must be replaced by the associated quantum fields; (b) the formal derivative 
$
d^{\mu}
$
goes in the true derivative in the coordinate space; (c) Wick ordering should be done to obtain well-defined operators.
Then
\be
~[Q, T^{I} ] = \partial_{\mu}T^{I\mu}
\label{gauge1}
\ee
where the equations of motion are automatically used because the quantum fields are on-shell

Finally we give the relation expressing gauge invariance in order $n$ of the perturbation theory. We define the operator 
$
\delta
$
on chronological products by:
\bea
\delta T(T^{I_{1}}(x_{1}),\dots,T^{I_{n}}(x_{n})) \equiv 
\sum_{m=1}^{n}~( -1)^{s_{m}}\partial_{\mu}^{m}T(T^{I_{1}}(x_{1}), \dots,T^{I_{m}\mu}(x_{m}),\dots,T^{I_{n}}(x_{n}))
\label{derT}
\eea
with
\be
s_{m} \equiv \sum_{p=1}^{m-1} |I_{p}|,
\ee
then we define the operator
\be
s \equiv d_{Q} - i \delta.
\label{s-n}
\ee

Gauge invariance in an arbitrary order is then expressed by
\be
sT(T^{I_{1}}(x_{1}),\dots,T^{I_{n}}(x_{n})) = 0.
\label{brst-n}
\ee
It is clear that this equation splits in independent equations according to the number of loops. For instance, in the second order we will have an equation for the one loop contributions and one for the tree contributions.
\newpage

\newpage

\section{Wick submonomials\label{submonomials}}
In a previous paper \cite{wick+hopf} we have emphasised the utility of the Wick submonomials. They can be used to write compactly the quantum anomalies. Here we extend the method to the general case of the standard model. First we define (in the classical context) the derivation
\be
\xi \cdot A \equiv (-1)^{|\xi| |A|}~{\partial \over \partial \xi}A
\label{derivative}
\ee
for any jet variables (fields and derivatives). In the particular case described by the theorem from the preceding section the non-zero submonomials are:
\be
B_{a\mu} \equiv \tilde{u}_{a,\mu} \cdot T = - f_{abc} u_{b}~v_{c\mu}
\label{Bamu}
\ee
\bea
C_{a\mu} \equiv v_{a\mu} \cdot T = f_{abc} (v_{b}^{\nu}~F_{c\nu\mu}
- u_{b}~\tilde{u}_{c,\mu}) + j_{a\mu}
\nonumber\\
+ f_{bca}^{\prime}~\Phi_{b}~d_{\mu}\Phi_{c}
- (f_{bac}^{\prime}~m_{a} + f_{bca}^{\prime}~m_{c})~\Phi_{b}~v_{c\mu}
\nonumber\\
+ f_{jba}^{\prime}~(\phi_{j}~d_{\mu}\Phi_{b} - d_{\mu}\phi_{j}~\Phi_{b})
- (f_{jba}^{\prime}~m_{b} + f_{jab}^{\prime}~m_{a})~\phi_{j}~v_{b\mu}
+ f_{jka}^{\prime}~\phi_{j}~d_{\mu}\phi_{k}
\label{Camu}
\eea

\be
D_{a} \equiv u_{a} \cdot T = f_{abc} v^{\mu}_{b}~d_{\mu}\tilde{u}_{c}
- f_{bca}^{\prime}~m_{c}~\Phi_{b}~\tilde{u}_{c}
\label{Da}
\ee

\be
E_{a\mu\nu} \equiv v_{a\mu,\nu} \cdot T = f_{abc}~v_{b\mu}~v_{c\nu}
\label{Eamunu}
\ee

If we define
\be
B_{a} \equiv {1 \over 2}~f_{abc}~u_{b}~u_{c}
\label{B}
\ee
we also have
\be
B_{a\nu\mu} \equiv \tilde{u}_{a,\nu} \cdot T_{\mu} = \eta_{\mu\nu}~B_{a}
\label{Banumu}
\ee
and
\be
E_{a\rho\sigma,\mu\nu} \equiv v_{a\rho,\sigma} \cdot T_{\mu\nu} = (\eta_{\mu\sigma}~\eta_{\nu\rho} - \eta_{\nu\sigma}~\eta_{\mu\rho})~B_{a}.
\label{Earhosigmamunu}
\ee

Next
\be
C_{a\nu,\mu} \equiv v_{a\nu} \cdot T_{\mu} = - f_{abc} u_{b}~F_{c\nu\mu}
- \eta_{\mu\nu}~( f_{bac}^{\prime}~m_{a}~\Phi_{b}~d_{\mu}~u_{c}
+ f_{jab}^{\prime}~m_{a}~\phi_{j}~u_{b})
\label{Camunu}
\ee

\bea
D_{a\mu} \equiv u_{a} \cdot T_{\mu} = - f_{abc} (v_{b}^{\nu}~F_{c\nu\mu}
- u_{b}~\tilde{u}_{c,\mu}) - j_{a\mu}
\nonumber\\
- f_{bca}^{\prime}~\Phi_{b}~( d_{\mu}\Phi_{c} - m_{c}~v_{c\mu})
\nonumber\\
- f_{jba}^{\prime}~\phi_{j}~( d_{\mu}\Phi_{b} - m_{b}~v_{b\mu})
+ f_{jba}^{\prime}~d_{\mu}\phi_{j}~\Phi_{b}
- f_{jka}^{\prime}~\phi_{j}~d_{\mu}\phi_{k}
\label{Damu}
\eea

\be
E_{a\rho\sigma,\mu} \equiv v_{a\rho,\sigma} \cdot T_{\mu}
= \eta_{\mu\sigma}~B_{a\rho} - \eta_{\mu\rho}~B_{a\sigma}
\label{Earhosigmamu}
\ee

\be
D_{a\mu\nu} \equiv u_{a} \cdot T_{\mu\nu} = f_{abc}~u_{b}~F_{c\nu\mu}
\label{Damunu}
\ee

\bea
G_{a} \equiv \Phi_{a} \cdot T =
f_{abc}^{\prime}~(d_{\mu}\Phi_{b}~v_{c}^{\mu}
- m_{b}~v_{b\mu}~v_{c}^{\mu} + m_{b}~\tilde{u}_{b}~u_{c} ) + j_{a}
\nonumber\\
- f_{jab}^{\prime}~d_{\mu}\phi_{j}~v_{b}^{\mu}
+ f_{jab}^{\prime\prime}~\phi_{j}~\Phi_{b}
+ {1\over 2}~f_{jka}^{\prime\prime}~\phi_{j}~\phi_{k}
\label{Ga}
\eea

\bea
G_{j} \equiv \phi_{j} \cdot T =
f_{jbc}^{\prime}~(d_{\mu}\Phi_{b}~v_{c}^{\mu}
- m_{b}~v_{b\mu}~v_{c}^{\mu} + m_{b}~\tilde{u}_{b}~u_{c} ) + d_{j}
\nonumber\\
+ f_{jkb}^{\prime}~d_{\mu}\phi_{k}~v_{b}^{\mu}
+ f_{jkb}^{\prime\prime}~\phi_{k}~\Phi_{b}
+ {1\over 2}~f_{jbc}^{\prime\prime}~\Phi_{b}~\Phi_{c}
+ {1\over 2}~\lambda_{jkl}~\phi_{k}~\phi_{l}
\label{Gj}
\eea

\be
H_{a\mu} \equiv \Phi_{a,\mu} \cdot T =
f_{bac}^{\prime}~\Phi_{b}~v_{c\mu}
+ f_{jac}^{\prime}~\phi_{j}~v_{c\mu}
\label{Hamu}
\ee

\be
H_{j\mu} \equiv \phi_{j,\mu} \cdot T =
- f_{jbc}^{\prime}~\Phi_{b}~v_{c\mu}
- f_{jkc}^{\prime}~\phi_{k}~v_{c\mu}
\label{Hjmu}
\ee

\be
K_{a} \equiv \tilde{u}_{a} \cdot T = f_{bac}^{\prime}~m_{a}~\Phi_{b}~u_{c}
+ f_{jac}^{\prime}~m_{a}~\phi_{j}~u_{c}
\label{Ka}
\ee

\be
G_{a\mu} \equiv \Phi_{a} \cdot T_{\mu} =
f_{abc}^{\prime}~(d_{\mu}\Phi_{b}~u_{c} - m_{b}~v_{b\mu}~u_{c} )
- f_{jac}^{\prime}~d_{\mu}\phi_{j}~u_{c}
\label{Gamu}
\ee

\be
G_{j\mu} \equiv \Phi_{a} \cdot T_{\mu} =
f_{jbc}^{\prime}~(d_{\mu}\Phi_{b}~u_{c} - m_{b}~v_{b\mu}~u_{c} )
- f_{kjc}^{\prime}~d_{\mu}\phi_{k}~u_{c}
\label{Gjmu}
\ee

\be
H_{a\nu,\mu} \equiv \Phi_{a,\nu} \cdot T_{\mu} = \eta_{\mu\nu}~H_{a}
\label{Hanumu}
\ee
where
\be
H_{a} \equiv - f_{abc}^{\prime}~\Phi_{b}~u_{c}
+ f_{jab}^{\prime}~\phi_{j}~u_{b}
\label{Ha}
\ee

\be
H_{j\nu,\mu} \equiv \phi_{j,\nu} \cdot T_{\mu} = \eta_{\mu\nu}~H_{j}
\label{Hjnumu}
\ee
where
\be
H_{j} \equiv - f_{jbc}^{\prime}~\Phi_{b}~u_{c}
- f_{jkb}^{\prime}~\phi_{k}~u_{b}
\label{Hj}
\ee

\be
\bar{V}_{A\alpha} \equiv \psi_{A\alpha} \cdot T =
- \bar{\psi}_{B\beta}~(t^{\epsilon}_{a})_{BA}~
(\gamma_{\mu}\gamma_{\epsilon})_{\beta\alpha}~v_{a}^{\mu}
- \bar{\psi}_{B\beta}~(s^{\epsilon}_{a})_{BA}~
(\gamma_{\epsilon})_{\beta\alpha}~\Phi_{a}
- \bar{\psi}_{B\beta}~(s^{\epsilon}_{j})_{BA}~
(\gamma_{\epsilon})_{\beta\alpha}~\phi_{j}
\label{barVAalpha}
\ee

\be
V_{A\alpha} \equiv \bar{\psi}_{A\alpha} \cdot T =
(t^{\epsilon}_{a})_{AB}~(\gamma_{\mu}\gamma_{\epsilon})_{\alpha\beta}~
\psi_{B\beta}~v_{a}^{\mu}
+ (s^{\epsilon}_{a})_{AB}~(\gamma_{\epsilon})_{\alpha\beta}~
\psi_{B\beta}~\Phi_{a}
+ (s^{\epsilon}_{j})_{AB}~(\gamma_{\epsilon})_{\alpha\beta}~
\psi_{B\beta}~\phi_{j}
\label{VAalpha}
\ee

\be
\bar{V}_{A\alpha,\mu} \equiv \psi_{A\alpha} \cdot T_{\mu} =
\bar{\psi}_{B\beta}~(t^{\epsilon}_{a})_{BA}~
(\gamma_{\mu}\gamma_{\epsilon})_{\beta\alpha}~u_{a}
\label{barVAalphamu}
\ee

\be
V_{A\alpha,\mu} \equiv \bar{\psi}_{A\alpha} \cdot T_{\mu} =
- (t^{\epsilon}_{a})_{AB}~(\gamma_{\mu}\gamma_{\epsilon})_{\alpha\beta}~
\psi_{B\beta}~u_{a}.
\label{VAalphamu}
\ee

There are a number of relations between these Wick submonomials which are useful in the computations from the following Sections.
\be
D_{a}^{\mu\nu} = - C_{a}^{[\mu\nu]}
\ee

\be
C_{a}^{\mu} + D_{a}^{\mu} + m_{a}~H_{a}^{\mu} = 0
\ee

\be
C_{a}^{\{\mu\nu\}} = - \eta^{\mu\nu}~K_{a}
\ee

\be
K_{a} = m_{a}~H_{a}.
\ee

The compact notation from the previous Section is also available:
\be
C_{a\mu} = f_{abc} (v_{b}^{\nu}~F_{c\nu\mu}
- u_{b}~\tilde{u}_{c,\mu}) + j_{a\mu}
\nonumber\\
+ f_{\beta\gamma a}^{\prime}~\Phi_{\beta}~d_{\mu}\Phi_{\gamma}
- h_{\beta ac}~\Phi_{\beta}~v_{c\mu}
\label{Camu-compact}
\ee
where
\be
h_{\beta ac} \equiv f_{\beta ac}^{\prime}~m_{a} + f_{\beta ca}^{\prime}~m_{c}
\ee

\be
D_{a} = f_{abc} v^{\mu}_{b}~d_{\mu}\tilde{u}_{c}
- f_{\beta ca}^{\prime}~m_{c}~\Phi_{\beta}~\tilde{u}_{c}
\label{Da-compact}
\ee
\be
C_{a\nu,\mu} = - f_{abc} u_{b}~F_{c\nu\mu}
- \eta_{\mu\nu}~f_{\beta ac}^{\prime}~m_{a}~\Phi_{\beta}~d_{\mu}~u_{c}
\label{Camunu-compact}
\ee

\be
D_{a\mu} = - f_{abc} (v_{b}^{\nu}~F_{c\nu\mu}
- u_{b}~\tilde{u}_{c,\mu}) - j_{a\mu}
- f_{\beta \gamma a}^{\prime}~\Phi_{\beta}~d_{\mu}\Phi_{\gamma}
+ f_{\beta ca}^{\prime}~m_{c}~\Phi_{\beta}~v_{c\mu}
\label{Damu-compact}
\ee

\be
G_{\alpha} \equiv =
f_{\alpha \beta c}^{\prime}~(d_{\mu}\Phi_{\beta}~v_{c}^{\mu}
+ f_{\alpha bc}^{\prime}~ m_{b}~( - v_{b\mu}~v_{c}^{\mu} + \tilde{u}_{b}~u_{c} ) + j_{\alpha} + {1\over 2}~ f_{\alpha\beta\gamma}~\Phi_{\beta}~\Phi_{\gamma}
\label{Ga-compact}
\ee

\be
H_{\alpha}^{\mu} = - f_{\alpha\beta c}^{\prime}~\Phi_{\beta}~v_{c\mu}
\label{Hamu-compact}
\ee

\be
K_{a} = f_{\beta ac}^{\prime}~m_{a}~\Phi_{\beta}~u_{c}
\label{Ka-compact}
\ee

\be
G_{\alpha}^{\mu} =
f_{\alpha\beta c}^{\prime}~d_{\mu}\Phi_{\beta}~u_{c}
- f_{\alpha bc}^{\prime}~m_{b}~v_{b}^{\mu}~u_{c}
\label{Gamu-compact}
\ee

\be
H_{\alpha} \equiv - f_{\alpha\beta c}^{\prime}~\Phi_{\beta}~u_{c}.
\label{Ha-compact}
\ee

Then we try to extend the structure (\ref{G1}) to the Wick submonomials defined above. We have for instance
\bea
s B_{a}^{\mu} \equiv d_{Q} B_{a}^{\mu} - i d_{\nu}B_{a}^{\mu\nu} =
d_{Q} B_{a}^{\mu} - i d^{\mu}B_{a} = 0
\nonumber\\
s B_{a}^{\mu\nu} \equiv d_{Q} B_{a}^{\mu\nu} = 0
\label{dQB}
\eea
but in other cases gauge invariance is ``broken". We fix this in the following way. We have the formal derivative
\be
\delta A \equiv d_{\mu}A^{\mu}
\ee
and also define the derivative
$
\delta^{\prime}
$
by
\bea
\delta^{\prime}C_{a}^{\mu} = - m_{a}^{2}~B_{a}^{\mu}
\nonumber\\
\delta^{\prime}D_{a} = m_{a}~G_{a}
\nonumber\\
\delta^{\prime}E_{a}^{\mu\nu} = C_{a}^{[\mu\nu]}
\nonumber\\
\delta^{\prime}C_{a}^{\mu\nu} =  m_{a}^{2}~B_{a}^{\mu\nu} \qquad
\Longleftrightarrow \qquad
\delta^{\prime}C_{a}^{[\mu\nu]} = 0, \qquad
\delta^{\prime}K_{a} = m_{a}^{2}~B_{a}
\nonumber\\
\delta^{\prime}D_{a}^{\mu} = - m_{a}~G_{a}^{\mu}
\nonumber\\
\delta^{\prime}G_{j} = - m_{j}^{2}H_{j}
\nonumber\\
\delta^{\prime}H_{a}^{\mu} = G_{a}^{\mu} + m_{a}~B_{a}^{\mu}
\nonumber\\
\delta^{\prime}H_{j}^{\mu} = G_{j}^{\mu}
\nonumber\\
\delta^{\prime}H_{a} = - m_{a}~B_{a}
\nonumber\\
\delta^{\prime}\psi_{A\alpha} \equiv - i (M t_{a}^{\epsilon})_{AB}~
(\gamma_{\epsilon})_{\alpha\beta}~\psi_{B\beta}~u_{a}
\nonumber\\
\delta^{\prime}\bar{\psi}}_{A\alpha} \equiv - i {\bar{\psi}_{B\beta}
(t_{a}^{- \epsilon} M)_{BA}~(\gamma_{\epsilon})_{\beta\alpha}~u_{a}.
\label{dprime}
\eea
and $0$ for the other Wick submonomials. It is easy to prove that
\be
\delta \delta^{\prime} + \delta^{\prime} \delta = 0,
\quad
(\delta^{\prime})^{2} = 0.
\ee

Finally we define
\be
s \equiv d_{Q} - i\delta, \qquad s^{\prime} \equiv s - i\delta^{\prime} = d_{Q} - i(\delta +\delta^{\prime}).
\label{sp}
\ee
and we have
\be
(s^{\prime})^{2} = 0.
\ee

Moreover we have the structure
\be
s^{\prime} A = 0
\ee
for all expressions
$
A = T^{I}, 
$
$
B_{a\mu}, C_{a\mu},
$
etc. and also for the basic jet variables
$
v_{a\mu}, u_{a}, \tilde{u}_{a}.
$
\newpage
\section{Distributions with Causal Support\label{distributions}}

In the Introduction we have mentioned the appearance of distributions with causal support. The basic example is the Pauli-Jordan causal distribution given by (\ref{pj}) + (\ref{pjpm}). It is known that the degree of singularity of this distribution is
$
\omega(D_{m}) = - 2;
$
this essentially means the the Fourier transform behaves at infinity as
$
\tilde{D}_{m}(p) \sim p^{-2}
$
which follows easily from (\ref{pjpm}). One has the causal decomposition
\be
D_{m} = D_{m}^{adv} - D_{m}^{ret}
\label{pjdec}
\ee
where the {\it advanced} and {\it retarded} distributions are defined by
\be
D_{m}^{adv}(x) \equiv \theta(x_{0})~D_{m}(x) , \qquad
D_{m}^{ret}(x)  \equiv - \theta(- x_{0})~D_{m}(x) .
\label{advret}
\ee
The multiplication of the distributions is allowed in this case and we obtain the well known formulas:
\be
D_{m}^{adv}(x) = - {1 \over (2 \pi)^{4}}~
\int {e^{ - i p\cdot x} \over p^{2} - m^{2} + i \epsilon p_{0}}, \qquad
D_{m}^{ret}(x) = - {1 \over (2 \pi)^{4}}~
\int {e^{ - i p\cdot x} \over p^{2} - m^{2} - i \epsilon p_{0}}
\label{adv+ret}
\ee
as it can be verified by direct computations. Another way of obtaining the prevoius formulas is the use the formulas
\be
\tilde{d}^{adv}(p) = {i \over 2\pi} \int_{-\infty}^{\infty} dt
{\tilde{d}(t p) \over (t - i0)^{\omega + 1}~(1 - t + i0)}
\label{central2}
\ee
if the degree of singularity is
$
\omega \geq 0
$
and
\be
\tilde{d}^{adv}(p) = {i \over 2\pi} \int_{-\infty}^{\infty} dt
{\tilde{d}(t p) \over 1 - t + i0}
\label{central1}
\ee
if
$
\omega < 0;
$
here
$
p \in V^{+}.
$
For the retarded distribution one has to make
$
i0 \mapsto -i0.
$
These are the so-called {\it central splitting} formulas \cite{Sc1}, \cite{Sc2}. Because we have
$
\omega(D_{m}) = - 2
$
we can apply the second formula and we obtain
\be
\tilde{D}_{m}^{adv}(p) = - {1 \over (2 \pi)^{2}}~{1 \over p^{2} - m^{2} + i \epsilon}, \quad
p \in V^{+}.
\label{adv}
\ee

We also have the {\it Feynman} and {\it anti-Feynman} distributions (also called {\it propagator} and {\it anti-propagator})
and given by:
\bea
D_{m}^{F} = D_{m}^{adv} -D_{m}^{(-)} = D_{m}^{ret} -  D_{m}^{(-)}.
\nonumber\\
\bar{D}_{m}^{F} = D_{m}^{(+)} - D_{m}^{adv} = - D_{m}^{ret} +  D_{m}^{(+)}
\label{FantiF}
\eea
with the explicit expressions
\be
D_{m}^{F}(x) = - {1 \over (2 \pi)^{4}}~
\int {e^{ - i k\cdot x} \over k^{2} - m^{2} + i \epsilon}, \qquad
{\bar D}_{m}^{F}(x) = - {1 \over (2 \pi)^{4}}~
\int {e^{ - i k\cdot x} \over k^{2} - m^{2} - i\epsilon}.
\label{F+antiF}
\ee
The advanced, retarded, Feynman and anti-Feynman distributions do not verify Klein-Gordon equation: an anomaly appears:
\be
(\square + m^{2})~D_{m}^{adv} = (\square + m^{2})~D_{m}^{ret}
= (\square + m^{2})~D_{m}^{F} = \delta, \qquad
(\square + m^{2})~\bar{D}_{m}^{F} = - \delta.
\label{KGano}
\ee

We remark that we can use the formula (\ref{central1}) and (\ref{central2}) with over-subtractions:
\be
\tilde{d}^{adv}(p) = {i \over 2\pi} \int_{-\infty}^{\infty} dt
{\tilde{d}(t p) \over (t - i0)^{\omega^{\prime} + 1}~(1 - t + i0)}
\label{central2a}
\ee
with
$
\omega^{\prime} > \omega \geq 0
$
and
\be
\tilde{d}^{adv}(p) = {i \over 2\pi} \int_{-\infty}^{\infty} dt
{\tilde{d}(t p) \over (t - i0)^{\omega^{\prime}}(1 - t + i0)},
\label{central1a}
\ee
with
$
\omega^{\prime} > 0
$
if
$
\omega < 0.
$
We expect to obtain expressions close to (\ref{adv}). Indeed, we have for
$
p \in V^{+}
$
\be
\tilde{D}_{m}^{adv(1)}(p) = {i \over 2\pi} \int_{-\infty}^{\infty} dt
{\tilde{d}(t p) \over (t - i0)(1 - t + i0)} =
- {1 \over (2 \pi)^{2}}~{ p^{2} \over m^{2}}~{1 \over p^{2} - m^{2} + i \epsilon}
\label{adv1}
\ee
and we have
\be
\tilde{D}_{m}^{adv(1)}(p) - \tilde{D}_{m}^{adv}(p) = - {1 \over (2 \pi)^{2}}~{1 \over  m^{2}}
\ee
which agrees with the fact that two causal splittings should difer by a polynomial. Similarly
\be
\tilde{D}_{m}^{adv(2)}(p) = {i \over 2\pi} \int_{-\infty}^{\infty} dt
{\tilde{d}(t p) \over (t - i0)^{3}(1 - t + i0)} =
- {1 \over (2 \pi)^{2}}~\Bigl({ p^{2} \over m^{2}}\Bigl)^{2}~{1 \over p^{2} - m^{2} + i \epsilon}
\label{adv2}
\ee
and
\be
\tilde{D}_{m}^{adv(n)}(p) = {i \over 2\pi} \int_{-\infty}^{\infty} dt
{\tilde{d}(t p) \over (t - i0)^{2n-1}(1 - t + i0)} =
- {1 \over (2 \pi)^{2}}~\Bigl({ p^{2} \over m^{2}}\Bigl)^{n}~{1 \over p^{2} - m^{2} + i \epsilon}.
\label{adv-n}
\ee

The next example of distribution with causal support is
\be
d_{m,m} \equiv (D_{m}^{(+)})^{2} - (D_{m}^{(-)})^{2}.
\label{DD}
\ee
The Fourier transform is given by
\be
\tilde{d}_{m,m}(p) = - {1 \over 4~(2\pi)^{3}}~\epsilon(p_{0})~
\theta(p^{2} - 4~m^{2})~\sqrt{1 - {4~m^{2} \over p^{2}}}
\ee
and one see that this distribution behaves at infinity as a constant, so it haves the degree of singularity
$
\omega(d_{m,m}) = 0.
$
In fact we need a slightly more general case namely
\be
d(D_{m_{1}},D_{m_{2}}) \equiv
D_{m_{1}}^{(+)}~D_{m_{2}}^{(+)} - D_{m_{1}}^{(-)}~D_{m_{2}}^{(-)}
\label{D1D2}
\ee
which also has causal support and
$
\omega(d(D_{m_{1}},D_{m_{2}})) = 0.
$
For simplicity we also use the notations
$
d_{m_{1},m_{2}} = d(D_{1},D_{2}) = d(D_{m_{1}},D_{m_{2}})
$
or even
$
d_{2}
$
if there is no possibility of confusion. The Fourier transform can be computed in this case also and it is
\be
\tilde{d}_{m_{1},m_{2}}(p) = {i \over 2\pi}~\epsilon(p_{0})~\rho(p^{2})
\ee
where
\be
\rho(\mu) \equiv {i \over 4~(2\pi)^{2}}~\theta(\mu - (m_{1} + m_{2})^{2})~
\Bigl[ 1 - {2 (m_{1}^{2} + m_{2}^{2}) \over \mu} + {(m_{1}^{2} - m_{2}^{2})^{2} \over \mu^{2}}\Bigl]^{1/2}.
\ee
Next, we generalize this type of distributions.
\begin{lemma}
Let $d$ a distribution such that the Fourier transform is of the type
\bea
\tilde{d}(p) = \epsilon(p_{0})~\theta(p^{2} - M^{2})~f(p^{2})
\eea
with $f$ some bounded function. Then this distribution has causal support. Consider the case when
$
M > 0
$
and
$
\omega(d) = 0
$
and define the distributions
$
d^{\prime}
$
and
$
d^{\prime\prime}
$
through their Fourier transforms:
\be
\tilde{d}^{\prime}(p) = {1\over p^{2}}~\tilde{d}(p), \qquad
\tilde{d}^{\prime\prime}(p) = {1\over (p^{2})^{2}}~\tilde{d}(p).
\ee
Then the distributions
$
d^{\prime}
$
and
$
d^{\prime\prime}
$
have also causal support. We also have
\be
\omega(d^{\prime}) = - 2, \quad \omega(d^{\prime\prime}) = - 4
\ee
and
\be
\square d^{\prime} = - d, \qquad
\square d^{\prime\prime} = - d^{\prime}
\label{d3}
\ee
\label{ddd}
\end{lemma}
It is useful to write the distribution
$
d_{m_{1},m_{2}}.
$
in the form
\be
d_{2} = \int d\mu~\rho(\mu)~D_{\sqrt{\mu}}.
\label{d0}
\ee
Then one can verify that we also have
\be
d_{2}^{\prime} = \int d\mu~{1 \over \mu}~\rho(\mu) D_{\sqrt{\mu}}.
\label{d1}
\ee
and
\be
d_{2}^{\prime\prime} = \int d\mu~{1 \over \mu^{2}}~\rho(\mu) D_{\sqrt{\mu}}.
\label{d2}
\ee
The causal support property is now obvious.
$\qed$

The formulas for the causal splitting should be obtained by the causal splitting of the causal distribution
$
D_{\sqrt{\mu}}(x).
$
However, we have noticed above that this causal splitting is not unique and we should use a splitting which keeps the integral over $\mu$ well defined. For instance we cannot split the formula (\ref{d0}) using
$
D_{\sqrt{\mu}} = D^{adv}_{\sqrt{\mu}} - D^{ret}_{\sqrt{\mu}}
$
because the integral over $\mu$ will not be convergent, but we can use the splitting
$
D_{\sqrt{\mu}} = D^{adv(1)}_{\sqrt{\mu}} - D^{ret(1)}_{\sqrt{\mu}}
$
according to (\ref{adv1}) and we obtain
\be
d_{2}^{adv} = \int d\mu~\rho(\mu) D^{adv(1)}_{\sqrt{\mu}} = - \square \int d\mu~{1 \over \mu}~\rho(\mu) D^{adv}_{\sqrt{\mu}}.
\label{dadv1}
\ee
In the case of
$
d_{2}^{\prime}
$
and
$
d_{2}^{\prime\prime}
$
one can obtain the causal splitting by multiplication with the Heaviside function
$
\theta(x_{0})
$
and we have
\be
d_{2}^{\prime adv} = \int d\mu~{1 \over \mu}~\rho(\mu) D^{adv}_{\sqrt{\mu}}.
\label{da1}
\ee
and
\be
d_{2}^{\prime\prime adv} = \int d\mu~{1 \over \mu^{2}}~\rho(\mu) D^{adv}_{\sqrt{\mu}}.
\label{da2}
\ee
As a result we have immediately
\be
\square d_{2}^{\prime adv} + d_{2}^{adv} = 0
\label{dd1}
\ee
and
\be
\square d_{2}^{\prime\prime adv} + d_{2}^{\prime adv} = c_{2}
\label{dd2}
\ee
where
\be
c_{n} \equiv \int d\mu~{1 \over \mu^{n}}~\rho(\mu).
\label{cn}
\ee

It is worth noting that there is no way to get rid of the anomaly
$
c_{2}
$
from the right hand side of (\ref{dd2}) which means that the causal splitting did not preserved the order of singularity.

Simplified notation will be used for the associated distributions:
$
d(D_{1},D_{2})^{\prime} = d(D_{m_{1}},D_{m_{2}})^{\prime}
$
and
$
d(D_{1},D_{2})^{\prime\prime} = d(D_{m_{1}},D_{m_{2}})^{\prime\prime}.
$
We first note that we have the symmetry properties
\be
d(D_{1},D_{2}) = d(D_{2},D_{1}), \qquad
d(D_{1},D_{2})^{\prime} = d(D_{2},D_{1})^{\prime}, \qquad
d(D_{1},D_{2})^{\prime\prime} = d(D_{2},D_{1})^{\prime\prime}.
\label{symD}
\ee

In computations we will usually encounter the case when some of the distributions
$
D_{1}, D_{2}
$
are replaced by derivatives. We want to express everything in terms of
$
d_{12},~d_{12}^{\prime},~d_{12}^{\prime\prime}.
$

We have:
\begin{prop}
The following formulas are true
\be
d(\partial_{\mu}D_{1},D_{2}) = {1\over 2}~\partial_{\mu}
[ d(D_{1},D_{2}) + (m_{1}^{2} - m_{2}^{2})~d(D_{1},D_{2})^{\prime} ]
\label{pD}
\ee
\bea
d(D_{1},\partial_{\mu}\partial_{\nu}D_{2}) =
\nonumber\\
{1\over 3}~(\partial_{\mu}\partial_{\nu} - \eta_{\mu\nu}\square)
[ d(D_{1},D_{2}) - (2 m_{1}^{2} - m_{2}^{2})~d(D_{1},D_{2})^{\prime}
+ (m_{1}^{2} - m_{2}^{2})^{2}~d(D_{1},D_{2})^{\prime\prime}]
\nonumber\\
+ {1\over 4}~\eta_{\mu\nu}~\square [ d(D_{1},D_{2}) - 2 (m_{1}^{2} - m_{2}^{2})~d(D_{1},D_{2})^{\prime}
+ (m_{1}^{2} - m_{2}^{2})^{2}~d(D_{1},D_{2})^{\prime\prime}]
\label{ppD}
\eea
\bea
d(\partial_{\mu}D_{1},\partial_{\nu}D_{2}) =
\nonumber\\
{1\over 6}~(\partial_{\mu}\partial_{\nu} - \eta_{\mu\nu}\square)
[ d(D_{1},D_{2}) + ( m_{1}^{2} + m_{2}^{2})~d(D_{1},D_{2})^{\prime}
- 2~(m_{1}^{2} - m_{2}^{2})^{2}~d(D_{1},D_{2})^{\prime\prime}]
\nonumber\\
+ {1\over 4}~\eta_{\mu\nu}~\square [ d(D_{1},D_{2})
- (m_{1}^{2} - m_{2}^{2})^{2}~d(D_{1},D_{2})^{\prime\prime}].
\label{pppD}
\eea
From the last formula we have
\be
d(\partial_{\mu}D_{1},\partial^{\mu}D_{2}) = {1\over 2}~
( \square + m_{1}^{2} + m_{2}^{2} )~d(D_{1},D_{2}).
\label{pppD1}
\ee
\end{prop}
{\bf Proof:} (i) From Lorentz covariance we must have
\be
d(\partial_{\mu}D_{1},D_{2}) = \partial_{\mu}A_{12}
\ee
with
$
A_{12}
$
a Lorentz invariant function. If we apply
$
\partial^{\mu}
$
we get
\be
\square A_{12} = - m_{1}^{2}~d(D_{1},D_{2})
+ d(\partial^{\mu}D_{1},\partial_{\mu}D_{2})
\label{sA}
\ee
On the other hand, if we apply
$\square$
to
$
d(D_{1},D_{2})
$
we immediately obtain
\be
\square~d(D_{1},D_{2}) = 2~d(\partial_{\mu}D_{1},\partial^{\mu}D_{2})
- (m_{1}^{2} + m_{2}^{2})~d(D_{1},D_{2})
\label{pppD2}
\ee
and from here formula (\ref{pppD1}). Inserting in formula (\ref{sA}) we get:
\be
\square [~2~A_{12} - d(D_{1},D_{2})
- ( m_{1}^{2} - m_{2}^{2})~d(D_{1},D_{2})^{\prime} ] = 0.
\ee
The equation
$
\square~f = 0
$
has the solution
$
f = \lambda~D_{0}
$
so we must have
\be
A_{12} = {1\over 2}~[ d(D_{1},D_{2})
+ ( m_{1}^{2} - m_{2}^{2})~d(D_{1},D_{2})^{\prime} ] + \lambda~D_{0}.
\ee
and from here
\be
d(\partial_{\mu}D_{1},D_{2}) - {1\over 2}~\partial_{\mu}
[ d(D_{1},D_{2}) + (m_{1}^{2} - m_{2}^{2})~d(D_{1},D_{2})^{\prime} ]
= \lambda~\partial_{\mu}D_{0}.
\label{pD1}
\ee
We prove that
$
\lambda = 0
$
in the following way: if all masses are null, that one considers the scaling behavior of left hand side which is
$
LHS(\lambda x) = \lambda^{-5}~LHS(x).
$
Because the scaling behavior of the right hand side is
$
RHS(\lambda x) = \lambda^{-2}~RHS(x)
$
we must have
$
\lambda = 0.
$
If at least one of the masses
$
m_{1}, m_{2}
$
is not null, then we consider the support properties in the momentum space of LHS and RHS of (\ref{pD1}). Then RHS has support in
$
k^{2} = 0
$
but the support of LHS does not contain this points, so again we must have
$
\lambda = 0.
$
In this way we get (\ref{pD}).

(ii) Next we have from Lorentz covariance
\be
d(D_{1},\partial_{\mu}\partial_{\nu}D_{2}) =
\partial_{\mu}\partial_{\nu}~B_{12} + \eta_{\mu\nu}~C_{12}
\ee
where
$
B_{12}, C_{12}
$
are Lorentz invariant functions. Contracting with
$
\eta^{\mu\nu}
$
we obtain
\be
C_{12} = - {1\over 4}~\square B_{12} - {m_{2}^{2}\over 4}~d(D_{1},D_{2})
\ee
so
\be
d(D_{1},\partial_{\mu}\partial_{\nu}D_{2}) =
~\left(\partial_{\mu}\partial_{\nu} - {1\over 4}~\eta_{\mu\nu}\square\right)B_{12} - \eta_{\mu\nu}~{m_{2}^{2}\over 4}~d(D_{1},D_{2}).
\label{ppD1}
\ee

If we apply
$
\partial^{\mu}\partial^{\nu}
$
to the left hand side we get
\be
\partial^{\mu}\partial^{\nu}d(D_{1},\partial_{\mu}\partial_{\nu}D_{2}) =
d(\partial^{\mu}\partial^{\nu}D_{1},\partial_{\mu}\partial_{\nu}D_{2})
- 2~m_{2}^{2}~d(\partial^{\mu}D_{1},\partial_{\mu}D_{2})
+ m_{2}^{4}~d(D_{1},D_{2}).
\ee
If we apply
$
\partial^{\mu}\partial^{\nu}
$
to the right hand side we get
\be
\partial^{\mu}\partial^{\nu}d(D_{1},\partial_{\mu}\partial_{\nu}D_{2}) =
{3\over 4}~\square^{2} B_{12} - {1\over 4}~m_{2}^{2}~d(D_{1},D_{2})
\ee
so by comparison
\be
3~\square^{2}B_{12} =
4~d(\partial^{\mu}\partial^{\nu}D_{1},\partial_{\mu}\partial_{\nu}D_{2})
- 8~m_{2}^{2}~d(\partial^{\mu}D_{1},\partial_{\mu}D_{2})
+ m_{2}^{2}~\square d(D_{1},D_{2})
+ 4 m_{2}^{4}~d(D_{1},D_{2}).
\label{B121}
\ee
The expression
$
d(\partial^{\mu}D_{1},\partial_{\mu}D_{2})
$
has been computed before (\ref{pppD1}). In the same way, we apply
$
\square^{2}
$
to
$
d(D_{1},D_{2})
$
and obtain
\bea
\square^{2}d(D_{1},D_{2}) =
4~d(\partial^{\mu}\partial^{\nu}D_{1},\partial_{\mu}\partial_{\nu}D_{2})
\nonumber\\
- 2~( m_{1}^{2} + m_{2}^{2})~\square d(D_{1},D_{2})
- ( m_{1}^{2} + m_{2}^{2})^{2}~d(D_{1},D_{2})
\eea
and from here we have
\bea
d(\partial^{\mu}\partial^{\nu}D_{1},\partial_{\mu}\partial_{\nu}D_{2}) =
\nonumber\\
{1\over 4}~[ \square^{2} + 2~( m_{1}^{2} + m_{2}^{2})~\square
+ ( m_{1}^{2} + m_{2}^{2})^{2} ]~d(D_{1},D_{2}).
\label{DDD}
\eea
Then we get
\be
\square^{2} [~3~B_{12} - d(D_{1},D_{2})
+ (2 m_{1}^{2} - m_{2}^{2})~d(D_{1},D_{2})^{\prime}
- ( m_{1}^{2} - m_{2}^{2})^{2} d(D_{1},D_{2})^{\prime\prime} ] = 0.
\ee
From here
\be
3~B_{12} = d(D_{1},D_{2})
- (2 m_{1}^{2} - m_{2}^{2})~d(D_{1},D_{2})^{\prime}
+ ( m_{1}^{2} - m_{2}^{2})^{2} d(D_{1},D_{2})^{\prime\prime} + \lambda~D_{0}
\ee
and one can prove that
$
\lambda = 0
$
as at point (i) of the proof, so
\be
3~B_{12} = d(D_{1},D_{2})
- (2 m_{1}^{2} - m_{2}^{2})~d(D_{1},D_{2})^{\prime}
+ ( m_{1}^{2} - m_{2}^{2})^{2} d(D_{1},D_{2})^{\prime\prime}.
\ee
If we substitute this expression for
$
B_{12}
$
in (\ref{ppD1}) we get the second formula from the statement.

(iii) Finally, as above, we have
\be
d(\partial_{\mu}D_{1},\partial_{\nu}D_{2}) =
\partial_{\mu}\partial_{\nu}~E_{12} + \eta_{\mu\nu}~F_{12}
\ee
where
$
E_{12}, F_{12}
$
are Lorentz invariant functions. Contracting with
$
\eta^{\mu\nu}
$
and using ({\ref{pppD1}) we obtain
\be
F_{12} = - {1\over 4}~\square E_{12}
+  {1\over 8}~( \square + m_{1}^{2} + m_{2}^{2})~d(D_{1},D_{2})
\ee
so
\be
d(\partial_{\mu}D_{1},\partial_{\nu}D_{2}) =
~\left(\partial_{\mu}\partial_{\nu} - {1\over 4}~\eta_{\mu\nu}\square\right)E_{12}
+ {1\over 8}~
\eta_{\mu\nu}~(\square + m_{1}^{2} + m_{2}^{2})~d(D_{1},D_{2}).
\label{pppD3}
\ee

If we apply
$
\partial^{\mu}\partial^{\nu}
$
to the left hand side we get
\be
\partial^{\mu}\partial^{\nu}d(\partial_{\mu}D_{1},\partial_{\nu}D_{2}) =
d(\partial^{\mu}\partial^{\nu}D_{1},\partial_{\mu}\partial_{\nu}D_{2})
- ( m_{1}^{2} + m_{2}^{2})~d(\partial^{\mu}D_{1},\partial_{\mu}D_{2})
+ m_{1}^{2}~m_{2}^{2}~d(D_{1},D_{2})
\ee

If we apply
$
\partial^{\mu}\partial^{\nu}
$
to the right hand side we get
\be
\partial^{\mu}\partial^{\nu}d(D_{1},\partial_{\mu}\partial_{\nu}D_{2}) =
{3\over 4}~\square F_{12}
- {1\over 8}~\square (\square + m_{1}^{2} + m_{2}^{2})~d(D_{1},D_{2})
\ee
so by comparison
\bea
3
~\square^{2}F_{12} =
4~d(\partial^{\mu}\partial^{\nu}D_{1},\partial_{\mu}\partial_{\nu}D_{2})
- 4~( m_{1}^{2} + m_{2}^{2})~d(\partial^{\mu}D_{1},\partial_{\mu}D_{2})
\nonumber\\
+ 4 m_{1}^{2}~m_{2}^{2}~d(D_{1},D_{2})
- {1\over 2}~\square (\square + m_{1}^{2} + m_{2}^{2})~d(D_{1},D_{2})
\label{F12}
\eea

If we use (\ref{pppD1}) and (\ref{DDD}) we obtain
\be
\square^{2} [ 6~F_{12} - d(D_{1},D_{2})
- (m_{1}^{2} + m_{2}^{2})~d(D_{1},D_{2})^{\prime}
+ 2~( m_{1}^{2} - m_{2}^{2})^{2} d(D_{1},D_{2})^{\prime\prime} ] = 0.
\ee
so
\be
F_{12} = {1\over 6}~d(D_{1},D_{2})
+ {1\over 6}~(m_{1}^{2} + m_{2}^{2})~d(D_{1},D_{2})^{\prime}
- {1\over 3}~( m_{1}^{2} - m_{2}^{2})^{2} d(D_{1},D_{2})^{\prime\prime}
+ \lambda~D_{0}.
\ee

As above, we can prove that
$
\lambda = 0
$
so
\be
F_{12} = {1\over 6}~d(D_{1},D_{2})
+ {1\over 6}~(m_{1}^{2} + m_{2}^{2})~d(D_{1},D_{2})^{\prime}
- {1\over 3}~( m_{1}^{2} - m_{2}^{2})^{2} d(D_{1},D_{2})^{\prime\prime}
\ee
and after some computation formula (\ref{pppD}) follows.
$\qed$

\newpage
\section{Second Order Gauge Invariance. Loop Contributions\label{loop}}

We concentrate now on the loop term from the chronological products. It will be a generalization of the second term from (\ref{wick}) to the case of the standard mode. Explicitly we have the following formulas.
\bea
T(T(x_{1})^{(1)}, T(x_{2})^{(1)}) =
: v_{a\mu}(x_{1})~v_{b\nu}(x_{2}):~
T_{00}(C_{a}^{\mu}(x_{1}), C_{b}^{\nu}(x_{2}))
\nonumber\\
+ {1\over 4}~: F_{a\nu\mu}(x_{1})~F_{b\sigma\rho}(x_{2}):~
T_{00}(E_{a}^{\mu\nu}(x_{1}), E_{b}^{\rho\sigma}(x_{2}))
\nonumber\\
+ {1\over 2}~: v_{a\mu}(x_{1})~F_{b\sigma\rho}(x_{2}):~
T_{00}(C_{a}^{\mu}(x_{1}),E_{b}^{\rho\sigma}(x_{2}))
+ ( 1 \longleftrightarrow 2)
\nonumber\\
- [ : u_{a}(x_{1})~\partial_{\mu}\tilde{u}_{b}(x_{2}):~
T_{00}(D_{a}(x_{1}), B_{b}^{\mu}(x_{2}))
+ ( 1 \longleftrightarrow 2) ]
\nonumber\\
- [ : u_{a}(x_{1})~\tilde{u}_{b}(x_{2}):~
T_{00}(D_{a}(x_{1}), K_{b}(x_{2}))
+ ( 1 \longleftrightarrow 2) ]
\nonumber\\
+ : v_{a\mu}(x_{1})~\Phi_{b}(x_{2}):~
T_{00}(C_{a}^{\mu}(x_{1}), G_{b}(x_{2}))
+ ( 1 \longleftrightarrow 2)
\nonumber\\
+ : v_{a\mu}(x_{1})~\partial_{\nu}\Phi_{b}(x_{2}):~
T_{00}(C_{a}^{\mu}(x_{1}), H_{b}^{\nu}(x_{2}))
+ ( 1 \longleftrightarrow 2)
\nonumber\\
+ : \Phi_{a}(x_{1})~\Phi_{b}(x_{2}):~
T_{00}(G_{a}(x_{1}), G_{b}(x_{2}))
\nonumber\\
+ : \Phi_{a}(x_{1})~\partial_{\nu}\Phi_{b}(x_{2}):~
T_{00}(G_{a}(x_{1}), H_{b}^{\nu}(x_{2}))
+ ( 1 \longleftrightarrow 2)
\nonumber\\
+ : \partial_{\mu}\Phi_{a}(x_{1})~\partial_{\nu}\Phi_{b}(x_{2}):~
T_{00}(H_{a}^{\mu}(x_{1}), H_{b}^{\nu}(x_{2}))
\nonumber\\
+ {1\over 2}~: F_{a\nu\mu}(x_{1})~\Phi_{b}(x_{2}):~
T_{00}(E_{a}^{\mu\nu}(x_{1}), G_{b}(x_{2}))
+ ( 1 \longleftrightarrow 2)
\nonumber\\
- [ :\bar{\psi}_{A\alpha}(x_{1})~\psi_{B\beta}(x_{2}):~
T_{00}(V_{A\alpha}(x_{1}), \bar{V}_{B\beta}(x_{2}))
+ ( 1 \longleftrightarrow 2) ]
\nonumber\\
+ : v_{a\mu}(x_{1})~\phi_{j}(x_{2}):~
T_{00}(C_{a}^{\mu}(x_{1}), G_{j}(x_{2}))
+ ( 1 \longleftrightarrow 2)
\nonumber\\
+ : v_{a\mu}(x_{1})~\partial_{\nu}\phi_{j}(x_{2}):~
T_{00}(C_{a}^{\mu}(x_{1}), H_{j}^{\nu}(x_{2}))
+ ( 1 \longleftrightarrow 2)
\nonumber\\
+ : \Phi_{a}(x_{1})~\phi_{k}(x_{2}):~
T_{00}(G_{a}(x_{1}), G_{k}(x_{2}))
+ ( 1 \longleftrightarrow 2)
\nonumber\\
+ : \partial_{\mu}\Phi_{a}(x_{1})~\phi_{k}(x_{2}):~
T_{00}(H_{a}^{\mu}(x_{1}), G_{k}(x_{2}))
+ ( 1 \longleftrightarrow 2)
\nonumber\\
+ : \Phi_{a}(x_{1})~\partial_{\nu}\phi_{j}(x_{2}):~
T_{00}(G_{a}(x_{1}), H_{j}^{\nu}(x_{2}))
+ ( 1 \longleftrightarrow 2)
\nonumber\\
+ : \partial_{\mu}\Phi_{a}(x_{1})~\partial_{\nu}\phi_{k}(x_{2}):~
T_{00}(H_{a}^{\mu}(x_{1}), H_{k}^{\nu}(x_{2}))
+ ( 1 \longleftrightarrow 2)
\nonumber\\
: \phi_{j}(x_{1})~\phi_{k}(x_{2}):~
T_{00}(G_{j}(x_{1}), G_{k}(x_{2}))
\nonumber\\
+ : \phi_{j}(x_{1})~\partial_{\nu}\phi_{k}(x_{2}):~
T_{00}(G_{j}(x_{1}), H_{k}^{\nu}(x_{2}))
+ ( 1 \longleftrightarrow 2)
\nonumber\\
+ : \partial_{\mu}\phi_{j}(x_{1})~\partial_{\nu}\phi_{k}(x_{2}):~
T_{00}(H_{j}^{\mu}(x_{1}), H_{k}^{\nu}(x_{2}))
\nonumber\\
+ {1\over 2}~: F_{a\nu\mu}(x_{1})~\phi_{k}(x_{2}):~
T_{00}(E_{a}^{\mu\nu}(x_{1}), G_{k}(x_{2}))
+ ( 1 \longleftrightarrow 2)
\label{t1-2}
\eea
\bea
T(T^{\mu}(x_{1})^{(1)}, T(x_{2})^{(1)}) =
- : u_{a}(x_{1})~v_{b\nu}(x_{2}):~
T_{00}(D_{a}^{\mu}(x_{1}), C_{b}^{\nu}(x_{2}))
\nonumber\\
- {1\over 2}~: u_{a}(x_{1})~F_{b\sigma\rho}(x_{2}):~
T_{00}(D_{a}^{\mu}(x_{1}),E_{b}^{\rho\sigma}(x_{2}))
\nonumber\\
- : v_{a\nu}(x_{1})~u_{b}(x_{2}):~
T_{00}(C_{a}^{\nu\mu}(x_{1}), D_{b}(x_{2}))
\nonumber\\
- : F_{a}^{\mu\nu}(x_{1})~u_{b}(x_{2}):~
T_{00}(B_{a\nu}(x_{1}), D_{b}(x_{2}))
\nonumber\\
- : u_{a}(x_{1})~\Phi_{b}(x_{2}):~
T_{00}(D_{a}^{\mu}(x_{1}), G_{b}(x_{2}))
\nonumber\\
:-  u_{a}(x_{1})~\partial_{\nu}\Phi_{b}(x_{2}):~
T_{00}(D_{a}^{\mu}(x_{1}), H_{b}^{\nu}(x_{2}))
\nonumber\\
- : \Phi_{a}(x_{1})~u_{b}(x_{2}):~
T_{00}(G_{a}^{\mu}(x_{1}), D_{b}(x_{2}))
\nonumber\\
- : \partial_{\nu}\Phi_{a}(x_{1})~u_{b}(x_{2}):~
T(H_{a}^{\mu,\nu}(x_{1}), D_{b}(x_{2}))
\nonumber\\
- : u_{a}(x_{1})~\phi_{j}(x_{2}):~
T_{00}(D_{a}^{\mu}(x_{1}), G_{j}(x_{2}))
\nonumber\\
- : u_{a}(x_{1})~\partial_{\nu}\phi_{j}(x_{2}):~
T_{00}(D_{a}^{\mu}(x_{1}), H_{j}^{\nu}(x_{2}))
\nonumber\\
- : \phi_{j}(x_{1})~u_{b}(x_{2}):~
T(G_{j}^{\mu}(x_{1}), D_{b}(x_{2}))
\nonumber\\
- : \partial_{\nu}\phi_{j}(x_{1})~u_{b}(x_{2}):~
T(H_{j}^{\mu,\nu}(x_{1}), D_{b}(x_{2}))
\label{tmu1-2}
\eea
\be
T(T^{\mu}(x_{1})^{(1)}, T^{\nu}(x_{2})^{(1)}) =
: u_{a}(x_{1})~u_{b}(x_{2}):~
T_{00}(D_{a}^{\mu}(x_{1}), D_{b}^{\nu}(x_{2}))
\label{tmu-nu-2}
\ee

\be
T(T^{\mu\nu}(x_{1})^{(1)}, T(x_{2})^{(1)}) =
: u_{a}(x_{1})~u_{b}(x_{2}):~
T_{00}(C_{a}^{[\mu\nu]}(x_{1}), D_{b}(x_{2}))
\label{tmunu1-2}
\ee

\be
T(T^{\mu\nu}(x_{1})^{(1)}, T^{\rho}(x_{2})^{(1)}) = 0,\qquad
T(T^{\mu\nu}(x_{1})^{(1)}, T^{\rho\sigma}(x_{2})^{(1)}) = 0.
\label{tmunurho-2}
\ee
We now study the gauge invariance condition (\ref{brst-n}) for the one-loop contributions described above. We have the following result:
\begin{thm}
The following relations are true:
\bea
sT(T(x_{1})^{(1)}, T(x_{2})^{(1)}) =
\nonumber\\
- [: u_{a}(x_{1})~v_{b\nu}(x_{2}):~
s^{\prime}T_{00}(D_{a}(x_{1}), C_{b}^{\nu}(x_{2}))
+ ( 1 \longleftrightarrow 2)]
\nonumber\\
+ {1\over 2}~: [ u_{a}(x_{1})~F_{b\sigma\rho}(x_{2}):~
s^{\prime}T_{00}(D_{a}(x_{1}),E_{b}^{\rho\sigma}(x_{2}))
+ ( 1 \longleftrightarrow 2) ]
\nonumber\\
+ : u_{a}(x_{1})~\Phi_{b}(x_{2}):~
s^{\prime}T_{00}(D_{a}(x_{1}), G_{b}(x_{2}))
+ ( 1 \longleftrightarrow 2)
\nonumber\\
+ : u_{a}(x_{1})~\partial_{\mu}\Phi_{b}(x_{2}):~
s^{\prime}T_{00}(D_{a}(x_{1}), H_{b}^{\mu}(x_{2}))
+ ( 1 \longleftrightarrow 2)
\nonumber\\
+ : u_{a}(x_{1})~\phi_{j}(x_{2}):~
s^{\prime}T_{00}(D_{a}(x_{1}), G_{j}(x_{2}))
+ ( 1 \longleftrightarrow 2)
\nonumber\\
+ : u_{a}(x_{1})~\partial_{\mu}\phi_{j}(x_{2}):~
s^{\prime}T_{00}(D_{a}(x_{1}), H_{j}^{\mu}(x_{2}))
+ ( 1 \longleftrightarrow 2)
\label{st1-2}
\eea
\be
sT(T^{\mu}(x_{1})^{(1)}, T(x_{2})^{(1)}) =
\nonumber\\
- : u_{a}(x_{1})~u_{b}(x_{2}):~
s^{\prime}T_{00}(D_{a}^{\mu}(x_{1}), D_{b}(x_{2}))
\label{stmu1-2}
\ee
\be
sT(T^{\mu}(x_{1})^{(1)}, T^{\nu}(x_{2})^{(1)}) = 0
\label{stmu-nu-2}
\ee
\be
sT(T^{\mu\nu}(x_{1})^{(1)}, T(x_{2})^{(1)}) = 0
\label{stmunu1-2}
\ee
\be
sT(T^{\mu\nu}(x_{1})^{(1)}, T^{\rho}(x_{2})^{(1)}) = 0.
\label{stmunurho-2}
\ee
\label{stt-loop}
\end{thm}
The proof is based on direct computations and we refer to
\cite{wick+hopf} for some details. One is reduced to numerical equations of the type
\be
s^{\prime}T_{00} = - i~(\delta + \delta^{\prime})T_{00}
\label{g2}
\ee
according to (\ref{sp}).

The explicit computations of the expressions of the type
$
T_{00}
$
are quite long. We will use the {\it canonical causal splitting}. By this, we mean that:

(a) we use  relations (\ref{2-scalars}), (\ref{2-diracs}) and
(\ref{2-massless+massive-vectors});

(b) we derive from here the canonical (anti)commutation relations
\be
[ \phi_{j}(x), \phi_{k}(y) ] = - i~\delta_{jk} D_{m_{j}}(x - y),
~j,k \in I_{3}
\label{ccr1}
\ee
\bea
\{ \psi_{A\alpha}(x_{1}), \bar{\psi}_{B\beta}(x_{2}) \} =
- i~\delta_{AB}~S_{A}(x_{1} - x_{2})_{\alpha\beta}
\nonumber \\
\{ \bar{\psi}_{A\alpha}(x_{1}), \psi_{B\beta}(x_{2}) \} =
- i~\delta_{AB}~S_{A}(x_{2} - x_{1})_{\beta\alpha},~
A, B \in I_{4}
\label{accr}
\eea
\bea
~[ v^{\mu}_{a}(x_{1}), v^{\nu}_{b}(x_{2})] =
i~\eta^{\mu\nu}~\delta_{ab}~D_{a}(x_{1} - x_{2}),
\nonumber \\
\{ u_{a}(x_{1}), \tilde{u}_{b}(x_{2}) \} = - i~\delta_{ab}~
D_{a}(x_{1} - x_{2}),
\nonumber\\
\{ \tilde{u}_{a}(x_{1}), u_{b}(x_{2}) \} = i~\delta_{ab}~
D_{a}(x_{1} - x_{2})
\nonumber\\
~[ \Phi_{a}(x_{1}), \Phi_{b}(x_{2}) ] = - i~\delta_{ab}~
D_{a}(x_{1} - x_{2}), \qquad a \in I_{1} \cup I_{2}.
\label{ccr2}
\eea

(c) We perform the causal splitting (which is unique) and get

\be
T(\phi_{j}(x), \phi_{k}(y)) = - i~\delta_{jk} D^{F}_{m_{j}}(x - y),
~j,k \in I_{3}
\label{cp1}
\ee
\bea
T(\psi_{A\alpha}(x_{1}), \bar{\psi}_{B\beta}(x_{2})) =
- i~\delta_{AB}~S^{F}_{A}(x_{1} - x_{2})_{\alpha\beta}
\nonumber \\
T(\bar{\psi}_{A\alpha}(x_{1}), \psi_{B\beta}(x_{2}))  =
- i~\delta_{AB}~S^{F}_{A}(x_{2} - x_{1})_{\beta\alpha},~
A, B \in I_{4}
\label{cp2}
\eea
\bea
T(v^{\mu}_{a}(x_{1}), v^{\nu}_{b}(x_{2})) =
i~\eta^{\mu\nu}~\delta_{ab}~D^{F}_{a}(x_{1} - x_{2}),
\nonumber \\
T(u_{a}(x_{1}), \tilde{u}_{b}(x_{2})) = - i~\delta_{ab}~
D^{F}_{a}(x_{1} - x_{2}),
\nonumber\\
T(\tilde{u}_{a}(x_{1}), u_{b}(x_{2})) = i~\delta_{ab}~
D^{F}_{a}(x_{1} - x_{2})
\nonumber\\
T(\Phi_{a}(x_{1}), \Phi_{b}(x_{2})) = - i~\delta_{ab}~
D^{F}_{a}(x_{1} - x_{2}), \qquad a \in I_{1} \cup I_{2}.
\label{cp3}
\eea

(d) We take
\be
T^{c}(\partial_{\mu}\xi(x_{1}),\xi^{\prime}(x_{2})) =
\partial^{1}_{\mu}T(\xi(x_{1}),\xi^{\prime}(x_{2})),
\quad
T^{c}(\partial_{\mu}\xi(x_{1}),\partial_{\nu}\xi^{\prime}(x_{2})) =
\partial^{1}_{\mu}\partial^{2}_{\nu}T(\xi(x_{1}),\xi^{\prime}(x_{2}))
\label{cp4}
\ee
for any of the basic fields
$
\xi, \xi^{\prime} = \phi_{j}, \psi_{A\alpha}, \bar{\psi}_{B\beta},
v_{a}^{\mu}, u_{a}, \tilde{u}_{a}, \Phi_{a}.
$
In this way we obtain by tedious computations all expressions of the type
$
T_{00}(\xi\cdot T^{I}(x_{1}),\xi^{\prime}\cdot T^{J}(x_{2})).
$
For simplicity we use the notation
$
d_{ab} \equiv d(D_{m_{a}},D_{m_{b}});
$
then we have a causal splitting
$
d_{ab} = d^{adv}_{ab} - d^{ret}_{ab}
$
and the associated Feynman propagators
$
d^{F}_{ab}, d^{\prime F}_{ab}, d^{\prime\prime F}_{ab}.
$
With these notations we have
\be
T_{00}(C_{a}^{\mu}(x_{1}),C_{b}^{\nu}(x_{2}))
= (\partial^{\mu}\partial^{\nu} - \eta^{\mu\nu}~\square) A^{F}_{ab}
+ \eta^{\mu\nu}~B^{\mu}_{ab}
\label{l1}
\ee
where
\bea
A^{F}_{ab} = {1\over 3}~
\left(f_{acd}~f_{bcd} + {1\over 2}~f^{\prime}_{cda}~f^{\prime}_{cdb}\right)~
[ d^{F}_{cd} - 2~(m_{c}^{2} + m_{d}^{2})~d^{\prime F}_{cd}
+ 4~(m_{c}^{2} - m_{d}^{2})^{2}~d^{\prime\prime F}_{cd} ]
\nonumber\\
+ {2\over 3}~
[ (t_{a})_{AB}~(t_{b})_{BA}
+ (t^{\prime}_{a})_{AB}~(t^{\prime}_{b})_{BA} ] ~
[ d^{F}_{AB} + (M_{A}^{2} + M_{B}^{2})~d^{\prime F}_{AB}
- 2~(M_{A}^{2} - M_{B}^{2})^{2}~d^{\prime\prime F}_{AB} ]
\nonumber\\
+ {1\over 3}~f^{\prime}_{jca}~f^{\prime}_{jcb}~
[ d^{F}_{jc} - 2~(m_{j}^{2} + m_{d}^{2})~d^{\prime F}_{jc}
+ 4~(m_{j}^{2} - m_{c}^{2})^{2}~d^{\prime\prime F}_{jc} ]
\nonumber\\
+ {1\over 3}~f^{\prime}_{jka}~f^{\prime}_{jkb}~
[ d^{F}_{jk} - 2~(m_{j}^{2} + m_{k}^{2})~d^{\prime F}_{jk}
+ 4~(m_{j}^{2} - m_{k}^{2})^{2}~d^{\prime\prime F}_{jk} ].
\label{l1a}
\eea
and
\bea
B^{F}_{ab} =
- \left(f_{acd}~f_{bcd}
+ {1\over 2}~f^{\prime}_{cda}~f^{\prime}_{cdb}\right)~
~(m_{c}^{2} - m_{d}^{2})^{2}~d^{\prime F}_{cd}
\nonumber\\
+ [ - f_{acd}~f_{bcd}~m_{d}^{2}
+ (f^{\prime}_{cda}~m_{d} + f^{\prime}_{cad}~m_{a})
(f^{\prime}_{cdb}~m_{d} + f^{\prime}_{cbd}~m_{b}) ]~d^{F}_{cd}
\nonumber\\
- 2~[ (M_{A} - M_{B})^{2}~(t_{a})_{AB}~(t_{b})_{BA}
+ (M_{A} + M_{B})^{2}~(t^{\prime}_{a})_{AB}~(t^{\prime}_{b})_{BA} ]
~d^{F}_{AB}
\nonumber\\
+ 2~(M_{A}^{2} - M_{B}^{2})^{2}~[ (t_{a})_{AB}~(t_{b})_{BA}
+ (t^{\prime}_{a})_{AB}~(t^{\prime}_{b})_{BA} ]
~d^{\prime F}_{AB}
\nonumber\\
- f^{\prime}_{jca}~f^{\prime}_{jcb}~(m_{j}^{2} - m_{c}^{2})~
d^{\prime F}_{jc}
- {1\over 2}~f^{\prime}_{jka}~f^{\prime}_{jkb}~d^{\prime F}_{jk}
\nonumber\\
+ (f^{\prime}_{jca}~m_{c} + f^{\prime}_{jac}~m_{a})
(f^{\prime}_{jcb}~m_{d} + f^{\prime}_{jbc}~m_{b})~d^{F}_{jc}
\label{l1b}
\eea

Next:
\be
T_{00}(C_{a}^{\mu}(x_{1}),E_{b}^{\rho\sigma}(x_{2}))
= {1\over 2}~f_{acd}~f_{bcd}~
(\eta^{\mu\sigma}\partial^{\rho} - \eta^{\mu\rho}\partial^{\sigma})~
d^{F}_{cd}
\label{l2}
\ee

\bea
T_{00}(G_{a}(x_{1}),C_{b}^{\nu}(x_{2}) =
- f^{\prime}_{acd}~f^{\prime}_{cdb}~m_{d}~\partial^{\nu}d^{F}_{cd}
\nonumber\\
- (m_{c}^{2} - m_{d}^{2})~f^{\prime}_{acd}~
( 3~f^{\prime}_{cdb}~m_{d} - 2~f^{\prime}_{cbd}~m_{b} )~
\partial^{\nu}d^{\prime F}_{cd}
\nonumber\\
+ 2 i~[ (M_{A} - M_{B})~(s_{a})_{AB}~(t_{b})_{BA}
+ (M_{a} + M_{B})~(s^{\prime}_{a})_{AB}~(t^{\prime}_{b})_{BA} ]~
\partial^{\nu}d^{F}_{AB}
\nonumber\\
- 2 i~(M_{A}^{2} - M_{B}^{2})~[(M_{A} + M_{B})(s_{a})_{AB}~(t_{b})_{BA}
+ (M_{A} - M_{B})~(s^{\prime}_{a})_{AB}~(t^{\prime}_{b})_{BA} ]~
\partial^{\nu}d^{\prime F}_{AB}
\nonumber\\
+ {1\over 2}~f^{\prime}_{jac}~
(f^{\prime}_{jcb}~m_{c} + f^{\prime}_{jbc}~m_{b})~
[  \partial^{\nu}d^{F}_{jc}
+ (m_{j}^{2} - m_{c}^{2})\partial^{\nu}d^{\prime F}_{cd} ]
\nonumber\\
- f^{\prime\prime}_{jac}~f^{\prime}_{jcb}~
(m_{j}^{2} - m_{c}^{2})\partial^{\nu}d^{\prime F}_{jc}
- {1\over 2}~f^{\prime\prime}_{jka}~f^{\prime}_{jkb}~
(m_{j}^{2} - m_{k}^{2})\partial^{\nu}d^{\prime F}_{jk}
\label{l3}
\eea

\bea
T_{00}(G_{j}(x_{1}),C_{b}^{\nu}(x_{2})) =
- {1\over 2}~f^{\prime}_{jcd}~
( f^{\prime}_{bac}~m_{a} + f^{\prime}_{bca}~m_{c} )~
\partial^{\nu}d^{F}_{bc}
\nonumber\\
+ (m_{b}^{2} - m_{c}^{2})~\left[ f^{\prime}_{jbc}~
( 2~f^{\prime}_{bac}~m_{a} - 3~f^{\prime}_{bca}~m_{c} )
- {1\over 2}~f^{\prime\prime}_{jbc}~f^{\prime}_{bca} \right]
\partial^{\nu}d^{\prime F}_{bc}
\nonumber\\
- {1\over 2}~f^{\prime}_{jkb}~
(f^{\prime}_{kba}~m_{b} + f^{\prime}_{kab}~m_{a})~
\partial^{\nu}d^{F}_{jb}
\nonumber\\
+ {1\over 2}~(m_{b}^{2} - m_{k}^{2})
[ f^{\prime}_{jkb}~( f^{\prime}_{kba}~m_{b} + f^{\prime}_{kab}~m_{a})
+ 2~f^{\prime\prime}_{jkb}~f^{\prime}_{kba} ]~
\partial^{\nu}d^{\prime F}_{kb} ]
\nonumber\\
+ {1\over 2}~(m_{l}^{2} - m_{k}^{2})~\lambda_{jkl}~f^{\prime}_{kla}~
\partial^{\nu}d^{\prime F}_{kl}
\nonumber\\
+ 2 i~[ (M_{A} - M_{B})~(s_{j})_{AB}~(t_{a})_{BA}
+ (M_{A} + M_{B})~(s^{\prime}_{j})_{AB}~(t^{\prime}_{a})_{BA} ]~
\partial^{\nu}d^{F}_{AB}
\nonumber\\
- 2 i~(M_{A} - M_{B})^{2}~[(M_{A} + M_{B})(s_{j})_{AB}~(t_{a})_{BA}
+ (M_{A} - M_{B})~(s^{\prime}_{j})_{AB}~(t^{\prime}_{a})_{BA} ]~
\partial^{\nu}d^{\prime F}_{AB}
\label{l4}
\eea

\be
T_{00}(H^{\mu}_{a}(x_{1}),C_{b}^{\nu}(x_{2})) =
\eta^{\mu\nu}[ f^{\prime}_{acd}~
( f^{\prime}_{cdb}~m_{d} + f^{\prime}_{cbd}~m_{b} )~d^{F}_{cd}
- f^{\prime}_{jac}~( f^{\prime}_{jcb}~m_{c} + f^{\prime}_{jbc}~m_{b} )~
d^{F}_{jc} ]
\label{l5}
\ee

\be
T_{00}(H^{\mu}_{j}(x_{1}),C_{b}^{\nu}(x_{2})) =
\eta^{\mu\nu}[ f^{\prime}_{jcd}~
( f^{\prime}_{cdb}~m_{d} + f^{\prime}_{cbd}~m_{b} )~d^{F}_{cd}
- f^{\prime}_{jkc}~( f^{\prime}_{kcb}~m_{c} + f^{\prime}_{kbc}~m_{b} )~
d^{F}_{kc} ]
\label{l6}
\ee

\be
T_{00}(D_{a}(x_{1}),B_{b}^{\mu}(x_{2})) =
- {1\over 2}~f_{acd}~f^{\prime}_{bcd}~\partial^{\mu}d^{F}_{cd}
\label{l7}
\ee

\be
T_{00}(D_{a}(x_{1}),C_{b}^{\mu\nu}(x_{2}) =
\eta^{\mu\nu}~( f^{\prime}_{cda}~f^{\prime}_{cbd}~m_{b}~m_{d}~d^{F}_{cd} + f^{\prime}_{jca}~f^{\prime}_{jbc}~m_{b}~m_{c}~d^{F}_{jc} )
\label{l8}
\ee

\bea
T_{00}(D_{a}(x_{1}),G_{b}^{\mu}(x_{2})) =
{1\over 2}~( 2~f^{\prime}_{cda}~m_{d} - f^{\prime}_{cad}~m_{a} )~
f^{\prime}_{bcd}~\partial^{\mu}d^{F}_{cd}
\nonumber\\
+ {1\over 2}~(m_{c}^{2} - m_{d}^{2})~f^{\prime}_{cad}~f^{\prime}_{bcd}~
m_{a}~\partial^{\mu}d^{\prime F}_{cd}
- {1\over 2}~f^{\prime}_{jca}~f^{\prime}_{jbc}~m_{c}~
\partial^{\mu}d^{F}_{jc}
\nonumber\\
- {1\over 2}~(m_{j}^{2} - m_{c}^{2})~f^{\prime}_{jca}~f^{\prime}_{jbc}
\partial^{\mu}d^{\prime F}_{jc}
\label{l9}
\eea

\bea
T_{00}(D_{a}(x_{1}),G_{j}^{\mu}(x_{2})) =
{1\over 2}~( 2~f^{\prime}_{cda}~m_{d} - f^{\prime}_{cad}~m_{a} )~
f^{\prime}_{jcd}~\partial^{\mu}d^{F}_{cd}
\nonumber\\
+ {1\over 2}~(m_{c}^{2} - m_{d}^{2})~f^{\prime}_{cad}~f^{\prime}_{jcd}~
m_{a}~\partial^{\mu}d^{\prime F}_{cd}
- {1\over 2}~f^{\prime}_{kca}~f^{\prime}_{kjc}~m_{c}~
\partial^{\mu}d^{F}_{kc}
\nonumber\\
- {1\over 2}~(m_{k}^{2} - m_{c}^{2})~m_{c}~
f^{\prime}_{kca}~f^{\prime}_{kjc}~\partial^{\mu}d^{\prime F}_{kc}
\label{l10}
\eea

\be
T_{00}(D_{a}(x_{1}),H_{b}(x_{2})) =
f^{\prime}_{cda}~f^{\prime}_{bcd}~m_{d}~d^{F}_{cd}
- f^{\prime}_{jca}~f^{\prime}_{jbc}~m_{c}~d^{F}_{jc}
\label{l11}
\ee

\be
T_{00}(D_{a}(x_{1}),H_{j}(x_{2})) =
f^{\prime}_{cda}~f^{\prime}_{jcd}~m_{d}~d^{F}_{cd}
- f^{\prime}_{kca}~f^{\prime}_{kjc}~m_{c}~d^{F}_{kc}
\label{l12}
\ee

\be
T_{00}(D_{a}(x_{1}),K_{b}(x_{2})) =
- f^{\prime}_{cda}~f^{\prime}_{cbd}~m_{b}~m_{d}~d^{F}_{cd}
- f^{\prime}_{jca}~f^{\prime}_{jbc}~m_{b}~m_{c}~d^{F}_{jc}
\label{l13}
\ee

\be
T_{00}(E_{a}^{\mu\nu}(x_{1}),E_{b}^{\rho\sigma}(x_{2})) =
- (\eta^{\mu\rho}~\eta^{\nu\sigma} - \eta^{\mu\sigma}~\eta^{\nu\rho} )~
f_{acd}~f_{bcd}~d^{F}_{cd}
\label{l14}
\ee

\be
T_{00}(E_{a}^{\mu\nu}(x_{1}), H_{b}^{\rho}(x_{2})) = 0
\label{l15}
\ee

\be
T_{00}(E_{a}^{\mu\nu}(x_{1}), H_{j}^{\rho}(x_{2})) = 0
\label{l16}
\ee

\be
T_{00}(E_{a}^{\mu\nu}(x_{1}), G_{b}(x_{2})) = 0
\label{l17}
\ee

\be
T_{00}(E_{a}^{\mu\nu}(x_{1}), G_{j}(x_{2})) = 0
\label{l18}
\ee

\bea
T_{00}(G_{a}(x_{1}), G_{b}(x_{2})) =
[ f^{\prime}_{acd}~f^{\prime}_{bcd}~m_{c}^{2}
\nonumber\\
- 2~( f^{\prime}_{acd}~m_{c} + f^{\prime}_{adc}~m_{d} )~
( f^{\prime}_{bcd}~m_{c} + f^{\prime}_{bdc}~m_{d} )
+ f^{\prime}_{acd}~f^{\prime}_{bdc}~m_{c}~m_{d}]~d^{F}_{cd}
\nonumber\\
+ ( m_{j}^{2}~f^{\prime}_{jac}~f^{\prime}_{jbc}
- f^{\prime\prime}_{jac}~f^{\prime\prime}_{jbc} )~d^{F}_{jc}
- {1\over 2}~f^{\prime\prime}_{jka}~f^{\prime\prime}_{jkb}~d^{F}_{jk}
\nonumber\\
+ 2~[ (s_{a})_{AB}~(s_{b})_{BA}
- (s^{\prime}_{a})_{AB}~(s^{\prime}_{b})_{BA} ]~\square d^{F}_{AB}
\nonumber\\
+ 2~[ (M_{A} + M_{B})^{2}~(s_{a})_{AB}~(s_{b})_{BA}
-  (M_{A} - M_{B})^{2}~(s^{\prime}_{a})_{AB}~(s^{\prime})_{BA} ]~
d^{F}_{AB}
\label{l19}
\eea

\bea
T_{00}(G_{a}(x_{1}), G_{j}(x_{2})) =
[ f^{\prime}_{abc}~f^{\prime}_{jbc}~m_{b}^{2}
\nonumber\\
- 2~( f^{\prime}_{abc}~m_{b} + f^{\prime}_{acb}~m_{c} )~
( f^{\prime}_{jbc}~m_{b} + f^{\prime}_{jcb}~m_{c} )
+ f^{\prime}_{abc}~f^{\prime}_{jcb}~m_{b}~m_{c}]~d^{F}_{bc}
\nonumber\\
+ ( - m_{k}^{2}~f^{\prime}_{kab}~f^{\prime}_{jkb}
+ f^{\prime\prime}_{kab}~f^{\prime\prime}_{jkb} )~d^{F}_{kb}
- {1\over 2}~f^{\prime\prime}_{kla}~\lambda_{jkl}~d^{F}_{kl}
\nonumber\\
+ 2~[ (s_{a})_{AB}~(s_{j})_{BA}
- (s^{\prime}_{a})_{AB}~(s^{\prime}_{j})_{BA} ]~\square d^{F}_{AB}
\nonumber\\
+ 2~[ (M_{A} + M_{B})^{2}~(s_{a})_{AB}~(s_{j})_{BA}
-  (M_{A} - M_{B})^{2}~(s^{\prime}_{a})_{AB}~(s^{\prime}_{j})_{BA} ]~
d^{F}_{AB}
\label{l20}
\eea

\bea
T_{00}(G_{j}(x_{1}), G_{k}(x_{2})) =
[ f^{\prime}_{jbc}~f^{\prime}_{kbc}~m_{b}^{2}
\nonumber\\
- 2~( f^{\prime}_{jbc}~m_{b} + f^{\prime}_{jcb}~m_{c} )~
( f^{\prime}_{kbc}~m_{b} + f^{\prime}_{kcb}~m_{c} )
+ f^{\prime}_{jbc}~f^{\prime}_{kcb}~m_{b}~m_{c}
- {1\over 2}~f^{\prime\prime}_{jbc}~f^{\prime\prime}_{kbc} ]~d^{F}_{bc}
\nonumber\\
+ ( m_{l}^{2}~f^{\prime}_{jlb}~f^{\prime}_{klb}
- f^{\prime\prime}_{jlb}~f^{\prime\prime}_{klb} )~d^{F}_{lb}
- {1\over 2}~\lambda_{jlm}~\lambda_{klm}~d^{F}_{lm}
\nonumber\\
+ 2~[ (s_{j})_{AB}~(s_{k})_{BA}
- (s^{\prime}_{j})_{AB}~(s^{\prime}_{k})_{BA} ]~\square d^{F}_{AB}
\nonumber\\
+ 2~[ (M_{A} + M_{B})^{2}~(s_{j})_{AB}~(s_{k})_{BA}
-  (M_{A} - M_{B})^{2}~(s^{\prime}_{j})_{AB}~(s^{\prime}_{k})_{BA} ]~
d^{F}_{AB}
\label{l21}
\eea

\bea
T_{00}(G_{a}(x_{1}), H_{b}^{\mu}(x_{2})) =
{1\over 2}~f^{\prime}_{acd}~f^{\prime}_{cbd}~\partial^{\mu}
[ d^{F}_{cd} + ( m_{c}^{2} - m_{d}^{2} )~d^{\prime F}_{cd} ]
\nonumber\\
- {1\over 2}~f^{\prime}_{jac}~f^{\prime}_{jbc}~\partial^{\mu}
[ d^{F}_{jc} + ( m_{j}^{2} - m_{c}^{2} )~d^{\prime F}_{jc} ]
\label{l22}
\eea

\bea
T_{00}(G_{a}(x_{1}), H_{j}^{\mu}(x_{2})) =
- {1\over 2}~f^{\prime}_{acd}~f^{\prime}_{jcd}~\partial^{\mu}
[ d^{F}_{cd} + ( m_{c}^{2} - m_{d}^{2} )~d^{\prime F}_{cd} ]
\nonumber\\
- {1\over 2}~f^{\prime}_{kac}~f^{\prime}_{kjc}~\partial^{\mu}
[ d^{F}_{kc} + ( m_{k}^{2} - m_{c}^{2} )~d^{\prime F}_{kc} ]
\label{l23}
\eea

\bea
T_{00}(G_{j}(x_{1}), H_{a}^{\mu}(x_{2})) =
{1\over 2}~f^{\prime}_{jbc}~f^{\prime}_{bac}~\partial^{\mu}
[ d^{F}_{bc} + ( m_{b}^{2} - m_{c}^{2} )~d^{\prime F}_{bc} ]
\nonumber\\
+ {1\over 2}~f^{\prime}_{jkc}~f^{\prime}_{kab}~\partial^{\mu}
[ d^{F}_{kb} + ( m_{k}^{2} - m_{b}^{2} )~d^{\prime F}_{kb} ]
\label{l24}
\eea

\bea
T_{00}(G_{j}(x_{1}), H_{k}^{\mu}(x_{2})) =
- {1\over 2}~f^{\prime}_{jbc}~f^{\prime}_{kbc}~\partial^{\mu}
[ d^{F}_{bc} + ( m_{b}^{2} - m_{c}^{2} )~d^{\prime F}_{bc} ]
\nonumber\\
+ {1\over 2}~f^{\prime}_{jlb}~f^{\prime}_{lkb}~\partial^{\mu}
[ d^{F}_{lb} + ( m_{l}^{2} - m_{b}^{2} )~d^{\prime F}_{lb} ]
\label{l25}
\eea

\be
T_{00}(H_{a}(x_{1}), H_{b}^{\nu}(x_{2})) =
\eta^{\mu\nu}~( f^{\prime}_{cad}~f^{\prime}_{cbd}~d^{F}_{cd}
+ f^{\prime}_{jac}~f^{\prime}_{jbc}~d^{F}_{jc} )
\label{l26}
\ee

\be
T_{00}(H_{a}(x_{1}), H_{k}^{\nu}(x_{2})) =
\eta^{\mu\nu}~( - f^{\prime}_{cad}~f^{\prime}_{kcd}~d^{F}_{cd}
+ f^{\prime}_{jac}~f^{\prime}_{jkc}~d^{F}_{jc} )
\label{l27}
\ee

\be
T_{00}(H_{j}(x_{1}), H_{k}^{\nu}(x_{2})) =
\eta^{\mu\nu}~( f^{\prime}_{jcd}~f^{\prime}_{kcd}~d^{F}_{cd}
+ f^{\prime}_{ljc}~f^{\prime}_{lkc}~d^{F}_{lc} ).
\label{l28}
\ee

Using these expressions we can compute the right hand sides of (\ref{g2}). After difficult but direct computations one arrives at

\begin{thm}
The following relations are true:
\bea
s^{\prime}T_{00}(D_{a}(x_{1}), C_{b}^{\nu}(x_{2})) = 0
\nonumber\\
s^{\prime}T_{00}(D_{a}(x_{1}), E_{b}^{\rho\sigma}(x_{2})) = 0
\nonumber\\
s^{\prime}T_{00}(D_{a}(x_{1}), H_{b}^{\mu}(x_{2})) = 0
\nonumber\\
s^{\prime}T_{00}(D_{a}(x_{1}), H_{j}^{\mu}(x_{2})) = 0
\eea
but
\bea
s^{\prime}T_{00}(D_{a}(x_{1}), G_{b}(x_{2})) =
- {3 \over 2~m_{b}}~( m_{c}^{2} - m_{d}^{2} )~f_{acd}~f_{bcd}~
(\square d^{\prime F}_{cd} + d^{F}_{cd} )
\nonumber\\
- {1 \over 2}~( m_{j}^{2} - m_{k}^{2} )~
f^{\prime}_{jka}~f^{\prime}_{jkb}~
(\square d^{\prime F}_{jk} + d^{F}_{jk} )
\nonumber\\
- 2 i~( M_{A}^{2} - M_{B}^{2} )~
[ (M_{A} + M_{B})~(s_{b})_{AB}~(t_{a})_{BA}
+  (M_{A} - M_{B})~(s^{\prime}_{b})_{AB}~(t^{\prime}_{a})_{BA} ]~
\nonumber\\
(\square d^{\prime F}_{AB} + d^{F}_{AB} )
\eea

\bea
s^{\prime}T_{00}(D_{a}(x_{1}), G_{j}(x_{2})) =
- ( m_{b}^{2} - m_{c}^{2} )~\left( 3~f^{\prime}_{jbc}~f_{abc}~m_{b}
+ {1\over 2}~f^{\prime\prime}_{jbc}~f^{\prime}_{bca} \right)
(\square d^{\prime F}_{bc} + d^{F}_{bc} )
\nonumber\\
+ ( m_{b}^{2} - m_{k}^{2} )~
f^{\prime\prime}_{jkb}~f^{\prime}_{kba}~
(\square d^{\prime F}_{kb} + d^{F}_{kb} )
\nonumber\\
- 2 i~( M_{A}^{2} - M_{B}^{2} )~
[ (M_{A} + M_{B})~(s_{j})_{AB}~(t_{a})_{BA}
+  (M_{A} - M_{B})~(s^{\prime}_{j})_{AB}~(t^{\prime}_{a})_{BA} ]~
\nonumber\\
(\square d^{\prime F}_{AB} + d^{F}_{AB} )
\eea

If we use the causal splitting from Section \ref{distributions} - see formula                                                                                                                                        (\ref{dd1})-  the we get zero also in the right hand sides of the previous two relations.
\label{loop-DCE}
Then it follows from Theorem \ref{stt-loop} that we have gauge invariance for the loop contributions:
\be
sT(T^{I}(x_{1})^{(1)}, T^{J}(x_{2})^{(1)})) = 0.
\label{lll}
\ee
\end{thm}

\newpage
\section{Second Order Gauge Invariance. Tree Contributions\label{tree}}

Now we study tree contributions. Similarly with the third term from (\ref{wick}) we have:
\bea
T(T(x_{1})^{(2)}, T(x_{2})^{(2)}) =
: C_{a}^{\mu}(x_{1})~C_{b}^{\nu}(x_{2}):~
T_{00}(v_{a\mu}(x_{1}), v_{b\nu}(x_{2}))
\nonumber\\
+ {1\over 4}~: E_{a}^{\nu\mu}(x_{1})~E_{b}^{\sigma\rho}(x_{2}):~
T_{00}(F_{a\mu\nu}(x_{1}), F_{b\rho\sigma}(x_{2}))
\nonumber\\
+ {1\over 2}~[ : E_{a}^{\mu\nu}(x_{1})~C_{b}^{\rho}(x_{2}):~
T_{00}(F_{a\nu\mu}(x_{1}),v_{b\rho}(x_{2}))
+ ( 1 \longleftrightarrow 2) ]
\nonumber\\
- [ : D_{a}(x_{1})~B_{b}^{\mu}(x_{2}):~
T_{00}(u_{a}(x_{1}), d_{\mu}\tilde{u}_{b}(x_{2}))
+ ( 1 \longleftrightarrow 2)]
\nonumber\\
- [ : D_{a}(x_{1})~K_{b}(x_{2}):~
T_{00}(u_{a}(x_{1}), \tilde{u}_{b}(x_{2}))
+ ( 1 \longleftrightarrow 2) ]
\nonumber\\
+ : G_{a}(x_{1})~G_{b}(x_{2}):~
T_{00}(\Phi_{a}(x_{1}), \Phi_{b}(x_{2}))
\nonumber\\
+ : G_{j}(x_{1})~G_{k}(x_{2}):~
T_{00}(\phi_{j}(x_{1}), \phi_{k}(x_{2}))
\nonumber\\
+ : H_{a}^{\mu}(x_{1})~H_{b}^{\nu}(x_{2}):~
T_{00}(d_{\mu}\Phi_{a}(x_{1}), d_{\nu}\Phi_{b}(x_{2}))
\nonumber\\
+ : H_{j}^{\mu}(x_{1})~H_{k}^{\nu}(x_{2}):~
T_{00}(d_{\mu}\phi_{j}(x_{1}), d_{\nu}\phi_{k}(x_{2}))
\nonumber\\
+ : G_{a}(x_{1})~H_{b}^{\nu}(x_{2}):~
T_{00}(\Phi_{a}(x_{1}), d_{\nu}\Phi_{b}(x_{2}))
+ ( 1 \longleftrightarrow 2)
\nonumber\\
+ : G_{j}(x_{1})~H_{k}^{\nu}(x_{2}):~
T_{00}(\phi_{j}(x_{1}), d_{\nu}\phi_{k}(x_{2}))
+ ( 1 \longleftrightarrow 2)
\nonumber\\
- [ : \bar{V}_{A\alpha}(x_{1})~V_{B\beta}(x_{2}):~
T_{00}(\psi_{A\alpha}(x_{1}), \bar{\psi}_{B\beta}(x_{2}))
+ ( 1 \longleftrightarrow 2) ]
\label{t2-2}
\eea
\bea
T(T^{\mu}(x_{1})^{(2)}, T(x_{2})^{(2)}) =
: C_{a}^{\nu\mu}(x_{1})~C_{b}^{\rho}(x_{2}):~
T_{00}(v_{a\nu}(x_{1}),v_{b\rho}(x_{2}))
\nonumber\\
+ {1 \over 2}~
: C_{a}^{\nu\mu}(x_{1})~E_{b}^{\rho\sigma}(x_{2}):~
T_{00}(v_{a\nu}(x_{1}), F_{b\sigma\rho}(x_{2}))
\nonumber\\
- : B_{a\nu}(x_{1})~C_{b}^{\rho}(x_{2}):~
T_{00}(F_{a}^{\nu\mu}(x_{1}), v_{b\rho}(x_{2}))
\nonumber\\
- {1 \over 2}~: B_{a\nu}(x_{1})~E_{b}^{\rho\sigma}(x_{2}):~
T_{00}(F_{a}^{\nu\mu}(x_{1}), F_{b\sigma\rho}(x_{2}))
\nonumber\\
- : D_{a}^{\mu}(x_{1})~B_{b}^{\nu}(x_{2}):~
T_{00}(u_{a}(x_{1}), d_{\nu}\tilde{u}_{b}(x_{2}))
\nonumber\\
- : D_{a}^{\mu}(x_{1})~K_{b}(x_{2}):~
T_{00}(u_{a}(x_{1}), \tilde{u}_{b}(x_{2}))
\nonumber\\
- : B_{a}(x_{1})~D_{b}(x_{2}):~
T_{00}(d^{\mu}\tilde{u}_{a}(x_{1}), u_{b}(x_{2}))
\nonumber\\
+ : G_{a}^{\mu}(x_{1})~G_{b}(x_{2}):~
T_{00}(\Phi_{a}(x_{1}), \Phi_{b}(x_{2}))
\nonumber\\
+ : G_{a}^{\mu}(x_{1})~H_{b}^{\nu}(x_{2}):~
T_{00}(\Phi_{a}(x_{1}), d_{\nu}\Phi_{b}(x_{2}))
\nonumber\\
+ : H_{a}^{\nu,\mu}(x_{1})~G_{b}(x_{2}):~
T_{00}(d_{\nu}\Phi_{a}(x_{1}), \Phi_{b}(x_{2}))
\nonumber\\
+ : H_{a}^{\nu,\mu}(x_{1})~H_{b}^{\rho}(x_{2}):~
T_{00}(d_{\nu}\Phi_{a}(x_{1}), d_{\rho}\Phi_{b}(x_{2}))
\nonumber\\
+ : G_{j}^{\mu}(x_{1})~G_{k}(x_{2}):~
T_{00}(\phi_{j}(x_{1}), \phi_{k}(x_{2}))
\nonumber\\
+ : G_{j}^{\mu}(x_{1})~H_{k}^{\nu}(x_{2}):~
T_{00}(\phi_{j}(x_{1}), d_{\nu}\phi_{k}(x_{2}))
\nonumber\\
+ : H_{j}^{\nu,\mu}(x_{1})~G_{k}(x_{2}):~
T_{00}(d_{\nu}\phi_{j}(x_{1}), \phi_{k}(x_{2}))
\nonumber\\
+ : H_{j}^{\nu,\mu}(x_{1})~H_{k}^{\rho}(x_{2}):~
T_{00}(d_{\nu}\phi_{j}(x_{1}), d_{\rho}\phi_{k}(x_{2}))
\nonumber\\
+ : \bar{V}_{A\alpha}^{\mu}(x_{1})~V_{B\beta}(x_{2}):~
T_{00}(\psi_{A\alpha}(x_{1}), \bar{\psi}_{B\beta}(x_{2}))
\nonumber\\
+ : V_{A\alpha}^{\mu}(x_{1})~\bar{V}_{B\beta}(x_{2}):~
T_{00}(\bar{\psi}_{A\alpha}(x_{1}), \psi_{B\beta}(x_{2}))
\label{tmu2-2}
\eea
\bea
T(T^{\mu}(x_{1})^{(2)}, T^{\nu}(x_{2})^{(2)}) =
: C_{a}^{\rho\mu}(x_{1})~C_{b}^{\sigma\nu}(x_{2}):~
T_{00}(v_{a\rho}(x_{1}),v_{b\sigma}(x_{2}))
\nonumber\\
- : C_{a}^{\rho\mu}(x_{1})~B_{b\sigma}(x_{2}):~
T_{00}(v_{a\rho}(x_{1}), F_{b}^{\sigma\nu}(x_{2}))
\nonumber\\
- : B_{a\rho}(x_{1})~C_{b}^{\sigma\nu}(x_{2}):~
T_{00}(F_{a}^{\rho\mu}(x_{1}), v_{b\sigma}(x_{2}))
\nonumber\\
+ : B_{a\rho}(x_{1})~B_{b\sigma}(x_{2}):~
T_{00}(F_{a}^{\rho\mu}(x_{1}), F_{b}^{\sigma\nu}(x_{2}))
\nonumber\\
+ : D_{a}^{\mu}(x_{1})~B_{b}(x_{2}):~
T_{00}(u_{a}(x_{1}), d^{\nu}\tilde{u}_{b}(x_{2}))
\nonumber\\
+ : B_{a}(x_{1})~D_{b}^{\nu}(x_{2}):~
T_{00}(d^{\mu}\tilde{u}_{a}(x_{1}), u_{b}(x_{2}))
\nonumber\\
+ : G_{a}^{\mu}(x_{1})~G_{b}^{\nu}(x_{2}):~
T_{00}(\Phi_{a}(x_{1}), \Phi_{b}(x_{2}))
\nonumber\\
+ : H_{a}^{\rho,\mu}(x_{1})~G_{b}^{\nu}(x_{2}):~
T_{00}(d_{\rho}\Phi_{a}(x_{1}), \Phi_{b}(x_{2}))
\nonumber\\
+ : G_{a}^{\mu}(x_{1})~H_{b}^{\rho,\nu}(x_{2}):~
T_{00}(\Phi_{a}(x_{1}), d_{\rho}\Phi_{b}(x_{2}))
\nonumber\\
+ : H_{a}^{\rho,\mu}(x_{1})~H_{b}^{\sigma,\nu}(x_{2}):~
T_{00}(d_{\rho}\Phi_{a}(x_{1}), d_{\sigma}\Phi_{b}(x_{2}))
\nonumber\\
+ : G_{j}^{\mu}(x_{1})~G_{k}^{\nu}(x_{2}):~
T_{00}(\phi_{j}(x_{1}), \phi_{k}(x_{2}))
\nonumber\\
+ : H_{j}^{\rho,\mu}(x_{1})~G_{k}^{\nu}(x_{2}):~
T_{00}(d_{\rho}\phi_{j}(x_{1}), \phi_{k}(x_{2}))
\nonumber\\
+ : G_{j}^{\mu}(x_{1})~H_{k}^{\rho,\nu}(x_{2}):~
T_{00}(\phi_{j}(x_{1}), d_{\rho}\phi_{k}(x_{2}))
\nonumber\\
+ : H_{j}^{\rho,\mu}(x_{1})~H_{k}^{\sigma,\nu}(x_{2}):~
T_{00}(d_{\rho}\phi_{j}(x_{1}), d_{\sigma}\phi_{k}(x_{2}))
\nonumber\\
- : \bar{V}_{A\alpha}^{\mu}(x_{1})~V_{B\beta}^{\nu}(x_{2}):~
T_{00}(\psi_{A\alpha}(x_{1}), \bar{\psi}_{B\beta}(x_{2}))
\nonumber\\
- : V_{A\alpha}^{\mu}(x_{1})~\bar{V}_{B\beta}^{\nu}(x_{2}):~
T_{00}(\bar{\psi}_{A\alpha}(x_{1}), \psi_{B\beta}(x_{2}))
\label{tmu-nu2-2}
\eea

\bea
T(T^{\mu\nu}(x_{1})^{(2)}, T(x_{2})^{(2)}) =
 : B_{a}(x_{1})~C_{b}^{\rho}(x_{2}):~
T_{00}(F^{\mu\nu}_{a}(x_{1}), v_{b\rho}(x_{2}))
\nonumber\\
+ {1 \over 2}~: B_{a}(x_{1})~E_{b}^{\rho\sigma}(x_{2}):~
T_{00}(F_{a}^{\mu\nu}(x_{1}), F_{b\sigma\rho}(x_{2}))
\nonumber\\
- : D_{a}^{\mu\nu}(x_{1})~B_{b}^{\rho}(x_{2}):~
T_{00}(u_{a}(x_{1}), d_{\rho}\tilde{u}_{b}(x_{2}))
\nonumber\\
- : D_{a}^{\mu\nu}(x_{1})~K_{b}(x_{2}):~
T_{00}(u_{a}(x_{1}), \tilde{u}_{b}(x_{2}))
\label{tmunu2-2}
\eea

\bea
T(T^{\mu\nu}(x_{1})^{(2)}, T^{\rho}(x_{2})^{(2)}) =
 : B_{a}(x_{1})~C_{b}^{\sigma\rho}(x_{2}):~
 T_{00}(F^{\mu\nu}_{a}(x_{1}), v_{b\sigma}(x_{2}))
\nonumber\\
- : B_{a}(x_{1})~B_{b\sigma}(x_{2}):~
T_{00}(F_{a}^{\mu\nu}(x_{1}), F_{b}^{\sigma\rho}(x_{2}))
\nonumber\\
+ : D_{a}^{\mu\nu}(x_{1})~B_{b}(x_{2}):~
T_{00}(u_{a}(x_{1}), d^{\rho}\tilde{u}_{b}(x_{2}))
\label{tmunu-rho2-2}
\eea

\be
T(T^{\mu\nu}(x_{1})^{(2)}, T^{\rho\sigma}(x_{2})^{(2)}) =
 : B_{a}(x_{1})~B_{b}(x_{2}):~
T_{00}(F^{\mu\nu}_{a}(x_{1}), F_{b}^{\sigma\rho}(x_{2})).
\label{tmunu-rhosigma2-2}
\ee

The expressions
$
T_{00}(\xi(x_{1}), \xi^{\prime}(x_{2})
$
from above are pure numerical so they must follow from some causal splitting of the (graded) commutator
$
D(\xi(x_{1}), \xi^{\prime}(x_{2})) =
[ \xi(x_{1}), \xi^{\prime}(x_{2}) ].
$
From
$
D = D^{adv} - D^{ret}
$
we can obtain the corresponding Feynman propagator
$
D^{F}.
$
We will use in the following the canonical splitting (\ref{cp1}) - (\ref{cp4}) and the previous formulas become:
\newpage
\bea
T^{c}(T(x_{1})^{(2)}, T(x_{2})^{(2)}) =
\nonumber\\
i~D^{F}_{a}( x_{1} - x_{2})~
: [ C_{a}^{\mu}(x_{1})~C_{a\mu}(x_{2}) - G_{a}(x_{1})~G_{a}(x_{2})
+ D_{a}(x_{1})~K_{a}(x_{2}) - K_{a}(x_{1})~D_{a}(x_{2}) ]:
\nonumber\\
+ i~\partial_{\mu}D^{F}_{a}( x_{1} - x_{2})~
: [ C_{a\nu}(x_{1})~E_{a}^{\mu\nu}(x_{2})
- E_{a}^{\mu\nu}(x_{1})~C_{a\nu}(x_{2})
\nonumber\\
- D_{a}(x_{1})~B_{a}^{\mu}(x_{2}) - B_{a}^{\mu}(x_{1})~D_{a}(x_{2})
+ G_{a}(x_{1})~H_{a}^{\mu}(x_{2}) - H_{a}^{\mu}(x_{1})~G_{a}(x_{2}) ]:
\nonumber\\
- i~\partial_{\mu}\partial_{\nu}D^{F}_{a}( x_{1} - x_{2})~
:[  E_{a}^{\mu\rho}(x_{1})~{E_{a}^{\nu}}_{\rho}(x_{2})
- H_{a}^{\mu}(x_{1})~H_{a}^{\nu}(x_{2}) ]:
\nonumber\\
- D^{F}_{j}( x_{1} - x_{2})~:G_{j}(x_{1})~G_{j}(x_{2}):
\nonumber\\
- i~\partial_{\mu}D^{F}_{j}( x_{1} - x_{2})~
: [ G_{j}(x_{1})~H_{j}^{\mu}(x_{2})
- H_{j}^{\mu}(x_{1})~G_{j}(x_{2}) ]:
\nonumber\\
+ \partial_{\mu}\partial_{\nu}D^{F}_{j}( x_{1} - x_{2})~
: H_{j}^{\mu}(x_{1})~H_{j}^{\nu}(x_{2}):
\nonumber\\
+ i~:\bar{V}_{A}(x_{1})~S^{F}_{A}( x_{1} - x_{2})~V_{A}(x_{2}):
+ :\bar{V}_{A}(x_{2})~S^{F}_{A}( x_{2} - x_{1})~V_{A}(x_{1}):
\label{t2-2a}
\eea

\bea
T^{c}(T^{\mu}(x_{1})^{(2)}, T(x_{2})^{(2)}) =
i~D^{F}_{a}( x_{1} - x_{2})~
:[ C_{a}^{\nu\mu}(x_{1})~C_{a\nu}(x_{2}):
\nonumber\\
- G_{a}^{\mu}(x_{1})~G_{a}(x_{2}) + D_{a}^{\mu}(x_{1})~K_{a}(x_{2}) ]:
\nonumber\\
+ i~\partial_{\nu}D^{F}_{a}( x_{1} - x_{2})~
:[ - D_{a}^{\mu}(x_{1})~B_{a}^{\nu}(x_{2})
+ C_{a}^{\rho\mu}(x_{1})~{E_{a}^{\nu}}_{\rho}(x_{2})
\nonumber\\
- : B_{a}^{\nu}(x_{1})~C_{a}^{\mu}(x_{2})
+ G_{a}^{\mu}(x_{1})~H_{a}^{\nu}(x_{2})]:
\nonumber\\
+ i~\partial^{\mu}D^{F}_{a}( x_{1} - x_{2})~
:[ B_{a\nu}(x_{1})~C_{a}^{\nu}(x_{2}) -  B_{a}(x_{1})~D_{a}(x_{2})
- H_{a}(x_{1})~G_{a}(x_{2}) ]:
\nonumber\\
+ i~\partial_{\nu}\partial_{\rho}D^{F}_{a}( x_{1} - x_{2})~
:B_{a}^{\nu}(x_{1})~E_{a}^{\mu\rho}(x_{2}):
\nonumber\\
+ i~\partial^{\mu}\partial_{\nu}D^{F}_{a}( x_{1} - x_{2})~
:[ B_{a\rho}(x_{1})~E_{a}^{\nu\rho}(x_{2})
+ H_{a}(x_{1})~H_{a}^{\nu}(x_{2}) ]:
\nonumber\\
+ i~D^{F}_{j}( x_{1} - x_{2})~:G_{j}^{\mu}(x_{1})~G_{j}(x_{2}):
\nonumber\\
+ i~\partial_{\nu}D^{F}_{j}( x_{1} - x_{2})~
:G_{j}^{\mu}(x_{1})~H_{j}^{\nu}(x_{2}):
- i~\partial^{\mu}D^{F}_{j}( x_{1} - x_{2})~
:H_{j}(x_{1})~G_{j}(x_{2}):
\nonumber\\
+ i~\partial^{\mu}\partial_{\nu}D^{F}_{j}( x_{1} - x_{2})~
:H_{j}(x_{1})~H_{j}^{\nu}(x_{2}):
\nonumber\\
+ i~:\bar{V}_{A}^{\mu}(x_{1})~S^{F}_{A}( x_{1} - x_{2})~V_{A}(x_{2}):
- :\bar{V}_{A}(x_{2})~S^{F}_{A}( x_{2} - x_{1})~V_{A}^{\mu}(x_{1}):
\label{tmu2-2a}
\eea

\bea
T^{c}(T^{\mu}(x_{1})^{(2)}, T^{\nu}(x_{2})^{(2)}) =
i~D^{F}_{a}( x_{1} - x_{2})~
:[ C_{a}^{\rho\mu}(x_{1})~{C_{a\rho}}^{\cdot\nu}(x_{2})
- G_{a}^{\mu}(x_{1})~G_{a}^{\nu}(x_{2}) ]:
\nonumber\\
+ i~\partial^{\mu}D^{F}_{a}( x_{1} - x_{2})~
:[ B_{a\rho}(x_{1})~C_{a}^{\rho\nu}(x_{2})
+ B_{a}(x_{1})~D_{a}^{\nu}(x_{2})
- H_{a}(x_{1})~G_{a}^{\nu}(x_{2}) ]:
\nonumber\\
+ i~\partial^{\nu}D^{F}_{a}( x_{1} - x_{2})~
:[ D_{a}^{\mu}(x_{1})~B_{a}(x_{2})
- C_{a}^{\rho\mu}(x_{1})~B_{a\rho}(x_{2})
+ G_{a}^{\mu}(x_{1})~H_{a}(x_{2}) ]:
\nonumber\\
+ i~\partial_{\rho}D^{F}_{a}( x_{1} - x_{2})~
:[ C_{a}^{\nu\mu}(x_{1})~B_{a}^{\rho}(x_{2})
- B_{a}^{\rho}(x_{1})~C_{a}^{\mu\nu}(x_{2}) ]:
\nonumber\\
- i~\eta^{\mu\nu}~
\partial_{\rho}\partial_{\sigma}D^{F}_{a}( x_{1} - x_{2})~
:B_{a}^{\rho}(x_{1})~B_{a}^{\sigma}(x_{2}):
\nonumber\\
- i~\partial^{\mu}\partial^{\nu}D^{F}_{a}( x_{1} - x_{2})~
:[ B_{a\rho}(x_{1})~B_{a}^{\rho}(x_{2})- H_{a}(x_{1})~H_{a}(x_{2}) ]:
\nonumber\\
+ i [ \partial^{\mu}\partial_{\rho}D^{F}_{a}( x_{1} - x_{2})~
:B_{a}^{\nu}(x_{1})~B_{a}^{\rho}(x_{2}):
+  \partial^{\nu}\partial_{\rho}D^{F}_{0}( x_{1} - x_{2})~
:B_{a}^{\rho}(x_{1})~B_{a}^{\mu}(x_{2}): ]
\nonumber\\
- i~D^{F}_{j}( x_{1} - x_{2})~G_{j}^{\mu}(x_{1})~G_{j}^{\nu}(x_{2}):
\nonumber\\
- i~\partial^{\mu}D^{F}_{j}( x_{1} - x_{2})~
 :H_{j}(x_{1})~G_{j}^{\nu}(x_{2}):
+ i~\partial^{\nu}D^{F}_{j}( x_{1} - x_{2})~
:G_{j}^{\mu}(x_{1})~H_{j}(x_{2}):
\nonumber\\
+ i~\partial^{\mu}\partial^{\nu}D^{F}_{j}( x_{1} - x_{2})~
:H_{j}(x_{1})~H_{j}(x_{2}):
\nonumber\\
+ i~:\bar{V}_{A}^{\mu}(x_{1})~S^{F}_{A}( x_{1} - x_{2})~V_{A}^{\nu}(x_{2}):
+ :\bar{V}_{A}^{\nu}(x_{2})~S^{F}_{A}( x_{2} - x_{1})~V_{A}^{\mu}(x_{1}):
\label{tmu-nu2-2a}
\eea

\bea
T^{c}(T^{\mu\nu}(x_{1})^{(2)}, T(x_{2})^{(2)}) =
i~D^{F}_{a}( x_{1} - x_{2})~
:D_{a}^{\mu\nu}(x_{1})~K_{a}(x_{2}):
\nonumber\\
+ i~[ \partial^{\mu}D^{F}_{a}( x_{1} - x_{2})~: B_{a}(x_{1})~C_{a}^{\nu}(x_{2}): - (\mu \leftrightarrow \nu)]
\nonumber\\
- i~\partial_{\rho}D^{F}_{a}( x_{1} - x_{2})~
:D_{a}^{\mu\nu}(x_{1})~B_{a}^{\rho}(x_{2}):
\nonumber\\
- i~[ \partial^{\mu}\partial_{\rho}D^{F}_{a}( x_{1} - x_{2})~
: B_{a}(x_{1})~E_{a}^{\nu\rho}(x_{2}): - (\mu \leftrightarrow \nu) ]
\label{tmunu2-2a}
\eea

\bea
T^{c}(T^{\mu\nu}(x_{1})^{(2)}, T^{\rho}(x_{2})^{(2)}) =
 i~[ \partial^{\mu}D^{F}_{a}( x_{1} - x_{2})~
:B_{a}(x_{1})~C_{a}^{\nu\rho}(x_{2}): - (\mu \leftrightarrow \nu) ]
\nonumber\\
+ i~\partial^{\rho}D^{F}_{0}( x_{1} - x_{2})~
:D_{a}^{\mu\nu}(x_{1})~B_{a}(x_{2}):
\nonumber\\
- i [ \partial^{\mu}\partial^{\rho}D^{F}_{0}( x_{1} - x_{2})~: B_{a}(x_{1})~B_{a}^{\nu}(x_{2}): - (\mu \leftrightarrow \nu )]
\nonumber\\
- i [ \eta^{\mu\rho}~ \partial^{\nu}\partial_{\sigma}D^{F}_{a}( x_{1} - x_{2})~: B_{a}(x_{1})~B_{a}^{\sigma}(x_{2}):
- (\mu \leftrightarrow \nu) ]
\label{tmunu-rho2-2a}
\eea

\bea
T^{c}(T^{\mu\nu}(x_{1})^{(2)}, T^{\rho\sigma}(x_{2})^{(2)}) =
- i~( \eta^{\nu\sigma} \partial^{\mu}\partial^{\rho} - \eta^{\nu\rho} \partial^{\mu}\partial^{\sigma} 
+ \eta^{\mu\rho} \partial^{\nu}\partial^{\sigma} - \eta^{\mu\sigma} \partial^{\nu}\partial^{\rho})D^{F}_{0}( x_{1} - x_{2})
\nonumber\\
: B_{a}(x_{1})~B_{a}(x_{2}): 
\label{tmunu-rhosigma2-2a}
\eea

From these formulas we can determine now if gauge invariance is true; in fact, we have anomalies, as it is well known:
\begin{thm}
The following formulas are true
\be
sT^{c}(T^{I}(x_{1})^{(2)}, T^{J}(x_{2})^{(2)}) =
\delta( x_{1} - x_{2})~{\cal A}^{c}(T^{I},T^{J})(x_{2})
+ \partial_{\lambda}\delta( x_{1} - x_{2})~
{\cal A}^{c,\lambda}(T^{I},T^{J})(x_{2})
\ee
where
\bea
{\cal A}^{c}(T,T) =
2~( B_{a\mu}~C_{a}^{\mu} - B_{a}~D_{a} - H_{a}~G_{a} - H_{j}~G_{j})
\nonumber\\
- \partial_{\mu}B_{a\nu}~E_{a}^{\mu\nu}
+ \partial_{\mu}E_{a}^{\mu\nu}~B_{a\nu}
- \partial_{\mu}H_{a}~H_{a}^{\mu} + \partial_{\mu}H_{a}^{\mu}~H_{a}
- \partial_{\mu}H_{j}~H_{j}^{\mu} + \partial_{\mu}H_{j}^{\mu}~H_{j}
+ {\cal A}_{\rm Dirac}
\nonumber\\
{\cal A}^{c,\lambda}(T,T) = 0
\label{st2-2a}
\eea

\bea
{\cal A}^{c}(T^{\mu},T) =
- B_{a}~C_{a}^{\mu} + D_{a}^{\mu}~B_{a} - C_{a}^{\nu\mu}~B_{a\nu}
+ G_{a}^{\mu}~H_{a} + G_{j}^{\mu}~H_{j}
- \partial_{\nu}B_{a}~E_{a}^{\mu\nu} + \partial_{\nu}B_{a}^{\nu}~B_{a}^{\mu}
\nonumber\\
+ \partial^{\mu}B_{a\nu}~B_{a}^{\nu} - \partial^{\mu}H_{a}~H_{a}
- \partial^{\mu}H_{j}^{\mu}~H_{j} + {\cal A}^{\mu}_{\rm Dirac}
\nonumber\\
{\cal A}^{c,\lambda}(T^{\mu},T) =
B_{a}~E_{a}^{\mu\lambda} + B_{a}^{\lambda}~B_{a}^{\mu}
\label{stmu2-2a}
\eea

\bea
{\cal A}^{c}(T^{\mu},T^{\nu}) =
- 2~B_{a}~C_{a}^{[\mu\nu]} - B_{a}^{\mu}~\partial^{\nu}B_{a}
- \partial^{\mu}B_{a}^{\nu}~B_{a}
+ \eta^{\mu\nu}~(\partial_{\rho}B_{a}~B_{a}^{\rho}
+ B_{a}~\partial_{\rho}B_{a}^{\rho} )
\nonumber\\
{\cal A}^{c,\lambda}(T^{\mu},T^{\nu}) =
\eta^{\mu\lambda}~A^{\nu} + \eta^{\nu\lambda}~A^{\mu} -
2~\eta^{\mu\nu}~A^{\lambda}
\label{stmu-nu2-2a}
\eea

\bea
{\cal A}^{c}(T^{\mu\nu},T) = - D^{\mu\nu}_{a}~B_{a}
+ B_{a}^{\mu}~\partial^{\nu}B_{a} - B_{a}^{\nu}~\partial^{\mu}B_{a}
\nonumber\\
{\cal A}^{c,\lambda}(T^{\mu\nu},T) =
\eta^{\mu\lambda}~A^{\nu} - \eta^{\nu\lambda}~A^{\mu}
\label{stmunu2-2a}
\eea

\bea
{\cal A}^{c}(T^{\mu\nu},T^{\rho}) =
\eta^{\nu\rho}~B_{a}~\partial^{\mu}B_{a}
- \eta^{\mu\rho}~B_{a}~\partial^{\nu}B_{a}
\nonumber\\
{\cal A}^{c,\lambda}(T^{\mu\nu},T^{\rho}) =
( \eta^{\mu\rho}~\eta^{\nu\lambda} - \eta^{\nu\rho}~\eta^{\mu\lambda} )
B_{a}~B_{a}
\label{stmunu-rho2-2a}
\eea

\bea
{\cal A}^{c}(T^{\mu\nu},T^{\rho\sigma}) = 0
\nonumber\\
{\cal A}^{c,\lambda}(T^{\mu\nu},T^{\rho\sigma}) = 0.
\label{stmunu-rhosigma2-2a}
\eea

Above, we have defined
\be
A^{\lambda} \equiv B_{a}~B_{a}^{\lambda}
\ee

\bea
{\cal A}_{\rm Dirac} \equiv
- 2~i~u_{a}~v_{b}^{\mu}~\bar{\psi} [ t_{a}^{\epsilon}, t_{b}^{\epsilon}] \otimes
\gamma_{\mu}\gamma_{\epsilon} \psi
\nonumber\\
- 2~i~u_{a}~\Phi_{b}~\bar{\psi}
( t_{a}^{- \epsilon}~s_{b}^{\epsilon} - s_{b}^{\epsilon}~t_{a}^{\epsilon}) \otimes \gamma_{\epsilon} \psi
- 2~i~u_{a}~\phi_{j}~\bar{\psi}
( t_{a}^{- \epsilon}~s_{j}^{\epsilon} - s_{j}^{\epsilon}~t_{a}^{\epsilon}) \otimes \gamma_{\epsilon} \psi
\nonumber\\
{\cal A}^{\mu}_{\rm Dirac} \equiv
- i~u_{a}~u_{b}~\bar{\psi} [ t_{a}^{\epsilon}, t_{b}^{\epsilon}] \otimes
\gamma^{\mu}\gamma_{\epsilon} \psi
\eea
and we have skipped the Wick product signs.
\label{sTT}
\end{thm}

The proof is done by direct computations. For some details, see \cite{wick+hopf}.
The expressions above are rather complicated; in particular we wonder if the form of the anomaly can be simplified such that there are no terms with derivatives on the delta function. In fact, this is possible if we redefine the chronological products with quasi-local terms (i.e. trivial coboundaries).

First, we derive some general result about such finite renormalizations.
\begin{prop}
Let us consider finite renormalizations of the type
\be
N(A_{1}(x_{1}),A_{2}(x_{2})) = \delta( x_{1} - x_{2})~N(A_{1},A_{2})(x_{2})
\label{R1}
\ee
where
$
A_{1}, A_{2}
$
are of the form
$
T^{I}
$
and verify the symmetry property
\be
N(A_{1},A_{2}) = (-1)^{|A_{1}||A_{2}|}~N(A_{2},A_{1}).
\ee
Then the corresponding gauge variation is
\bea
sN(T^{I}(x_{1}),T^{J}(x_{2})) \equiv
\nonumber\\
d_{Q}N(T^{I}(x_{1}),T^{J}(x_{2}))
- i~\partial_{\mu}^{1}N(T^{I\mu}(x_{1}),T^{J}(x_{2}))
+ (-1)^{|I|}~\partial_{\mu}^{2}N(T^{I}(x_{1}),T^{J\mu}(x_{2}))
\nonumber\\
= \delta( x_{1} - x_{2})~R(T^{I},T^{J})(x_{2})
+ \partial_{\lambda}\delta( x_{1} - x_{2})~R^{\lambda}(T^{I},T^{J})(x_{2})
\label{R2}
\eea
where
\bea
R(T,T) = d_{Q}N(T,T) - i~\partial_{\mu}N(T^{\mu},T)
\nonumber\\
R^{\lambda}(T,T) = 0.
\eea

\bea
R(T^{\mu},T) = d_{Q}N(T^{\mu},T) - i~\partial_{\nu}N(T^{\mu},T^{\nu})
\nonumber\\
R^{\lambda}(T^{\mu},T) = - i~[ N(T^{\mu\lambda},T) + N(T^{\mu},T^{\lambda}) ]
\eea

\bea
R(T^{\mu},T^{\nu}) = d_{Q}N(T^{\mu},T^{\nu})
+ i~\partial_{\rho}N(T^{\nu\rho},T^{\mu})
\nonumber\\
R^{\lambda}(T^{\mu},T^{\nu}) = - i~[ N(T^{\mu\lambda},T^{\nu}) + N(T^{\nu\lambda},T^{\mu}) ]
\eea

\bea
R(T^{\mu\nu},T) = d_{Q}N(T^{\mu\nu},T) - i~\partial_{\rho}N(T^{\mu\nu},T^{\rho})
\nonumber\\
R^{\lambda}(T^{\mu\nu},T) = i~N(T^{\mu\nu},T^{\lambda})
\eea

\bea
R(T^{\mu\nu},T^{\rho}) = d_{Q}N(T^{\mu\nu},T^{\rho})
- i~\partial_{\sigma}N(T^{\mu\nu},T^{\rho\sigma})
\nonumber\\
R^{\lambda}(T^{\mu\nu},T^{\rho}) = i~N(T^{\mu\nu},T^{\rho\lambda})
\eea

\bea
R(T^{\mu\nu},T^{\rho\sigma}) = d_{Q}N(T^{\mu\nu},T^{\rho\sigma})
\nonumber\\
R^{\lambda}(T^{\mu\nu},T^{\rho\sigma}) = 0.
\eea
\end{prop}

The proof is done by simple computations. Next we choose a specific form for the finite renormalizations.
\begin{prop}
Let us consider the finite renormalizations
\bea
N(d_{\nu}v_{a\mu},d_{\sigma}v_{b\rho}) = {i \over 2} \eta_{\mu\rho}~\eta_{\nu\sigma}~\delta_{ab}
\nonumber\\
N(d_{\mu}\Phi_{a},d_{\nu}\Phi_{b}) = - i~\eta_{\mu\nu}~\delta_{ab}
\nonumber\\
N(d_{\mu}\phi_{j},d_{\nu}\phi_{k}) = - i~\eta_{\mu\nu}~\delta_{jk}.
\label{vv}
\eea
Then the expressions
$
N(A_{1},A_{2})
$
are
\bea
N(T,T) = i~\left({1 \over 2}~E_{a}^{\mu\nu}~E_{a\mu\nu}
- H_{a}^{\mu}~H_{a\mu} - H_{j}^{\mu}~H_{j\mu} \right)
\nonumber\\
N(T^{\mu},T) = N(T, T^{\mu}) = - i~( B_{a\nu}~E_{a}^{\mu\nu}
+ H_{a}~H_{a}^{\mu} + H_{j}~H_{j}^{\mu} )
\nonumber\\
N(T^{\mu},T^{\nu}) = i~B_{a}^{\mu}~B_{a}^{\nu}
\nonumber\\
N(T^{\mu\nu},T) =  - i~B_{a}~E_{a}^{\mu\nu}
\nonumber\\
N(T^{\mu\nu},T^{\rho}) = i~( \eta^{\mu\rho}~B_{a}~B_{a}^{\nu}
-  \eta^{\nu\rho}~B_{a}~B_{a}^{\mu} )
\nonumber\\
N(T^{\mu\nu},T^{\rho\sigma}) =
i~( \eta^{\mu\rho}~\eta^{\nu\sigma} - \eta^{\mu\sigma}~\eta^{\nu\rho} )~
B_{a}~B_{a}.
\label{N-TT}
\eea
\end{prop}
{\bf Proof:}
The proof is again done by simple computations. We note that from the first relation (\ref{vv}) we have:
\be
N(F_{a}^{\mu\nu},F_{b}^{\rho\sigma}) = i~(\eta^{\mu\rho}~\eta^{\nu\sigma} - \eta^{\nu\rho}~\eta^{\mu\sigma})~\delta_{ab}.
\label{N-FF}
\ee
We select from the expressions (\ref{t2-2}) - (\ref{tmunu-rhosigma2-2}) the terms with the factors
$
T_{00}(F_{a\mu\nu}(x_{1}), F_{b\rho\sigma}(x_{2})),
$
$
T_{00}(d_{\mu}\Phi_{a}(x_{1}), d_{\nu}\Phi_{b}(x_{2}))
$
and
$
T_{00}(d_{\mu}\phi_{j}(x_{1}), d_{\nu}\phi_{k}(x_{2})).
$
If we substitute the expressions (\ref{vv}) and (\ref{N-FF}) we obtain finite renormalizations of the type (\ref{R1}); the explicit expressions are in the statement of the Proposition.
$\qed$

If we modify the chronological products (\ref{t2-2}) - (\ref{tmunu-rhosigma2-2})
with the finite renormalizations just described, the form of the anomalies simplifies considerably.

\begin{thm}
Let us define
\be
T^{\rm ren}(T^{I}(x_{1})^{(2)}, T^{J}(x_{2})^{(2)}) \equiv
T^{c}(T^{I}(x_{1})^{(2)}, T^{J}(x_{2})^{(2)}) +
N(T^{I}(x_{1}), T^{J}(x_{2}))
\label{Tren}
\ee
Then
\be
sT^{\rm ren}(T^{I}(x_{1})^{(2)}, T^{J}(x_{2})^{(2)}) =
\delta( x_{1} - x_{2})~{\cal A}(T^{I},T^{J})(x_{2})
\ee
where
\bea
{\cal A}(T,T) =
- 2~( B_{a\mu}~D_{a}^{\mu} + B_{a}~D_{a}) + E_{a}^{\mu\nu}~D_{a\mu\nu}
\nonumber\\
- 2~( H_{a}~G_{a} + H_{j}~G_{j})
+ 2( H_{a}^{\mu}~G_{a\mu} + H_{j}^{\mu}~G_{j\mu} ) + {\cal A}_{\rm Dirac}
\label{st2-2a1}
\eea

\be
{\cal A}(T^{\mu},T) =
2~( B_{a}~D_{a}^{\mu} + B_{a\nu}~D_{a}^{\mu\nu} - H_{a}~G_{a}^{\mu}
- H_{j}~G_{j}^{\mu} ) + {\cal A}^{\mu}_{\rm Dirac}
\label{stmu2-2a1}
\ee

\be
{\cal A}(T^{\mu},T^{\nu}) = 2~B_{a}~D_{a}^{\mu\nu}
\label{stmu-nu2-2a2}
\ee

\be
{\cal A}(T^{\mu\nu},T) = - 2~B_{a}~D^{\mu\nu}_{a}
\label{stmunu2-2a2}
\ee

\be
{\cal A}(T^{I},T^{J}) = 0, \qquad |I| + |J| \geq 3.
\label{stmunu-rho2-2a3}
\ee

Here the expressions
$
{\cal A}_{\rm Dirac}
$
and
$
{\cal A}^{\mu}_{\rm Dirac}
$
are the same as in Theorem \ref{sTT} and we have skip the Wick product signs as there.

The anomalies described above do verify the Wess-Zumino consistency relation
\be
\bar{s}{\cal A}(T^{I}(x_{1}), T^{J}(x_{2})) = 0, \qquad
\bar{s} \equiv d_{Q} + i~\delta
\ee
which is equivalent to
\bea
d_{Q}{\cal A}(T, T) + i~\partial_{\mu}{\cal A}(T^{\mu}, T) = 0
\nonumber\\
d_{Q}{\cal A}(T^{\mu}, T) - i~\partial_{\nu}{\cal A}(T^{\mu}, T^{\nu}) = 0
\nonumber\\
d_{Q}{\cal A}(T^{\mu\nu}, T) + {\cal A}(T^{\mu}, T^{\nu}) = 0
\nonumber\\
d_{Q}{\cal A}(T^{\mu}, T^{\nu}) = 0, \qquad
d_{Q}{\cal A}(T^{\mu\nu}, T) = 0.
\label{WZ}
\eea
\label{sTT1}
\end{thm}

Using the expressions of the Wick submonomials from Section \ref{submonomials} we can derive the explicit form of the three remaining anomalies from the previous theorem. Let us first define the following expressions:
\bea
f_{abcd} \equiv f_{eac}~f_{ebd} + f_{ecb}~f_{ead} + f_{eba}~f_{ecd}
\nonumber\\
f^{\prime}_{abcd} \equiv f_{eab}~f^{\prime}_{cde}
- f^{\prime}_{eca}~f^{\prime}_{edb} + f^{\prime}_{ecb}~f^{\prime}_{eda}
- f^{\prime}_{jca}~f^{\prime}_{jdb} + f^{\prime}_{jcb}~f^{\prime}_{jda}
\nonumber\\
f^{\prime}_{abcj} \equiv f_{eab}~f^{\prime}_{jce}
+ f^{\prime}_{jea}~f^{\prime}_{ecb} - f^{\prime}_{jeb}~f^{\prime}_{eca}
+ f^{\prime}_{jka}~f^{\prime}_{kcb} - f^{\prime}_{jkb}~f^{\prime}_{kca}
\nonumber\\
f^{\prime}_{abjk} \equiv f_{eab}~f^{\prime}_{jke}
- f^{\prime}_{jea}~f^{\prime}_{keb} + f^{\prime}_{jeb}~f^{\prime}_{kea}
- f^{\prime}_{jma}~f^{\prime}_{kmb} + f^{\prime}_{jmb}~f^{\prime}_{kma}
\nonumber\\
g_{abcj} \equiv 2~( f^{\prime}_{eba}~f^{\prime\prime}_{jec}
+ f^{\prime}_{eca}~f^{\prime\prime}_{jeb}
+ f^{\prime}_{kba}~f^{\prime\prime}_{jkc}
+ f^{\prime}_{kca}~f^{\prime\prime}_{jkb}
- f^{\prime}_{jka}~f^{\prime\prime}_{kbc} )
\nonumber\\
g_{abjk} \equiv 2~( f^{\prime}_{cba}~f^{\prime\prime}_{jkc}
- f^{\prime}_{jca}~f^{\prime\prime}_{kcb}
- f^{\prime}_{kca}~f^{\prime\prime}_{jkb}
- f^{\prime}_{jma}~f^{\prime\prime}_{kmb}
- f^{\prime}_{kma}~f^{\prime\prime}_{jmb}
+ f^{\prime}_{mba}~\lambda_{jkm})
\nonumber\\
g_{ajkm} \equiv - 6~{\cal S}_{jkm}( f^{\prime}_{jba}~f^{\prime\prime}_{kmb}
- f^{\prime}_{jna}~f^{\prime\prime}_{kmn} )
\nonumber\\
g_{abcd} \equiv 6~{\cal S}_{bcd}( f^{\prime}_{jca}~f^{\prime\prime}_{jbd} ).
\eea
We remark that we have the following symmetry properties:
\bea
f_{abcd} = f_{[abc]d}, \quad f^{\prime}_{abcd} = f^{\prime}_{[ab][cd]},
\quad
f^{\prime}_{abcj} = f^{\prime}_{[ab]cj}, \quad
f^{\prime}_{abjk} = f^{\prime}_{[ab][jk]},
\nonumber\\
g_{abcj} = g_{a\{bc\}j}, \quad g_{abjk} = g_{ab\{jk\}},
\quad
g_{ajkm} = g_{a\{jkm\}}, \quad g_{abcd} = g_{a\{bcd\}}.
\eea

Then the expressions
$
{\cal A}(^{I}T,T^{J})
$
are explicitly:
\bea
{\cal A}(T,T) = f_{abcd}~( - u_{a}~v_{b}^{\mu}~v_{c}^{\nu}~F_{d\mu\nu}
+ u_{a}~u_{b}~v_{c}^{\mu}~d_{\mu}\tilde{u}_{d} )
\nonumber\\
- 2~f^{\prime}_{abcd}~u_{a}~v_{b}^{\mu}~\Phi_{c}~d_{\mu}\Phi_{d}
- 2~f^{\prime}_{abjk}~u_{a}~v_{b}^{\mu}~\phi_{j}~d_{\mu}\phi_{k}
\nonumber\\
+ 2~f^{\prime}_{abcj}~u_{a}~v_{b}^{\mu}~
( \Phi_{c}~d_{\mu}\phi_{j} - d_{\mu}\Phi_{c}~\phi_{j} )
\nonumber\\
- 2 i~u_{a}~v_{b}^{\mu}~\bar{\psi} ( [ t_{a}^{\epsilon}, t_{b}^{\epsilon} ]
- i~f_{abc}~t_{c}^{\epsilon} ) \otimes \gamma_{\mu}\gamma_{\epsilon} \psi
\nonumber\\
- 2 i~u_{a}~\Phi_{b}~\bar{\psi} ( t_{a}^{- \epsilon}~s_{b}^{\epsilon}
- s_{b}^{\epsilon}~t_{a}^{\epsilon} + i~f^{\prime}_{cba}~s_{c}^{\epsilon}
+ i~f^{\prime}_{jba}~s_{j}^{\epsilon}) \otimes \gamma_{\epsilon} \psi
\nonumber\\
- 2 i~u_{a}~\phi_{j}~\bar{\psi} ( t_{a}^{- \epsilon}~s_{j}^{\epsilon}
- s_{j}^{\epsilon}~t_{a}^{\epsilon} - i~f^{\prime}_{jba}~s_{c}^{\epsilon}
+ i~f^{\prime}_{kja}~s_{k}^{\epsilon}) \otimes \gamma_{\epsilon} \psi
\nonumber\\
+ f^{\prime}_{abcd}~m_{c}~( 2~u_{a}~v_{b}^{\mu}~v_{c\mu}~\Phi_{d}
+ u_{a}~u_{b}~\tilde{u}_{c}~\Phi_{d} )
\nonumber\\
+ f^{\prime}_{abcj}~m_{c}~( 2~u_{a}~v_{b}^{\mu}~v_{c\mu}~\phi_{j}
+ u_{a}~u_{b}~\tilde{u}_{c}~\phi_{j} )
\nonumber\\
+ {1\over 6}~g_{abcd}~u_{a}~\Phi_{b}~\Phi_{c}~\Phi_{d}
+ {1\over 2}~g_{abcj}~u_{a}~\Phi_{b}~\Phi_{c}~\phi_{j}
\nonumber\\
+ {1\over 2}~g_{abjk}~u_{a}~\Phi_{b}~\phi_{j}~\phi_{k}
+ {1\over 6}~g_{ajkm}~u_{a}~\phi_{j}~\phi_{k}~\phi_{m}
\eea

\bea
{\cal A}(T^{\mu},T) = - f_{abcd}~\left( u_{a}~u_{b}~v_{c\nu}~F_{d}^{\mu\nu}
+ {1\over 3}~u_{a}~u_{b}~u_{c}~d^{\mu}\tilde{u}_{d} \right)
\nonumber\\
+ f^{\prime}_{abcd}~u_{a}~u_{b}~\Phi_{c}~d^{\mu}\Phi_{d}
- f^{\prime}_{abjk}~u_{a}~u_{b}~\phi_{j}~d^{\mu}\phi_{k}
\nonumber\\
+ f^{\prime}_{abcj}~u_{a}~u_{b}~
( \Phi_{c}~d^{\mu}\phi_{j} - d^{\mu}\Phi_{c}~\phi_{j} )
\nonumber\\
- i~u_{a}~u_{b}~\bar{\psi} ( [ t_{a}^{\epsilon}, t_{b}^{\epsilon} ]
- i~f_{abc}~t_{c}^{\epsilon} ) \otimes \gamma^{\mu}\gamma_{\epsilon} \psi
\nonumber\\
- f^{\prime}_{abcd}~m_{d}~u_{a}~u_{b}~v_{c}^{\mu}~\Phi_{d}
+ f^{\prime}_{abcj}~m_{c}~u_{a}~u_{b}~v_{c}^{\mu}~\phi_{j}
\eea

\be
{\cal A}(T^{\mu},T^{\nu}) = - {\cal A}(T^{\mu\nu},T) =
{1\over 3}~f_{abcd}~u_{a}~u_{b}~u_{c}~F_{d}^{\mu\nu}.
\ee

\newpage
\begin{thm}
The anomalies given in the previous theorem can be eliminated if and only if
the constants appearing in (\ref{Tint}) verify the following constraints:

(i)
\be
f_{abcd} = 0
\ee
i.e the Jacobi identity
\be
f_{eab}~f_{ecd} + f_{ebc}~f_{ead} + f_{eca}~f_{ebd}  = 0
\label{Jacobi}
\ee

(ii)
\be
f^{\prime}_{abcd} = 0, \quad f^{\prime}_{abcj} = 0, \quad f^{\prime}_{abjk} = 0
\ee
which can be written using the compact notation from (\ref{Tint1}) as follows:
we define the matrices
$
T^{\prime}_{a},\quad a \in I_{1} \cup I_{2}
$
according to
\be
(T^{\prime}_{a})_{cd} = - f^{\prime}_{cda},\quad c, d \in I_{2} \cup I_{3}
\ee
and we have the representation property
\be
[ T^{\prime}_{a}, T^{\prime}_{b} ] = f_{abc}~T^{\prime}_{c},\quad a \in I_{1} \cup I_{2}.
\ee

(iii)
\bea
[ t_{a}^{\epsilon}, t_{b}^{\epsilon} ] = i~f_{abc}~t_{c}^{\epsilon}, \quad
a, b \in I_{1} \cup I_{2}
\nonumber\\
t_{a}^{- \epsilon}~s_{b}^{\epsilon} - s_{b}^{\epsilon}~t_{a}^{\epsilon} =
- i~( f^{\prime}_{cba}~s_{c}^{\epsilon}
+ f^{\prime}_{jba}~s_{j}^{\epsilon} ), \quad a \in I_{1} \cup I_{2},
\quad b \in I_{2}
\nonumber\\
t_{a}^{- \epsilon}~s_{j}^{\epsilon} - s_{j}^{\epsilon}~t_{a}^{\epsilon} = i~(f^{\prime}_{jba}~s_{b}^{\epsilon}
+ f^{\prime}_{kja}~s_{k}^{\epsilon} ), \quad a \in I_{1} \cup I_{2}.
\eea
If we use the conventions at the end of Section \ref{ym}
then we can write the previous relations in the compact way:
\be
t_{a}^{- \epsilon}~s_{\beta}^{\epsilon} - s_{\beta}^{\epsilon}~t_{a}^{\epsilon} = i~f^{\prime}_{\beta\gamma a}~s_{\gamma}^{\epsilon}.
\ee

(iv)
\be
g_{abcd} = 0, \quad g_{abcj} = 0, \quad g_{abjk} = 0, \quad a \in I_{1}
\ee
and
\be
{1 \over m_{a}}~g_{abcd} = {1 \over m_{b}}~g_{bacd},
\quad a, b \in I_{1} \cup I_{2}, \quad c, d \in I_{2} \cup I_{3}
\ee
where we have used compact notations. We also must perform the finite renormalization
\be
R_{s}(T,T) \equiv {i \over 24}~h_{\alpha\beta\gamma\delta}~
\Phi_{\alpha}~\Phi_{\beta}~\Phi_{\gamma}~\Phi_{\delta}.
\ee
where we have defined the completely symmetric tensor:
\be
h_{\alpha\beta\gamma\delta} \equiv {1 \over m_{\alpha}}~g_{\alpha\beta\gamma\delta}.
\ee
\label{ren-TT}
\end{thm}
{\bf Proof:} The gauge invariance condition will be true if the anomalies described above are coboundaries
\be
{\cal A}(T^{I}(x_{1}), T^{J}(x_{2})) = sB(T^{I}(x_{1}), T^{J}(x_{2}))
\ee
where
$
B
$
is quasi-local:
\be
B(T^{I}(x_{1}), T^{J}(x_{2})) = \delta(x_{1} - x_{2})~B(T^{I},T^{J})(x_{2})
\ee
with the expressions
$
B(T^{I},T^{J})
$
quadri-linear in the fields and verifying the restrictions
\be
\omega( B(T^{I},T^{J})) \leq 4, \qquad
gh( B(T^{I},T^{J})) = |I| + |J|.
\ee
In particular we must have
\be
{\cal A}(T, T) = d_{Q} B(T,T) - \partial_{\mu}B(T^{\mu},T).
\label{AB}
\ee
We have the generic forms
\be
B(T,T) = \sum B_{j}, \quad B(T^{\mu},T) = \sum B^{\mu}_{j}
\ee
where:
\bea
B_{1} = {1\over 4}~f^{1}_{abcd}~v_{a\mu}~v_{b\nu}~v_{c}^{\mu}~v_{d}^{\nu},
\quad f^{1}_{abcd} = f^{1}_{cbad} = f^{1}_{adcb} = f^{1}_{cdab} = f^{1}_{badc}
\nonumber\\
B_{2} = {1\over 2}~f^{2}_{abcd}~u_{a}~\tilde{u}_{b}~v_{c}^{\mu}~v_{d}^{\nu},
\quad f^{2}_{abcd} = f^{2}_{abdc}
\nonumber\\
B_{3} = {1\over 4}~f^{3}_{abcd}~u_{a}~u_{b}~\tilde{u}_{c}~\tilde{u}_{d},
\quad f^{3}_{abcd} = - f^{3}_{bacd} = - f^{3}_{abdc}
\nonumber\\
B_{4} = {1\over 4}~f^{4}_{abcd}~v_{a\mu}~v_{b\nu}~\Phi_{c}~\Phi_{d},
\quad f^{4}_{abcd} = f^{4}_{bacd} = f^{4}_{abdc}
\nonumber\\
B_{5} = {1\over 2}~f^{5}_{abcd}~u_{a}~\tilde{u}_{b}~\Phi_{c}~\Phi_{d},
\quad f^{5}_{abcd} = f^{5}_{abdc}
\nonumber\\
B_{6} = {1\over 24}~f^{6}_{abcd}~\Phi_{a}~\Phi_{b}~\Phi_{c}~\Phi_{d},
\quad f^{6}_{abcd} = f^{6}_{\{abcd\}}
\nonumber\\
B_{7} = {1\over 2}~f^{4}_{abcj}~v_{a\mu}~v_{b}^{\mu}~\Phi_{c}~\phi_{j},
\quad f^{7}_{abcj} = f^{7}_{bacj}
\nonumber\\
B_{8} = f^{8}_{abcj}~u_{a}~\tilde{u}_{b}~\Phi_{c}~\phi_{j},
\nonumber\\
B_{9} = {1\over 6}~f^{9}_{abcj}~\Phi_{a}~\Phi_{b}~\Phi_{c}~\phi_{j},
\quad f^{9}_{abcj} = f^{9}_{\{abc\}j}
\nonumber\\
B_{10} = {1\over 4!}~f^{10}_{abjk}~v_{a\mu}~v_{b}^{\mu}~\phi_{j}~\phi_{k},
\quad f^{10}_{abjk} = f^{10}_{bajk} = f^{10}_{abkj}
\nonumber\\
B_{11} = {1\over 2}~f^{2}_{abjk}~u_{a}~\tilde{u}_{b}~\phi_{j}~\phi_{k},
\quad f^{11}_{abjk} = f^{11}_{abkj}
\nonumber\\
B_{12} = {1\over 4}~f^{12}_{abjk}~\Phi_{a}~\Phi_{b}~\phi_{j}~\phi_{k},
\quad f^{12}_{abcj} = f^{12}_{bajk} = f^{12}_{abkj}
\nonumber\\
B_{13} = {1\over 6}~f^{13}_{ajkl}~\Phi_{a}~\phi_{j}~\phi_{k}~\phi_{l},
\quad f^{13}_{ajkl} = f^{13}_{a\{jkl\}}
\nonumber\\
B_{14} = {1\over 24}~f^{14}_{jklm}~\phi_{j}~\phi_{k}~\phi_{l}~\phi_{m},
\quad f^{14}_{jklm} = f^{14}_{\{jklm\}}.
\eea
and
\bea
B_{1}^{\mu} = {1\over 2}~g^{1}_{abcd}~u_{a}~v_{b}^{\mu}~v_{c}^{\nu}~v_{d\nu},
\quad g^{1}_{abcd} = g^{1}_{abdc}
\nonumber\\
B_{2}^{\mu} = {1\over 2}~g^{2}_{abcd}~u_{a}~u_{b}~v_{c}^{\mu}~\tilde{u}_{d}^,
\quad g^{2}_{abcd} = - g^{2}_{bacd}
\nonumber\\
B_{3}^{\mu} = {1\over 2}~g^{3}_{abcd}~u_{a}~v_{b}^{\mu}~\Phi_{c}~\Phi_{d},
\quad g^{3}_{abcd} = g^{3}_{abdc}
\nonumber\\
B_{4}^{\mu} = g^{4}_{abcd}~u_{a}~v_{b}^{\mu}~\Phi_{c}~\Phi_{d},
\nonumber\\
B_{5}^{\mu} = {1\over 2}~g^{5}_{abjk}~u_{a}~v_{b}^{\mu}~\phi_{j}~\phi_{k},
\quad g^{5}_{abjk} = g^{5}_{abkj}.
\eea

If we introduce these expressions in (\ref{AB}) we obtain the following system of equations:
\bea
f^{1}_{abcd} - {1\over 2}~g^{1}_{acbd} = 0 \qquad
(d_{\mu}u_{a}~v_{b\nu}~v_{c}^{\mu}~v_{d}^{\nu} )
\nonumber\\
f^{2}_{abcd} - g^{2}_{abcd} = 0 \qquad
(u_{a}~d_{\nu}v_{b}^{\nu}~v_{c}^{\mu}~v_{d\mu} )
\nonumber\\
g^{1}_{abcd} + 2i~f_{abcd} = 0 \qquad
(u_{a}~v_{b}^{\mu}~v_{c}^{\nu}~d_{\mu}v_{d\nu} )
\nonumber\\
f^{2}_{abcd} - g^{2}_{acdb} = 0 \qquad
(u_{a}~\tilde{u}_{b}~d_{\mu}u_{c}~v_{d}^{\mu} )
\nonumber\\
f^{3}_{abcd} + g^{2}_{abcd} = 0 \qquad
(u_{a}~u_{b}~d_{\mu}v_{c}^{\mu}~\tilde{u}_{d} )
\nonumber\\
g^{2}_{abcd} - 2i~f_{abcd} = 0 \qquad
(u_{a}~u_{b}~d_{\mu}\tilde{u}_{c}~v_{d}^{\mu} )
\nonumber\\
f^{4}_{abcd} - g^{3}_{abcd} = 0 \qquad
(d_{\mu}u_{a}~v_{b}^{\mu}~\Phi_{c}~\Phi_{d} )
\nonumber\\
f^{5}_{abcd} - g^{3}_{abcd} = 0 \qquad
(u_{a}~d_{\mu}v_{b}^{\mu}~\Phi_{c}~\Phi_{d} )
\nonumber\\
g^{3}_{abcd} - 2i~f^{\prime}_{abcd} = 0 \qquad
(u_{a}~v_{b}^{\mu}~\Phi_{c}~d_{\mu}\Phi_{d} )
\nonumber\\
f^{7}_{abcj} - g^{4}_{abcj} = 0 \qquad
(d_{\mu}u_{a}~v_{b}^{\mu}~\Phi_{c}~\phi_{j} )
\nonumber\\
f^{8}_{abcj} - g^{4}_{abcj} = 0 \qquad
(u_{a}~d_{\mu}v_{b}^{\mu}~\Phi_{c}~\phi_{j} )
\nonumber\\
g^{4}_{abcd} + 2i~f^{\prime}_{abcj} = 0 \qquad
(u_{a}~v_{b}^{\mu}~d_{\mu}\Phi_{c}~\phi_{j} )
\nonumber\\
g^{4}_{abcd} - 2i~f^{\prime}_{abcj} = 0 \qquad
(u_{a}~v_{b}^{\mu}~\Phi_{c}~d_{\mu}\Phi_{d} )
\nonumber\\
f^{10}_{abjk} - g^{5}_{abjk} = 0 \qquad
(d_{\mu}u_{a}~v_{b}^{\mu}~\phi_{j}~\phi_{k} )
\nonumber\\
f^{11}_{abjk} - g^{5}_{abjk} = 0 \qquad
(u_{a}~d_{\mu}v_{b}^{\mu}~\phi_{j}~\phi_{k} )
\nonumber\\
g^{5}_{abjk} + 2i~f^{\prime}_{abjk} = 0 \qquad
(u_{a}~v_{b}^{\mu}~d_{\mu}\phi_{j}~\phi_{k}^{\mu} )
\nonumber\\
f^{2}_{adbc}~m_{d} + f^{4}_{cbad} = 0 \qquad
(u_{a}~v_{b}^{\mu}~v_{c\mu}~\Phi_{d} )
\nonumber\\
f^{3}_{abdc}~m_{d} + 2~{\cal A}_{ab} ( f^{5}_{acbd}~m_{b} ) = 0 \qquad
(u_{a}~u_{b}~\tilde{u}_{c}~\Phi_{d} )
\nonumber\\
3~{\cal S}_{bcd} ( f^{5}_{abcd}~m_{b} ) + f^{6}_{abcd}~m_{a} + i~g_{abcd} = 0 \qquad
(u_{a}~\Phi_{b}~\Phi_{c}~\Phi_{d} )
\nonumber\\
f^{7}_{bcaj}~m_{a} = 0 \qquad
(u_{a}~v_{b}^{\mu}~v_{c\mu}~\phi_{j} )
\nonumber\\
2~{\cal S}_{bc} ( f^{8}_{abcj}~m_{b} ) + f^{9}_{abcj}~m_{a} +i~g_{abcj} = 0 \qquad
(u_{a}~\Phi_{b}~\Phi_{c}~\phi_{j} )
\nonumber\\
{\cal A}_{ab} ( f^{8}_{acbj}~m_{b} )  = 0 \qquad
(u_{a}~u_{b}~\tilde{u}_{c}~\phi_{j} )
\nonumber\\
f^{11}_{abjk}~m_{b} + f^{12}_{abjk}~m_{a} + i~g_{abjk} = 0 \qquad
(u_{a}~\Phi_{b}~\phi_{j}~\phi_{k} )
\nonumber\\
f^{13}_{ajkl}~m_{a} + i~g_{ajkl} = 0 \qquad
(u_{a}~\phi_{j}~\phi_{k}~\phi_{l} ).
\eea

Now from the third relation of the system we obtain that
$
f_{abcd} = ( c\leftrightarrow d)
$
but on the other hand it is completely anti-symmetric in
$
a,b,c
$
so we immediately derive that
$
f_{abcd} = 0.
$
One easily derives now that
$
g^{j}_{abcd} = 0, \quad j = 1,2, \qquad
f^{j}_{abcd} = 0, \quad j = 1,2,3.
$
From the ninth relation of the system we have that
$
f^{\prime}_{abcd} = ( c\leftrightarrow d)
$
but on the other hand it is anti-symmetric in
$
c, d
$
so we immediately derive that
$
f^{\prime}_{abcd} = 0.
$
It easily follows that
$
g^{3}_{abcd} = 0, \quad
f^{j}_{abcd} = 0, \quad j = 4,5
$
and
$
f^{\prime}_{abcj} = 0.
$
Now we have
$
g^{4}_{abcd} = 0, \quad
f^{j}_{abcd} = 0, \quad j = 7,8.
$
From conflicting symmetry properties we derive as above that
$
f^{\prime}_{abjk} = 0
$
and then we have
$
g^{5}_{abcd} = 0, \quad
f^{j}_{abcd} = 0, \quad j = 10, 11.
$
The remaining equation can be used to determine
$
f^{j}_{abcd} = 0, \quad j = 6, 9, 12, 13
$
leading to the expression from the statement. The Dirac anomaly must be trivially null. From the relations from the statement it follows that we also have
$
{\cal A}(T^{\mu},T) = {\cal A}(T^{\mu},T^{\nu}) = {\cal A}(T^{\mu\nu},T) = 0.
$
$\qed$

\section{Conclusions}
We have used the method of Wick submonomials from \cite{wick+hopf} for the standard model. The new points are: (i) a simple and compact form of the anomaly
in both loop and tree sector; (ii) the elimination of the loop anomaly follows from the causal splitting from Lemma \ref{ddd}.

\end{document}